%% file: SESM.tex
\documentclass[a4paper,11pt]{article}

\usepackage[utf8]{inputenc}
\usepackage[numbers,sort&compress]{natbib}

\usepackage[margin=1.0in]{geometry}
\usepackage{bold-extra} 

\usepackage{subfig} 
\usepackage{graphicx}
\usepackage{booktabs}
\usepackage{amsmath}
\usepackage{hyperref}
\usepackage{placeins}
\usepackage{slashed}
\usepackage[colorinlistoftodos, shadow]{todonotes}
\usepackage{titlesec}
\usepackage{gensymb}
\usepackage{textcomp}
\usepackage{amssymb}
\usepackage{comment}
\usepackage{multirow} 

\usepackage[capitalise]{cleveref}
\crefmultiformat{equation}{Eqs.~(#2#1#3)}%
{ and~(#2#1#3)}{, (#2#1#3)}{, and~(#2#1#3)}
\crefname{chapter}{Chap.}{Chap.}
\crefname{section}{Sect.}{Sects.}
\Crefname{chapter}{Chapter}{Chapters}
\Crefname{section}{Section}{Sections}

\usepackage[nolist]{acronym}

\usepackage{axodraw}
\usepackage{relsize}

\begin{acronym}
\acro{BHS}{Böhm, Hollik ans Spiesberger}
\acro{BSM}{Beyond the Standard Model}
\acro{CKM}{Cabibbo$-$Kobayashi$-$Maskawa}
\acro{CT}{counterterm}
\acro{EW}{electroweak}
\acro{FCNC}{flavour-changing neutral currents}
\acro{HXSWG}{LHC Higgs Cross Section Working Group}
\acro{IR}{Infrared}
\acro{LHC}{Large Hadron Collider}
\acro{LO}{leading order}
\acro{MS}{Minimal Subtraction}
\acro{NLO}{next-to-leading order}
\acro{OS}{on-shell}
\acro{PDG}{Particle Data Group}
\acro{QCD}{quantum chromodynamics}
\acro{RGE}{renormalization group equation}
\acro{SB}{symmetry breaking}
\acro{SM}{Standard Model}
\acro{SESM}{Singlet Extension of the Standard Model}
\acrodefplural{SESM}{Singlet Extensions of the Standard Model}
\acro{THDM}{Two-Higgs-Doublet Model}
\acro{UV}{ultraviolet}
\acro{vev}{vacuum expectation value}
\end{acronym}

\allowdisplaybreaks

\def\citere#1{\mbox{Ref.~\cite{#1}}}
\def\citeres#1{\mbox{Refs.~\cite{#1}}}

\newcommand{\lsim}
{\mathrel{\raisebox{-.3em}{$\stackrel{\displaystyle <}{\sim}$}}}

\def\al{\alpha}
\def\be{\beta}

\def\de{\delta}

\def\la{\lambda}
\def\si{\sigma}

\def\De{\Delta}

\newcommand{\im}{\mathrm{i}}

\newcommand{\GeV}{\unskip\,\mathrm{GeV}}
\newcommand{\MeV}{\unskip\,\mathrm{MeV}}

\def\mathswitch#1{\relax\ifmmode#1\else$#1$\fi}
\def\mathswitchr#1{\relax\ifmmode{\mathrm{#1}}\else$\mathrm{#1}$\fi}
\def\mathswitchit#1{\relax\ifmmode{#1}\else$#1$\fi}

\newcommand{\Pf}{f}
\newcommand{\Pfb}{\bar{\Pf}}
\newcommand{\Pq}{\text{q}}
\newcommand{\Pqb}{\bar{\Pq}}
\newcommand{\PF}{F}
\newcommand{\PFb}{\bar{\PF}}

\newcommand{\Ph}{\text{h}}
\newcommand{\PH}{\text{H}}
\newcommand{\Pu}{\text{u}}
\newcommand{\Pub}{\bar{\Pu}}
\newcommand{\Pd}{\text{d}}
\newcommand{\Pdb}{\bar{\Pd}}
\newcommand{\Ps}{\text{s}}

\newcommand{\Pc}{\text{c}}
\newcommand{\Pcb}{\bar{\Pc}}
\newcommand{\Pt}{\text{t}}

\newcommand{\Pb}{\text{b}}

\newcommand{\Pl}{\text{l}}
\newcommand{\Pnu}{\nu}
\newcommand{\Pnue}{\Pnu_\Pe}
\newcommand{\Pnueb}{\bar{\Pnu}_\Pe}
\newcommand{\Pnumu}{\Pnu_\Pmu}
\newcommand{\Pnumub}{\bar{\Pnu}_\Pmu}
\newcommand{\Pe}{\text{e}}
\newcommand{\Pep}{{\Pe^+}}
\newcommand{\Pem}{{\Pe^-}}
\newcommand{\Pmu}{\mu}
\newcommand{\Pmup}{{\Pmu^+}}
\newcommand{\Pmum}{{\Pmu^-}}
\newcommand{\Ptau}{\tau}
\newcommand{\PV}{\text{V}}

\newcommand{\PW}{\text{W}}
\newcommand{\PZ}{\text{Z}}
\newcommand{\PA}{\text{A}}
\newcommand{\Pg}{\text{g}}
\newcommand{\Ff}{f}
\newcommand{\FF}{F}
\newcommand{\FS}{S}
\newcommand{\Fh}{h}
\newcommand{\FH}{H}
\newcommand{\FV}{V}
\newcommand{\FB}{B}
\newcommand{\FW}{W}
\newcommand{\FZ}{Z}
\newcommand{\FA}{A}

\newcommand{\Mf}{m_\Pf}
\newcommand{\Mfs}{\Mf^2}

\newcommand{\MW}{M_\PW}
\newcommand{\MWs}{\MW^2}

\newcommand{\MZ}{M_\PZ}
\newcommand{\MZs}{\MZ^2}
\newcommand{\MH}{M_\PH}
\newcommand{\Mh}{M_\Ph}

\newcommand{\MHs}{\MH^2}
\newcommand{\Mhs}{\Mh^2}

\newcommand{\me}{m_\Pe}
\newcommand{\mmu}{m_\Pmu}
\newcommand{\mtau}{m_\Ptau}
\newcommand{\mdq}{m_\Pd}
\newcommand{\muq}{m_\Pu}
\newcommand{\msq}{m_\Ps}
\newcommand{\mcq}{m_\Pc}
\newcommand{\mbq}{m_\Pb}
\newcommand{\mtq}{m_\Pt}

\newcommand{\GammaW}{\Gamma_\PW}
\newcommand{\GammaZ}{\Gamma_\PZ}
\newcommand{\Gammah}{\Gamma_\Ph}
\newcommand{\GammaH}{\Gamma_\PH}

\newcommand{\scrs}{\scriptscriptstyle}
\newcommand{\sw}{s_{\IW}}
\newcommand{\cw}{c_{\IW}}
\newcommand{\sa}{s_\alpha}
\newcommand{\ca}{c_\alpha}
\newcommand{\ta}{t_\alpha}
\newcommand{\sws}{\sw^2}
\newcommand{\cws}{\cw^2}
\newcommand{\sas}{\sa^2}
\newcommand{\cas}{\ca^2}
\newcommand{\sta}{s_{2 \alpha}}
\newcommand{\cta}{c_{2 \alpha}}
\newcommand{\tta}{t_{2 \alpha}}

\newcommand{\Gf}{G_\mu}
\newcommand{\alem}{\alpha_\text{em}}
\newcommand{\IW}{{\scrs \text{W}}}
\newcommand{\IY}{{\scrs \text{Y}}}
\newcommand{\thetaw}{\theta_\IW}
\newcommand{\mur}{\mu_\text{r}}
\newcommand{\murs}{\mur^2}

\newcommand{\alphas}{\alpha_\text{s}}

\def\ie{i.e.\ }

\renewcommand{\O}{{\cal O}}
\newcommand{\ME}{{\cal M}}

\newcommand{\rT}{\text{T}}

\renewcommand{\L}{{\cal L}}

\newcommand{\MSbar}{{\overline{\text{MS}}}}
\newcommand{\MSb}{$\overline{\text{MS}}$}

\def\Re{\mathop{\mathrm{Re}}\nolimits}

\def\sgn{\mathop{\mathrm{sgn}}\nolimits}

\newcommand{\BHMa}{BHM200}
\newcommand{\BHMap}{\BHMa$^+$}
\newcommand{\BHMam}{\BHMa$^-$}
\newcommand{\BHMapm}{\BHMa$^\pm$}
\newcommand{\BHMb}{BHM400}
\newcommand{\BHMc}{BHM600}
\newcommand{\BHMd}{BHM800}

\newcommand{\collier}{{\sc Collier}}
\newcommand{\prophecy}{\textsc{Prophecy4f}}
\newcommand{\mathematica}{{\sc Mathematica}}
\newcommand{\feynrules}{{\sc FeynRules}}
\newcommand{\feynarts}{{\sc FeynArts}}
\newcommand{\formcalc}{{\sc FormCalc}}

\newcommand{\fortran}{\textsc{Fortran}}

\newcommand{\lsd}{\lambda_{12}}
\newcommand{\DeUV}{\De_\text{UV}}
\newcommand{\SUtwo}{\text{SU}(2)_\IW}
\newcommand{\Uone}{\text{U}(1)_\IY}

\def\draftdate{\relax}
\def\mda{\relax}
\def\mua{\relax}
\def\mla{\relax}
\def\draft{
\def\thtystars{******************************}
\def\sixtystars{\thtystars\thtystars}
\typeout{}
\typeout{\sixtystars**}
\typeout{* Draft mode!
         For final version remove \protect\draft\space in source file *}
\typeout{\sixtystars**}
\typeout{}
\def\draftdate{\today}
\def\mua{\marginpar[\boldmath\hfil$\uparrow$]%
                   {\boldmath$\uparrow$\hfil}%
                    \typeout{marginpar: $\uparrow$}\ignorespaces}
\def\mda{\marginpar[\boldmath\hfil$\downarrow$]%
                   {\boldmath$\downarrow$\hfil}%
                    \typeout{marginpar: $\downarrow$}\ignorespaces}
\def\mla{\marginpar[\boldmath\hfil$\rightarrow$]%
                   {\boldmath$\leftarrow $\hfil}%
                    \typeout{marginpar: $\leftrightarrow$}\ignorespaces}
\def\Mua{\marginpar[\boldmath\hfil$\Uparrow$]%
                   {\boldmath$\Uparrow$\hfil}%
                    \typeout{marginpar: $\uparrow$}\ignorespaces}
\def\Mda{\marginpar[\boldmath\hfil$\Downarrow$]%
                   {\boldmath$\Downarrow$\hfil}%
                    \typeout{marginpar: $\downarrow$}\ignorespaces}
\def\Mla{\marginpar[\boldmath\hfil$\Rightarrow$]%
                   {\boldmath$\Leftarrow $\hfil}%
                    \typeout{marginpar: $\leftrightarrow$}\ignorespaces}
\def\muua{\marginpar[\boldmath\hfil$\upuparrows$]%
                   {\boldmath$\upuparrows$\hfil}%
                    \typeout{marginpar: $\upuparrows$}\ignorespaces}
\def\mdda{\marginpar[\boldmath\hfil$\downdownarrows$]%
                   {\boldmath$\downdownarrows$\hfil}%
                    \typeout{marginpar: $\downdownarrows$}\ignorespaces}
\def\mlla{\marginpar[\boldmath\hfil$\leftleftarrows$]%
                   {\boldmath$\leftleftarrows $\hfil}%
                    \typeout{marginpar: $\leftleftarrows$}\ignorespaces}                    
\overfullrule 5pt
\oddsidemargin -15mm
\marginparwidth 29mm
}

\numberwithin{equation}{section}

\begin{document}

\thispagestyle{empty}
\def\thefootnote{\fnsymbol{footnote}}
\setcounter{footnote}{1}
\null
\draftdate\hfill FR-PHENO-2017-024
\vfill
\begin{center}
  {\Large {\boldmath\bf {Precision calculations for $\Ph \to \PW\PW/\PZ\PZ \to 4$ fermions 
\\[0.5em]
in a Singlet Extension of the Standard Model with \textsc{Prophecy4f}}
\par} \vskip 2.5em
{\large
{\sc Lukas Altenkamp, Michele Boggia and Stefan Dittmaier  }\\[2ex]
{\normalsize \it 
Albert-Ludwigs-Universit\"at Freiburg, Physikalisches Institut, \\
79104 Freiburg, Germany
}\\[2ex]
}}
\par \vskip 1em
\end{center}\par
\vskip .0cm \vfill {\bf Abstract:} 
\par 
We consider an extension of the Standard Model by a real singlet scalar field
with a $\mathbb{Z}_2$-symmetric Lagrangian and 
spontaneous symmetry breaking with vacuum expectation value for the singlet.
Considering the lighter of the two scalars of the theory to be the $125 \, \GeV$ Higgs particle, we parametrize the scalar sector by the mass of the heavy Higgs boson, 
a mixing angle $\al$, and a scalar Higgs self-coupling $\lsd$.
Taking into account theoretical constraints from perturbativity and vacuum stability, we compute next-to-leading-order electroweak and QCD corrections to the decays $\Ph \to \PW\PW/\PZ\PZ \to 4 \,$fermions of the light Higgs boson for some scenarios proposed in the literature.
We formulate
two renormalization schemes and investigate the conversion of the input parameters between the schemes, 
finding sizeable effects.
Solving the renormalization-group equations for the \MSb{} parameters~$\al$ and~$\lsd$, we observe a significantly reduced scale and scheme dependence in the next-to-leading-order results.
For some scenarios suggested in the literature, 
the total decay width for the process $\Ph \to 4 \Pf$ is computed as a function of the mixing angle and compared to the width of a corresponding
Standard Model Higgs boson, 
revealing deviations below~$10 \%$.
Differential distributions do not show significant distortions by effects beyond the Standard Model.
The calculations are implemented in the Monte Carlo generator \prophecy{}, which is ready for applications in data analyses in the framework of the singlet extension.
\par
\vskip 1cm
\noindent
January 2018
\par
\null
\setcounter{page}{0}
\clearpage
\def\thefootnote{\arabic{footnote}}
\setcounter{footnote}{0}

\section{Introduction}
\label{sec:intro}

The discovery of a Higgs boson~\cite{Aad:2012tfa,Chatrchyan:2012xdj} at Run~1 of the \ac{LHC} was a milestone in the experimental exploration of \ac{EW} interaction.
Precision studies of the Higgs particle are now needed in order to further
explore the nature of the \ac{EW} symmetry breaking mechanism.
Measurements from the ATLAS and CMS collaborations at \ac{LHC} Runs~1 and~2 are compatible, within the current accuracy, with the \ac{SM}, in which the symmetry breaking is modeled by the Higgs mechanism and driven by an $\SUtwo$ scalar doublet in the Lagrangian.
Since there are observed phenomena that cannot be explained within the standard framework, such as the existence of dark matter,
massive neutrinos, and the baryonic asymmetry of the universe, we believe that the \ac{SM} is not the ultimate theory.
The hope is to observe deviations from the \ac{SM} in the next years of data taking at the \ac{LHC}, as experimental uncertainties will decrease with increasing luminosity.
In case deviations will show up, theoretical predictions at the highest possible accuracy will be required for physics \ac{BSM} in order to properly confront predictions with the experimental findings.
If new resonances are observed, precise predictions within \ac{BSM} theories will be necessary in order to find out to which model extensions they might belong.
In case no new particles are found, experimental and theoretical accuracy will help to test the viability of \ac{BSM} theories as well.

Several strategies were proposed to make steps towards the next \ac{SM} (see e.g.\ the reviews~\cite{deFlorian:2016spz,Boggia:2017hyq} and references therein), which includes the study of specific models, the use of effective field theories based on the \ac{SM} gauge group, and simplified models. Among these, Higgs sector extensions are of particular interest, as these can be considered both as complete $\SUtwo \times \Uone$ symmetric models (featuring a non-minimal \ac{EW} symmetry breaking mechanism) and as simplified models (where additional scalars can interact with hypothetical \ac{BSM} sectors).
The simplest way to enlarge the \ac{SM} Higgs sector is by adding a gauge-singlet field, which is neutral under the gauge symmetry of the \ac{SM}.
This extension, despite its simplicity, can provide interesting phenomenology. It was initially proposed by Silveira and Zee to motivate the presence of dark matter~\cite{Silveira:1985rk} and 
introduced---in different variants---in \citeres{Hill:1987ea,Veltman:1989vw} to analyze 
the high-energy and the heavy-Higgs-mass limits as well as the 
(non-)decoupling properties in radiative corrections to the self-interactions of W~bosons.

The key feature of this extension is that the additional field interacts with \ac{SM} matter only through couplings to the $\SUtwo$ Higgs doublet.
The form of the Lagrangian is determined by requiring gauge invariance, renormalizability, 
and optional extra symmetries. Different scenarios can be realized with a zero or a non-vanishing \ac{vev} for the singlet field,
which could be real or complex.
Under certain conditions, the additional scalar provides the ``Higgs portal''
for a dark matter candidate or is a dark-matter candidate itself, 
as discussed in \citeres{Silveira:1985rk,%
McDonald:1993ex,%
Bento:2000ah,%
Burgess:2000yq,%
Davoudiasl:2004be,
Schabinger:2005ei,%
vanderBij:2006ne,%
Patt:2006fw,%
Cerdeno:2006ha,%
Kusenko:2006rh,%
Barger:2007im,%
Barger:2008jx,%
He:2008qm,%
Gonderinger:2009jp,%
Andreas:2010dz,
Mambrini:2011ik,%
He:2011gc}.
On the other hand, the additional singlet might act as initiator of a first-order \ac{EW} phase 
transition~\cite{Profumo:2007wc,Espinosa:2011ax,Barger:2011vm}. 
The interesting phenomenology of 
{\acp{SESM}} influenced search strategies for the 
Higgs boson (and vice versa) already before the Higgs-boson discovery
(see, e.g., \citeres{Datta:1997fx,%
OConnell:2006rsp,%
Englert:2011yb,%
Mambrini:2011ik,%
Djouadi:2011aa}); 
nowadays data from EW precision physics, LHC Higgs measurements, and
dark matter searches lead to strong constraints on 
\acp{SESM}~\cite{Gupta:2011gd,%
Coimbra:2013qq,%
Pruna:2013bma,%
Costa:2014qga,%
Robens:2015gla,%
Falkowski:2015iwa,
Robens:2016xkb,%
Casas:2017jjg}.

We consider the most simple variant of a \ac{SESM} which comprises one real singlet field 
with a $\mathbb{Z}_2$-symmetric Lagrangian in the unbroken phase and assign a non-vanishing \ac{vev} to the gauge singlet.
The non-vanishing \ac{vev} leads to mixing between
the singlet scalar and the Higgs boson contained in the $\SUtwo$ doublet, a feature that is quite generic in
more comprehensive SM extensions, which renders this SESM variant a very useful prototype for a simplified model.
In comparison to that, the \ac{SESM} with a vanishing \ac{vev} is phenomenologically less interesting
and will, therefore, not be considered in this paper.
In the SESM, 
the single CP-even Higgs boson of the \ac{SM} is replaced by two CP-even Higgs bosons.
The \ac{SM} coupling strength is shared by the two Higgs bosons, i.e.\ the Higgs bosons couple with the \ac{SM} strength weighted by the sine or cosine of a mixing angle. The mass of the additional Higgs boson, the Higgs mixing angle, and one coupling factor of the scalar self-interactions parametrize the extended sector.
In our phenomenological study, the lightest of the two Higgs bosons of the theory is considered to be the $125\, \GeV$ resonance observed at the LHC, but our theoretical approach is not restricted to this case.

Our goal is to perform \ac{NLO} computations within the \ac{SESM}, including both \ac{EW} and \ac{QCD} corrections.
To this end, it is necessary to renormalize the theory.
Recently, the renormalization of \ac{SM} extensions has been subject of discussion, 
since very often there is no obvious formulation of \ac{OS} renormalization conditions 
to define \ac{BSM} parameters that are fully based on physical S-matrix elements.
To define these parameters, it is customary to use renormalization conditions 
in the modified Minimal Subtraction (\MSb) scheme.
{While a consistent use of \ac{OS} conditions based on physical S-matrix elements
guarantees a gauge-independent parametrization of physical observables by renormalized
input parameters, making}
use of both \ac{OS} and \MSb{} conditions in the definition of a renormalization scheme can lead to gauge-dependent renormalization constants 
if the (gauge-dependent) tadpoles are not treated properly.
In \citere{Fleischer:1980ub}, Fleischer and Jegerlehner proposed a renormalization scheme to avoid this problem in the \ac{SM}, followed by other approaches such as, e.g., described in \citere{Actis:2006ra}.
Recently, these strategies have been applied to \acp{THDM} in \citeres{Krause:2016oke,Denner:2016etu,Altenkamp:2017ldc,Denner:2017vms}.

Different renormalization schemes for the SESM with a real singlet scalar were already considered in 
{\citeres{Kanemura:2015fra,Bojarski:2015kra,Kanemura:2016lkz,%
Kanemura:2017wtm,Denner:2017vms}.}
Among these proposals there is no convincing scheme that is fully based on
\ac{OS} conditions. Most schemes are based on ad hoc or on \MSb{} conditions,
and many variants still suffer from gauge dependence issues.
In this paper, we build on \ac{OS} conditions as far as possible and take 
\MSb{} conditions for those parameters for which no distinguished \ac{OS} conditions
are available.%
\footnote{In our work we do not consider schemes based on the ``pinch technique'',
as, e.g., suggested in \citere{Kanemura:2017wtm}.
Following the arguments of \citeres{Denner:1994nn,Denner:1994xt}
we consider the ``pinch technique'' just as one of many physically
equivalent choices to fix the gauge arbitrariness in off-shell
quantities (related to the 't~Hooft--Feynman gauge of the
quantum fields in the background-field gauge)
rather than singling out ``its gauge-invariant part'' in any sense.}
We formulate two renormalization schemes for the SESM, using two different ways
to treat tadpole contributions, one of them 
based on the FJ variant~\cite{Fleischer:1980ub},
similar to a proposal made in \citere{Denner:2017vms}.
We analyze, in both cases, the dependence of our NLO
results on the renormalization scale $\mur$, which is due to \MSb{} definitions of the Higgs 
mixing angle and the Higgs self-coupling $\lsd$.
We study the conversion of input parameters between the two schemes and compare the results 
obtained in the two schemes to inspect the perturbative consistency of the 
chosen region for $\mur$.
Parameter conversions between different renormalization schemes,
and the corresponding scheme dependence of NLO results, were not yet discussed for the previously
proposed schemes and their applications. 
In a situation where no distinguished renormalization scheme has yet emerged,
discussions of renormalization scale and renormalization scheme dependences are very important
in applications.

In this work, we compute decay observables for the decays of the $125 \, \GeV$ Higgs boson of 
the \ac{SESM} into four fermions via intermediate (off-shell)~$\PW\PW$ or~$\PZ\PZ$ states.
These processes played a central role in the discovery of the Higgs boson, and are very important channels in Higgs couplings analyses.
\ac{NLO} computations for these processes were performed, including both \ac{EW} and \ac{QCD} corrections, in the \ac{SM} with the Monte Carlo generator \prophecy{}~\cite{Bredenstein:2006ha,Bredenstein:2006nk,Bredenstein:2006rh} (and matched to a QED parton shower in \citere{Boselli:2015aha}), as well as in presence of a fourth fermion generation~\cite{Denner:2011vt} and in the \ac{THDM}~\cite{Denner:2017vms,Altenkamp:2017kxk},
but there are no corresponding results available yet in the SESM.

Using the \mathematica{} package \feynrules{}~\cite{Christensen:2008py,Alloul:2013bka}, we have implemented the Feynman rules for the \ac{SESM} into a \feynarts{}~\cite{Hahn:2000kx} model file, which includes one-loop \acp{CT} for both renormalization schemes.
The model file has been used to compute, with the \mathematica{} package \formcalc{}~\cite{Hahn:1998yk}, the matrix elements for the decays of the light Higgs into four fermions (including \ac{EW} and \ac{QCD} \ac{NLO} corrections) and to produce \fortran{} routines to extend the Monte Carlo program \prophecy{}.
Finally, we have
used the extended \prophecy{} version to compute partial widths and 
to generate differential distributions for a selection of benchmark scenarios, proposed in \citeres{Robens:2016xkb,deFlorian:2016spz}.
Other computations of NLO EW corrections relevant for phenomenology in \acp{SESM} can be found in
{\citeres{Bojarski:2015kra,%
Kanemura:2015fra,%
Kanemura:2016lkz,%
Kanemura:2017wtm,%
Denner:2017vms}.}

The paper is organized as follows. In \cref{sec:SESM}, we describe the Lagrangian of the \ac{SESM} considered in this work and discuss the basic features of the model.
We set up the renormalization procedure introducing renormalization transformations for fields and parameters in \cref{sec:loopLag} and fixing the renormalization constants
in \cref{sec:renorm}.
In \cref{sec:impl}, we briefly describe the capabilities of the Monte Carlo program \prophecy{} and discuss the computation of decay observables for the processes $\Ph \to \PW\PW/\PZ\PZ \to 4\,$fermions.
In \cref{sec:scenarios}, we provide the input parameters and discuss the benchmark scenarios used to produce the numerical results shown in \cref{sec:num}.
In \cref{sec:conclusions} we draw our conclusions. Additional numerical results are reported in the appendices.

\section{Singlet Extension of the Standard Model}
\label{sec:SESM}
%
Singlet extensions of the \ac{SM} add one or more scalar singlet fields to the \ac{SM} Higgs sector. In general, the scalar fields can be complex, but we here consider the simplest variant of a \ac{SESM} with one additional real field. The Lagrangian of the model can be easily obtained by modifying the scalar sector of the \ac{SM}. 
As the extension does not modify strong interactions, we present only the contributions to the total \ac{SESM} Lagrangian that are relevant for the \ac{EW} sector, while the \ac{QCD} part can be taken over from any standard reference, such as \citere{Ellis:1991qj}.
For convenience, we split the \ac{EW} Lagrangian of the model as follows:
\begin{equation}
\label{eq:totLag}
 \L_{\text{SESM}} = \L_{\text{Gauge}} + \L_{\text{Fermion}} + \L_{\text{Higgs}} + \L_{\text{Yukawa}} + \L_{\text{Fix}} + \L_{\text{Ghost}}.
\end{equation}
By gauge symmetry and renormalizability constraints, only small deviations
from the \ac{SM} are allowed: 
{The additional scalar field, which has mass dimension~1
and is neutral under $\SUtwo \times \Uone$ gauge transformations, 
can be coupled to the \ac{SM} fields only through gauge-invariant terms
with mass dimension of at most~3, i.e.\ the singlet scalar can only
couple to $\Phi^\dagger \Phi$, 
with $\Phi$ denoting the \ac{SM}-like Higgs doublet.}
Moreover, the Higgs Lagrangian contains a kinetic term and self-interaction terms of the singlet field in addition to the terms that are already present in the \ac{SM}.
Since the singlet enters only the Higgs Lagrangian $\L_\text{Higgs}$, the gauge, fermionic, 
Yukawa, gauge-fixing, and ghost terms of \cref{eq:totLag} 
only need little adaption from the \ac{SM}. For these contributions we make use of the formulation of \citere{Denner:1991kt}.
In \cref{ssec:higgsLag} we introduce the conventions used for the Higgs Lagrangian, then we briefly discuss the gauge (\cref{ssec:gaugeSec}) and the fermion sectors (\cref{ssec:fermionSec}). Finally, we define our input parameter set in \cref{ssec:inputParam}.

\subsection{Higgs Lagrangian}
\label{ssec:higgsLag}

\subsubsection{Mass spectrum}
In complex \acp{SESM}, the singlet field is supposed to be responsible for the symmetry breaking in a hypothetical hidden sector, where interactions of the hidden sector are governed by an exotic $\text{U}(1)$ 
or even higher-rank gauge 
symmetry~\cite{Schabinger:2005ei,Patt:2006fw,Barger:2008jx,Coimbra:2013qq,Costa:2014qga}. 
Considering our \ac{SESM} as a downgrade of such a more comprehensive theory which still carries salient features of the more complete theory, we enforce a $\mathbb{Z}_2$ symmetry on the Higgs Lagrangian (under sign change of the singlet field).
Moreover, we require a non-vanishing \ac{vev} for the singlet.
The model obtained in this way is very simple, but still phenomenologically interesting, as it involves mixing of the new singlet scalar with the Higgs field of the \ac{SM} scalar sector, which is a generic feature of many \ac{SM} extensions.
The most general renormalizable, gauge- and $\mathbb{Z}_2$-invariant Higgs Lagrangian in presence of one real singlet and one doublet is given by~\cite{Barger:2007im,OConnell:2006rsp}
\begin{equation}
\label{eq:scalarLag}
\begin{split}
\L_\text{Higgs} &= \L_\text{Higgs,kin} - V(\Phi,\sigma),
\\
\L_\text{Higgs,kin} &= (D_\mu \Phi)^\dagger (D^\mu \Phi)
      + \frac{1}{2} (\partial_\mu \sigma) (\partial^\mu \sigma),
\\
V(\Phi,\sigma) &= - \mu_2^2 \Phi^\dagger \Phi
               + \frac{\lambda_2}{4} (\Phi^\dagger \Phi)^2
	       + \lambda_{12} \sigma^2 \Phi^\dagger \Phi
	       - \mu_1^2 \sigma^2
	       + \lambda_1 \sigma^4,
\end{split}
\end{equation}
where $\Phi$ is the complex \ac{SM} scalar doublet with hypercharge $Y_{\IW,\Phi} = 1$, and $\sigma$ is the real singlet field.
Splitting off the \acp{vev} $v_2$ and $v_1$, we parametrize $\Phi$ and $\si$ by
\begin{equation}
\label{eq:doubletSingletDef}
 \Phi =
 \begin{pmatrix}
 \phi^+ \\ \frac{1}{\sqrt{2}} (v_2 + \Fh_2 + \im \chi)
 \end{pmatrix},
 \quad
 \sigma = v_1 + \Fh_1,
\end{equation}
{where $\phi^+$, $\phi^-=(\phi^+)^\dagger$, and $\chi$
denote the would-be Goldstone-boson fields for the $\PW^\pm$ and
$\PZ$~bosons.}
The covariant derivative, which is relevant for the couplings of $\Phi$ to the \ac{EW} gauge bosons, is given by
\begin{equation}
\label{eq:covDerDef}
 D_\mu = \partial_\mu 
{- \im g_2 I_\IW^a \FW_\mu^a} 
+ \im g_1 \frac{Y_\IW}{2} \FB_\mu.
\end{equation}
In \cref{eq:covDerDef}, $g_2$, $I_\IW^a$ ($a = 1,2,3$), and $\FW_\mu^a$ are, respectively, the gauge coupling, the generators, and the gauge fields of the weak isospin $\SUtwo$ group; $g_1$, $Y_\IW$, and $\FB_\mu$ are the gauge coupling, the generator, and gauge field  of the weak hypercharge $\Uone$ group.%
\footnote{The negative sign in the $\SUtwo$ term 
of the covariant derivative $D_\mu$ 
is used by B\"ohm, Hollik and Spiesberger~\cite{Bohm:1986rj}, while a
positive sign is, e.g., 
used by Gunion, Haber, Kane and Dawson~\cite{Gunion:1989we}. We make use of the former by default, but in our implementations both conventions are supported.}
For convenience, in \cref{eq:doubletSingletDef}, we introduce the subscripts $1$ and $2$ to label \acp{vev} and fields for the singlet and the doublet sectors, respectively.
The role of $v_2$ is the same as in the \ac{SM}, since only $\Phi$ couples to the \ac{EW} gauge bosons.
The \ac{vev} of the singlet, $v_1$, can be vanishing or non-vanishing, providing different phenomenology. As stated before, we consider the case $v_1 \neq 0$, which is phenomenologically more interesting. Two non-vanishing \acp{vev} can only arise for $\mu_2^2, \mu_1^2 > 0$ and if the following vacuum stability conditions are fulfilled,
\begin{equation}
\label{eq:vacuumStabCond}
\lambda_2 > 0, \quad
\lambda_1 > 0, \quad \lambda_2 \lambda_1 - \lambda_{12}^2 > 0.
\end{equation}

Expanding the potential $V$ with the decomposition~\eqref{eq:doubletSingletDef} and keeping terms containing the fields $\Fh_2, \Fh_1$ up to second order leads to
\begin{equation}
V_2 = - t_2 \Fh_2 - t_1 \Fh_1 + \frac{1}{2} \begin{pmatrix} \Fh_2, \Fh_1 \end{pmatrix} {\cal M}^2_\text{Higgs} \begin{pmatrix} \Fh_2 \\ \Fh_1 \end{pmatrix},
\end{equation}
with the tadpole parameters
\begin{equation}
\label{eq:tadpolesh1h2}
t_2 =  \frac{v_2}{4} \left( 4 \mu_2^2 - 4 v_1^2 \lsd - v_2^2 \lambda_2 \right),
\qquad
t_1 = v_1 \left( 2 \mu_1^2 - v_2^2 \lsd - 4 v_1^2 \lambda_1 \right),
\end{equation}
and the non-diagonal mass matrix
\begin{equation}
{\cal M}^2_\text{Higgs} =
\begin{pmatrix}
v_1^2 \lsd + \frac{3 v_2^2 \lambda_2}{4} - \mu_2^2 	& 2 v_1 v_2 \lsd \\
2 v_1 v_2 \lsd 						& v_ 2^2 \lsd + 12 v_1^2 \lambda_1 - 2 \mu_1^2
\end{pmatrix}.
\end{equation}
In order to work with fields related to mass eigenstates, we diagonalize the mass matrix by a rotation about an angle $\alpha$ with $-\pi/2 \le \al \le \pi/2$,
\begin{equation}
\label{eq:higgsFieldsRot}
 \begin{pmatrix}
 \Fh \\ \FH
 \end{pmatrix}
 =
 \begin{pmatrix}
 \ca & - \sa \\ \sa & \ca
 \end{pmatrix}
 \begin{pmatrix}
 \Fh_2 \\ \Fh_1
 \end{pmatrix},
\end{equation}
using the abbreviations $\ca \equiv \cos \alpha$ and $\sa \equiv \sin \alpha$.
The potential (up to second power in $\Fh, \FH$) takes the form
\begin{equation}
V_2 = - t_{\Ph} \Fh - t_{\PH} \FH + \frac{1}{2} \begin{pmatrix} \Fh, \FH \end{pmatrix} \widetilde{{\cal M}}^2_\text{Higgs} \begin{pmatrix} \Fh \\ \FH \end{pmatrix},
\end{equation}
with the tadpole parameters $t_\Ph, t_\PH$ for the fields $\Fh$ and $\FH$ given by
\begin{equation}
\label{eq:tadpoleshH}
t_\Ph = \ca t_2 - \sa t_1,
\qquad
t_\PH = \sa t_2 + \ca t_1.
\end{equation}
Diagonalizing $\widetilde{{\cal M}}^2_\text{Higgs}$ fixes the mixing angle by
\begin{equation}
\label{eq:mixingAngle}
\begin{split}
\cos(2 \al) &= \pm \frac{ 16 v_1^2 \la_1 - v_2^2 \la_2 }{\sqrt{\left( 8 v_1 v_2 \lsd \right)^2 + \left( 16 v_1^2 \la_1 - v_2^2 \la_2 \right)^2} },
\\
\sin(2 \al) &= \pm \frac{8 v_1 v_2 \lsd}{\sqrt{\left( 8 v_1 v_2 \lsd \right)^2 + \left( 16 v_1^2 \la_1 - v_2^2 \la_2 \right)^2} },
\end{split}
\end{equation}
with eigenvalues
\begin{equation}
\begin{split}
\Mhs &= \frac{1}{2} \la_2 v_2^2 - 2 \lsd v_1 v_2 \ta,
\\
\MHs &= \frac{1}{2} \la_2 v_2^2 + \frac{2 \lsd v_1 v_2}{\ta},
\end{split}
\end{equation}
where $\ta \equiv \sa / \ca$ is the tangent of the mixing angle.
We choose the upper sign in \eqref{eq:mixingAngle} to enforce the hierarchy $\Mh < \MH$ of the two Higgs-boson masses, i.e.\ we take $0 < \al < \frac{\pi}{2}$ if $\lsd$ is positive or $- \frac{\pi}{2} < \al < 0$ if $\lsd$ is negative.
The squared Higgs-boson masses, which are equal to the eigenvalues of $\widetilde{{\cal M}}^2_\text{Higgs}$, can be written as
\begin{equation}
\label{eq:higgsMasses}
\begin{split}
\Mhs &= \frac{1}{4} v_2^2 \la_2 + 4 v_1^2 \la_1 - \frac{1}{4} \sqrt{\left( 8 v_1 v_2 \lsd \right)^2 + \left( 16 v_1^2 \la_1 - v_2^2 \la_2 \right)^2},
\\
\MHs &= \frac{1}{4} v_2^2 \la_2 + 4 v_1^2 \la_1 + \frac{1}{4} \sqrt{\left( 8 v_1 v_2 \lsd \right)^2 + \left( 16 v_1^2 \la_1 - v_2^2 \la_2 \right)^2}.
\end{split}
\end{equation}

We express the original parameters $\mu_2^2, \mu_1^2, \lambda_2, \lambda_1, \lsd$ in terms of the physical masses $\Mh, \MH$, the mixing angle $\al$, the \ac{vev} $v_2$, and the dimensionless coupling $\lsd$, in order to have five free parameters that are more suited as phenomenological input. Note that the \ac{vev} $v_2$ is fixed by \ac{EW} symmetry breaking (see \cref{ssec:gaugeSec}).
To derive the Lagrangian and the Feynman rules, we express all original parameters of the Higgs potential in terms of the new ones. In detail, we solve the equations $t_2 = t_1 = 0$ (see \cref{eq:tadpolesh1h2}) and Eqs.~\eqref{eq:mixingAngle},~\eqref{eq:higgsMasses} for the original Higgs parameters, resulting in
\begin{equation}
\label{eq:inversionRel}
\begin{split}
v_1 &= \frac{\sta \left( \MHs - \Mhs \right)}{4 v_2 \lsd},
\\
\mu_2^2 &= \frac{1}{2} \left( \cas \Mhs + \sas \MHs \right) + \frac{\sta^2 \left( \MHs - \Mhs \right)^2}{16 \lsd v_2^2},
\\
\mu_1^2 &= \frac{1}{4} \left( \cas \MHs + \sas \Mhs \right) + \frac{1}{2} \lsd v_2^2,
\\
\lambda_2 &= \frac{2 \left(c_{\alpha }^2 \Mhs + s_{\alpha }^2 \MHs \right)}{v_2^2},
\\
\lambda_1 &= \frac{2 \left(c_{\alpha }^2 \MHs + s_{\alpha }^2 \Mhs \right) \lsd^2 v_2^2}{ \sta^2 \left( \MHs - \Mhs \right)^2},
\end{split}
\end{equation}
where the shorthand notations~$s_{n \al} \equiv \sin \left( n \al \right), c_{n \al} \equiv \cos \left( n \al \right)$ are used.
Using the parameterization \eqref{eq:inversionRel}, the vacuum stability conditions \eqref{eq:vacuumStabCond} are automatically fulfilled, while $\mu_2^2 > 0$ and $\mu_1^2 > 0$ lead to a restriction on the parameter space of the theory by
\begin{equation}
\label{eq:vacuumStabCond2}
\lsd > 0 \quad \text{or} \quad
 - \frac{\cas \MHs + \sas \Mhs}{2 v_2^2} < \lsd < - \frac{ \sta^2 (\MHs - \Mhs)^2}{8 v_2^2 (\cas \Mhs + \sas \MHs)}.
\end{equation}

\subsubsection{Couplings}
We insert the expressions 
{\eqref{eq:doubletSingletDef} and \eqref{eq:higgsFieldsRot} as well as \cref{eq:tadpolesh1h2} with $t_2 = t_1 = 0$ into the potential of \cref{eq:scalarLag} obtaining}
\begin{equation}
\label{eq:potScalarCoupl}
\begin{split}
V &=
  \text{const.}
  + \frac{1}{2} \Mhs \Fh^2
  + \frac{1}{2} \MHs \FH^2
  + c_{\Ph \Ph \Ph} \Fh^3
  + c_{\Ph \Ph \PH} \Fh^2 \FH
  + c_{\Ph \PH \PH} \Fh \FH^2
  + c_{\PH \PH \PH} \FH^3
  \\ & \quad
  + c_{\Ph \Ph \Ph \Ph} \Fh^4
  + c_{\Ph \Ph \Ph \PH} \Fh^3 \FH
  + c_{\Ph \Ph \PH \PH} \Fh^2 \FH^2
  + c_{\Ph \PH \PH \PH} \Fh \FH^3
  + c_{\PH \PH \PH \PH} \FH^4
  \\ & \quad
  +  \frac{1}{2} \left[
    c_{\Ph \phi \phi} \Fh
    + c_{\PH \phi \phi} \FH
    + c_{\Ph \Ph \phi \phi} \Fh^2
    + c_{\Ph \PH \phi \phi} \Fh \FH
    + c_{\PH \PH \phi \phi} \FH^2
  \right] \left( 2 \phi^+ \phi^- + \chi^2 \right)
  \\ & \quad
  + \frac{\la_2}{16} \left( 2 \phi^+ \phi^- + \chi^2 \right)^2
  ,
\end{split}
\end{equation}
with the coupling constants given by
\begin{align}
\label{eq:scalarCoupl}
c_{\Ph\Ph\Ph} &= \frac{\ca^3}{4} \la_2 v_2 + \frac{\sta}{2} \lsd \left( \sa v_2 - \ca v_1 \right) - 4 \sa^3 \la_1 v_1,
\notag \\
c_{\Ph\Ph\PH} &= \frac{3 \ca \sta}{8} \la_2 v_2 + \lsd \left[ \sa \left( - 2 \cas + \sas \right) v_2 + \ca \left( \cas - 2 \sas \right) v_1 \right] + 6 \sa \sta \la_1 v_1,
\notag \\
c_{\Ph\PH\PH} &= \frac{3 \sa \sta}{8} \la_2 v_2 + \lsd \left[ \ca \left( \cas - 2 \sas \right) v_2 + \sa \left( 2 \cas - \sas \right) v_1 \right] - 6 \ca \sta \la_1 v_1,
\notag \\
c_{\PH\PH\PH} &= \frac{\sa^3}{4} \la_2 v_2 + \frac{\sta}{2} \lsd \left( \ca v_2 + \sa v_1 \right) + 4 \ca^3 \la_1 v_1,
\notag \\
c_{\Ph\Ph\Ph\Ph} &= \frac{\ca^4}{16} \la_2 + \frac{\sta^2}{8} \lsd + \sa^4 \la_1,
\notag \\
c_{\Ph\Ph\Ph\PH} &= \frac{\cas \sta}{8} \la_2 - \frac{\cta \sta}{2} \lsd - 2 \sas \sta \la_1,
\notag \\
c_{\Ph\Ph\PH\PH} &= \frac{3 \sta^2}{32} \la_2 + \frac{2 \cta^2 - \sta^2}{4} \lsd + \frac{3 \sta^2}{2} \la_1,
\\
c_{\Ph\PH\PH\PH} &= \frac{\sta \sas}{8} \la_2 + \frac{\cta \sta}{2} \lsd - 2 \cas \sta \la_1,
\notag \\
c_{\PH\PH\PH\PH} &= \frac{\sa^4}{16} \la_2 + \frac{\sta^2}{8} \lsd + \ca^4 \la_1,
\notag \\
c_{\Ph \phi \phi} &= \frac{\ca}{2} \la_2 v_2 - 2 \sa \lsd v_1,
\notag \\
c_{\PH \phi \phi} &= \frac{\sa}{2} \la_2 v_2 + 2 \ca \lsd v_1,
\notag \\
c_{\Ph\Ph \phi \phi} &= \frac{\cas}{4} \la_2 + \sas \lsd,
\notag \\
c_{\Ph\PH \phi \phi} &= \frac{\sta}{4} \la_2 - \sta \lsd,
\notag \\
c_{\PH\PH \phi \phi} &= \frac{\sas}{4} \la_2 + \cas \lsd.
\notag
\end{align}
The couplings for the full Higgs Lagrangian~\eqref{eq:scalarLag} can be obtained from the Higgs Lagrangian of the \ac{SM} following a simple procedure: Firstly, all the couplings coming from the \ac{SM} Higgs potential are removed  and replaced with the couplings reported in \cref{eq:potScalarCoupl,eq:scalarCoupl}; then the couplings of the light Higgs $\Fh$ (heavy Higgs $\FH$) to the other fields are obtained by rescaling the \ac{SM} Higgs couplings by a factor of $\ca$ ($\sa$).
Thanks to its simplicity, many tree-level computations in the \ac{SESM} can be easily obtained from the \ac{SM} results via rescaling coupling factors appropriately. However, when considering multi-Higgs processes or loop contributions, some care has to be taken.

\subsection{Gauge sector}
\label{ssec:gaugeSec}
In the total Lagrangian~\eqref{eq:totLag} we use the gauge, gauge-fixing, and ghost Lagrangians of \citere{Denner:1991kt} (with corresponding generalized Higgs couplings in the latter).
Since the singlet field $\sigma$ does not couple to the gauge bosons, the gauge-boson mass terms are generated through the interactions of the \ac{EW} gauge bosons with the \ac{vev} $v_2$ of the Higgs doublet in the same way as in the \ac{SM}.
After a rotation about the \ac{EW} mixing angle $\thetaw$ (quantified here by $\cw \equiv \cos \thetaw$ and $\sw \equiv \sin \thetaw$) of the gauge fields into fields related to mass and charge eigenstates, the usual relation among the \ac{EW} coupling constants,
\begin{equation}
\label{eq:g1g2}
e = g_2 \sw = g_1 \cw,
\end{equation}
ensures the presence of a massless field $\FA_\mu$, the photon, coupling to fermions as in pure QED via the electric unit charge $e$.
To avoid confusion with the mixing angle $\al$, we denote the electromagnetic coupling constant with $\alem = e^2/(4 \pi)$.
Moreover, the \ac{vev} $v_2$ is related to the $\PW$- and $\PZ$-boson masses $\MW$ and $\MZ$ and to the \ac{EW} mixing angle as follows, 
\begin{equation}
\label{eq:gaugeMasses}
\MW = \cw \MZ = \frac{g_2 v_2}{2}.
\end{equation}

\subsection{Fermion sector}
\label{ssec:fermionSec}
The form of the fermion and Yukawa Lagrangians, $\L_\text{Fermion}$ and $\L_\text{Yukawa}$, of \cref{eq:totLag} is the same as in the \ac{SM} (see \citere{Denner:1991kt}).
In contrast to the \ac{SM} case, the Higgs doublet contains a mixture of the fields $\Fh$ and $\FH$ (according to \cref{eq:doubletSingletDef,eq:higgsFieldsRot}, $\Fh_2 = \ca \Fh + \sa \FH$).
Consequently the Yukawa Lagrangian provides two copies of the \ac{SM} Higgs couplings to fermions, one rescaled by a factor $\ca$ for the light field $\Fh$, and another rescaled by $\sa$ for the heavy field $\FH$.
The free parameters of the fermion sector are the elements of the Yukawa matrices, which are related to the fermion masses and to the elements of the \ac{CKM} matrix.
In the \ac{SESM} the \ac{CKM} matrix can be treated exactly as in the \ac{SM} case (see e.g.\ \citere{Denner:1991kt}).

\subsection{Input parameters}
\label{ssec:inputParam}
{The Lagrangian of the gauge and Higgs sector,
$\L_{\text{Gauge}} + \L_{\text{Higgs}}$, contains the seven free
input parameters
\begin{equation}
\label{eq:freeParam}
\left\{
g_1, g_2, \mu_1^2, \mu_2^2, \lambda_1, \lambda_2, \lambda_{12}
\right\}.
\end{equation}
Building on the standard procedure in the SM, we derive 
the gauge couplings $g_1$, $g_2$, and one combination of parameters of the Higgs
potential (viz.\ the vev $v_2$) via the relations \eqref{eq:g1g2}
and \eqref{eq:gaugeMasses} using 
$e$, $\MW$, and $\MZ$ as input parameters, which are fixed by
measured values.%
\footnote{Actually, $e$ is derived from the Fermi 
constant as described below.}
The remaining four parameters of the SESM Higgs sector 
(or better, the remaining independent parameter combinations) 
are derived via Eq.~\eqref{eq:inversionRel}
by taking $\Mh$, $\MH$, the angle $\alpha$, and $\lambda_{12}$ as
input parameters. Since we identify $\Ph$ with the observed
Higgs state, $\Mh$ corresponds to the measured Higgs-boson
mass, while $\MH$, $\alpha$, and $\lambda_{12}$ parametrize the
extension of the Higgs sector without being tightly constrained.
Note that this procedure fixes all parameters contained
in $\L_{\text{Fix}} + \L_{\text{Ghost}}$ as well.

The additional free parameters of the fermionic sector,
contained in $\L_{\text{Fermion}}+\L_{\text{Yukawa}}$, 
all originate from the Yukawa coupling matrices as in the SM
and are derived from the 
fermion masses $\Mf$ and the \ac{CKM} matrix elements~$V_{ij}$
as usual.}
The input parameter set for the theory used in this work is then given by:
\begin{equation}
\label{eq:inputParam}
\left\{
\Mh, \MH, \MW, \MZ, e, \lsd, \alpha, \Mf, V_{ij}
\right\}.
\end{equation}

\section{The counterterm Lagrangian}
\label{sec:loopLag}
%
Moving beyond \ac{LO} by including \ac{NLO} effects, the relations between the parameters in the Lagrangian and observables change and, in order to restore the physical meaning of the parameters, it is necessary to renormalize the theory. From now on, we denote the \textit{bare} quantities introduced in the previous section with a subscript ``$0$'', in order to distinguish them from the \textit{renormalized} quantities defined in this section. 
We split bare parameters into renormalized parts and corresponding renormalization constants additively and split bare fields multiplicatively into renormalized parts and field renormalization constants.
After performing this renormalization transformation in the Lagrangian, all parts containing renormalization constants define the \ac{CT} Lagrangian $\de \L$, which at \ac{NLO} is linear in all renormalization constants.
We use dimensional regularization to treat \ac{UV} divergences.
In the \ac{SESM}, the renormalization of the \ac{QCD} sector is the same as it is in the \ac{SM}, so that we will not describe it. For the \ac{EW} sector, in analogy to \cref{eq:totLag}, the \ac{CT} Lagrangian can be divided into several contributions,
\begin{equation}
\label{eq:ctLag}
\delta \L_{\text{SESM}} =
  \delta \L_{\text{Gauge}}
  + \delta \L_{\text{Fermion}}
  + \delta \L_{\text{Higgs,kin}}
  - \delta V
  + \delta \L_{\text{Yukawa}}
  .
\end{equation}
Among the various components of \cref{eq:ctLag}, we focus on the contributions deriving from the renormalization of $\L_\text{Higgs}$, \ie $\delta \L_{\text{Higgs,kin}}$ and $\delta V$. For the renormalization of the gauge and fermion sectors we proceed as described in \citere{Denner:1991kt}.
Note that in \cref{eq:ctLag} there is no counterpart of $\L_\text{Fix}$ of \cref{eq:totLag}, as the gauge-fixing term is introduced in the Lagrangian after the renormalization. Moreover, for our purpose it is not necessary to compute $\delta \L_\text{Ghost}$, since the ghost \acp{CT} enter only beyond the one-loop level.

\subsection{Counterterms from the Higgs potential}
\label{ssec:CThiggsPot}
In the previous section we have replaced the parameters appearing in the Lagrangian 
\eqref{eq:scalarLag} by the input parameters $\Mh, \MH, \lsd$.
Moreover, we introduced a field rotation about the mixing angle $\alpha$ in order to work with the fields $\Fh$ and $\FH$, which have diagonal propagators in lowest order. 
For the proper treatment of tadpoles at NLO, the bare tadpole terms
$t_{\Ph,0}$ and $t_{\PH,0}$ should be restored in the tree-level relations of the previous section,
most notably in Eqs.~\eqref{eq:inversionRel},
\eqref{eq:potScalarCoupl}, and \eqref{eq:scalarCoupl}.
We perform the following renormalization transformations on the free parameters of the scalar potential,
\begin{align}
\label{eq:renormTrasfParam}
M_{\Ph,0}^2 &= \Mhs + \delta \Mhs,
&
M_{\PH,0}^2 &= \MHs + \delta \MHs,
\notag \\
\la_{12,0} &= \lsd + \delta \lsd,
&
\al_0 &= \al + \delta \al,
\\
t_{\Ph,0} &= t_\Ph + \delta t_\Ph,
&
t_{\PH,0} &= t_\PH + \delta t_\PH,
\notag
\end{align}
while the Higgs fields~$\Fh, \FH$ are renormalized in terms of a matrix transformation,
\begin{equation}
\label{eq:renormTrasfHiggs}
\begin{pmatrix} \Fh_0 \\ \FH_0 \end{pmatrix} =
\begin{pmatrix}
1 + \frac{1}{2} \delta Z_{\Ph\Ph} & \frac{1}{2} \delta Z_{\Ph\PH} \\
\frac{1}{2} \delta Z_{\PH\Ph} & 1 + \frac{1}{2} \delta Z_{\PH\PH} \\
\end{pmatrix}
\begin{pmatrix} \Fh \\ \FH \end{pmatrix}.
\end{equation}
Instead of using the renormalization constant $\de \al$ for the angle $\al$, it is often more handy to use the the following derived renormalization constants for trigonometric functions of $\al$,
\begin{equation}
\begin{split}
c_{\al,0} = \ca + \de \ca,
\qquad
s_{\al,0} = \sa + \de \sa,
\qquad
t_{\al,0} = t_\al + \de t_\al,
\end{split}
\end{equation}
which are related to $\de \al$ as follows,
\begin{equation}
\label{eq:trigFuncCT}
\de \ca = - \sa \de \al,
\qquad
\de \sa = - \frac{\ca}{\sa} \de \ca = \ca \de \al,
\qquad
\de t_\al = t_\al \left( \frac{\de \sa}{\sa} - \frac{\de \ca}{\ca} \right) = \frac{\de \al}{\cas}.
\end{equation}
Formally all renormalization constants count as~$\O(\alem)$ corrections, and terms beyond~$\O(\alem)$ will be dropped.
To determine the \acp{CT} of all couplings, we have to express the renormalization constants of the original parameters in terms of the renormalization constants of the chosen independent input parameters.
Defining renormalization transformations for the original parameters by
\begin{align}
\mu_{2,0}^2 &= \mu_2^2 + \de \mu_2^2, &
\mu_{1,0}^2 &= \mu_1^2 + \de \mu_1^2,
\notag \\
\la_{2,0} &= \la_2 + \de \la_2, &
\la_{1,0} &= \la_1 + \de \la_1,
\\
v_{2,0} &= v_2 + \de v_2, &
v_{1,0} &= v_1 + \de v_1,
\notag \\
t_{2,0} &= t_2 + \de t_2, &
t_{1,0} &= t_1 + \de t_1,
\notag
\end{align}
the renormalization constants, in terms of the renormalization constants defined in \cref{eq:renormTrasfParam}, are given by
\begin{equation}
\label{eq:lagParamCT}
\begin{split}
\de \mu_2^2 &=
{\frac{3 \de t_2}{2 v_2}}
- v_1^2 \de \lsd - 2 \lsd v_1^2 \frac{\de v_2}{v_2} + \frac{1}{2} \left( \sas + \frac{v_1}{v_2} \sta  \right) \de \MHs \\&\quad + \frac{1}{2} \left( \cas - \frac{v_1}{v_2} \sta  \right) \de \Mhs + \lsd v_1 \left( \sta v_2 + 2 \cta v_1 \right) \frac{\de \ta}{\ta},
\\
\de \mu_1^2 &=
{\frac{3 \de t_1}{4 v_1}}
+ \frac{v_2^2}{2} \de \lsd + \lsd v_2 \de v_2 + \frac{\cas}{4} \de \MHs + \frac{\sas}{4} \de \Mhs + \lsd v_2 v_1 \frac{\de \ca}{\sa},
\\
\de \la_2 &=
{\frac{2 \de t_2}{v_2^3}} +
\frac{2}{v_2^2} \left( \cas \de \Mhs + \sas \de \MHs \right) - 2 \la_2 \frac{\de v_2}{v_2} + \frac{8 \lsd v_1 \cas}{v_2} \de \ta,
\\
\de \la_1 &=
{\frac{\de t_1}{8 v_1^3}} +
2 \la_1 \left( \frac{\de \lsd}{\lsd} + \frac{\de v_2}{v_2} \right)
	     - \left( \sas + \frac{2\MHs}{\Mhs-\MHs} \right) \frac{\de \Mhs}{8 v_1^2}
	     \\&\quad
	     - \left(  \cas + \frac{2\Mhs}{\MHs-\Mhs} \right) \frac{\de \MHs}{8 v_1^2}
+ \left( \frac{8 \la_1}{\tta} + \frac{\lsd v_2}{v_1} \right) \frac{\de \ca}{2 \sa},
\\
\frac{\de v_1}{v_1} &= \frac{\de \MHs - \de \Mhs}{\MHs - \Mhs} + \frac{\de \sta}{\sta} - \frac{\de v_2}{v_2} - \frac{\de \lsd}{\lsd},
\end{split}
\end{equation}
where $\de v_2$ is determined in the renormalization of the gauge sector below.
In some places we kept the dependent parameters $v_2, v_1, \la_2$, and $\la_1$, in order to keep the result compact.
Note that to derive \cref{eq:lagParamCT} non-vanishing tadpole contributions in \cref{eq:inversionRel} had to be restored.
However, at \ac{NLO} the relations given in \cref{eq:inversionRel}, which are valid for vanishing tadpoles, can be consistently used in all coefficients of renormalization constants in $\de \L$, independent of any detail of the renormalization of the tadpoles, 
since tadpole terms are of~$\O(\alem)$ or even zero.
The tadpole renormalization constants $\de t_2$ and $\de t_1$ are related to $\de t_\PH$ and $\de t_\Ph$ by
\begin{equation}
\de t_\Ph = \ca \de t_2 - \sa \de t_1,
\qquad
\de t_\PH = \sa \de t_2 + \ca \de t_1.
\end{equation}

Applying the renormalization transformations~\eqref{eq:renormTrasfParam} and~\eqref{eq:renormTrasfHiggs} to the bare Higgs potential~\eqref{eq:scalarLag} and keeping terms linear in the renormalization constants leads to $V + \de V$, where $V$ denotes now the renormalized potential, which has the same analytic form as the bare potential, but contains renormalized quantities (including tadpole parameters) instead of the bare ones.
The \ac{CT} potential $\de V$, which is of $\O(\alem)$, reads
\begin{equation}
\label{eq:CTpotential}
\begin{split}
\de V &=
  - \de t_\Ph \Fh - \de t_\PH \FH
  + \frac{1}{2} \left( \de \Mhs + \Mhs \de Z_{\Ph\Ph} \right) \Fh^2
  + \frac{1}{2} \left( \de \MHs + \MHs \de Z_{\PH\PH} \right) \FH^2
  \\ & \quad
  + \frac{1}{2} \left( \Mhs \de Z_{\Ph\PH} + \MHs \de Z_{\PH\Ph} \right) \Fh \FH
  + \text{interaction terms},
\end{split}
\end{equation}
where the interaction terms can be obtained with the same procedure used to derive the terms linear and bilinear in the fields $\Fh, \FH$. 
Note that the CTs to the scalar self-interaction terms contain explicit tadpole renormalization constants
$\de t_{\Ph,\PH}$ after the renormalization transformation.
As, e.g., discussed in~\citere{Altenkamp:2017ldc}, the \ac{UV}-divergent parts of the field renormalization constants appearing in the kinetic terms of \cref{eq:CTpotential} are not all independent. Indeed, a renormalization transformation for the fields~$\Phi$ and~$\si$
\begin{equation}
\label{eq:renormTrasfh1h2}
\Phi_0 = \Phi \left( 1 + \frac{1}{2} \de Z_\Phi \right),
\qquad
\si_0 = \si \left( 1 + \frac{1}{2} \de Z_\si \right),
\end{equation}
would be sufficient to absorb the \ac{UV} divergences of all Higgs field renormalization constants. In this sense, the field renormalization condition~\eqref{eq:renormTrasfHiggs} is non-minimal. Considering the relation between bare parameters given in \cref{eq:higgsFieldsRot} and applying the renormalization transformations of \cref{eq:renormTrasfParam,eq:renormTrasfHiggs,eq:renormTrasfh1h2}, the \ac{UV}-divergent parts of the Higgs field renormalization constants are related as follows,
\begin{equation}
\label{eq:renConstRelations}
\begin{split}
\left. \de Z_{\Ph\Ph} \right|_\text{UV} &= \cas \left. \de Z_{\Phi} \right|_\text{UV} + \sas \left. \de Z_{\si} \right|_\text{UV},
\\
\left. \de Z_{\PH\PH} \right|_\text{UV} &= \sas \left. \de Z_{\Phi} \right|_\text{UV} + \cas \left. \de Z_{\si} \right|_\text{UV},
\\
\left. \de Z_{\Ph\PH} \right|_\text{UV} &= \sa \ca \left( \left. \de Z_{\Phi} \right|_\text{UV} - \left. \de Z_{\si} \right|_\text{UV} \right) - 2 \left. \de \al \right|_\text{UV},
\\
\left. \de Z_{\PH\Ph} \right|_\text{UV} &= \sa \ca \left( \left. \de Z_{\Phi} \right|_\text{UV} - \left. \de Z_{\si} \right|_\text{UV} \right) + 2 \left. \de \al \right|_\text{UV}.
\end{split}
\end{equation}
We will use some of these expressions to compute \ac{UV}-divergent contributions to specific renormalization constants.
Moreover, these expressions can be used to check internal consistency after the application of the renormalization conditions, which will be discussed in \cref{sec:renorm}.

\subsection{Counterterms from the Higgs kinetic term and from the gauge and fermion sectors}
Starting from the kinetic term $\L_\text{Higgs,kin}$ of the Higgs Lagrangian~\eqref{eq:scalarLag}, written in terms of bare parameters and fields, we derive the corresponding \ac{CT} Lagrangian $\de \L_\text{Higgs,kin}$.
For this purpose, we apply the renormalization transformations given in \cref{ssec:CThiggsPot}, supplemented by the renormalization transformations that are relevant for the gauge sector.
The bare $\PW$- and $\PZ$-boson squared masses $M_{\PW,0}^2$ and $M_{\PZ,0}^2$, and the bare {electric} 
charge $e_0$ are transformed according to
\begin{equation}
\label{eq:renormTrasfGaugeParam}
\begin{split}
 M_{\PW,0}^2 &= \MWs + \delta \MWs, \\
 M_{\PZ,0}^2 &= \MZs + \delta \MZs, \\
 e_0 &= (1 + \delta Z_e) \, e. \\
\end{split}
\end{equation}
{Following the ``complete on-shell renormalization'' of the
gauge sector~\cite{Denner:1991kt}, the}
gauge-boson fields $\FW, \FZ$, and $\FA$ are renormalized by
\begin{equation}
\label{eq:renormTrasfGaugeFields}
\begin{split}
\FW^\pm_0 &= \left(1 + \frac{1}{2} \delta Z_\PW \right) \FW^\pm,
\\
\begin{pmatrix} \FZ_0 \\ \FA_0 \end{pmatrix} &= 
\begin{pmatrix} 1 + \frac{1}{2} \delta Z_{\PZ\PZ} & \frac{1}{2} \delta Z_{\PZ\PA} \\ \frac{1}{2} \delta Z_{\PA\PZ} & 1 + \frac{1}{2} \delta Z_{\PA\PA} \end{pmatrix}
\begin{pmatrix} \FZ \\ \FA \end{pmatrix},
\end{split}
\end{equation}
i.e.\ we apply a matrix-valued renormalization transformation to the photon$-\PZ$ system.
{This has the advantage that no further wave-function
or $\gamma$--Z~mixing corrections need to be applied for external
electroweak gauge bosons $\PW$, $\gamma$, $\PZ$ if the corresponding
field renormalization constants are fixed appropriately, as done 
in \Cref{se:gaugefieldren} below.}
Inserting \cref{eq:renormTrasfParam,eq:renormTrasfHiggs,eq:renormTrasfGaugeParam,eq:renormTrasfGaugeFields} into $\L_\text{Higgs,kin}$ yields the \ac{CT} Lagrangian
\begin{equation}
\label{eq:CTHiggsKin}
\begin{split}
\de \L_\text{Higgs,kin} &=
  \frac{1}{2} \de Z_{\Ph\Ph} \left( \partial_\mu \Fh \right) \left( \partial^\mu \Fh \right)
  + \frac{1}{2} \de Z_{\PH\PH} \left( \partial_\mu \FH \right) \left( \partial^\mu \FH \right)
  \\ & \quad
  + \frac{1}{2} \left( \de Z_{\Ph\PH} + \de Z_{\PH\Ph} \right) \left( \partial_\mu \Fh \right) \left( \partial^\mu \FH \right)
  + {\frac{1}{2} \left( \de \MZs + \MZs \de Z_{\PZ\PZ} \right) \FZ_\mu \FZ^\mu}
  \\ & \quad
  + {\left( \de \MWs + \MWs \de Z_\PW \right) \FW^+_\mu \FW^{-,\mu}
  + \frac{1}{2} \MZs \de Z_{\PZ\PA} \FA_\mu \FZ^\mu}
  \\ & \quad
  + \frac{1}{2 \MZ} \left( \de \MZs + \MZs \de Z_{\PZ\PZ} \right) \PZ_\mu \partial^\mu \chi
  + {\frac{\MZ}{2} \de Z_{\PZ\PA} \FA_\mu \partial^\mu \chi}
  \\ & \quad
  + \frac{\im}{2 \MW} \left( \de \MWs + \MWs \de Z_{\PW} \right) \left( \PW^-_\mu \partial^\mu \phi^+ - \PW^+_\mu \partial^\mu \phi^- \right)
  \\ & \quad
  + \text{interaction terms},
\end{split}
\end{equation}
where, for the sake of brevity, we again do not spell out the interactions explicitly.
It is important to note that the scalar$-$vector mixing terms appearing in \cref{eq:CTHiggsKin}, in contrast to what happens in the bare Lagrangian, are not canceled by the gauge-fixing terms, since the gauge-fixing contribution $\L_\text{Fix}$ is introduced after renormalization, i.e.\ directly in terms of renormalized quantities.

The complete set of renormalization transformations necessary to renormalize the gauge and fermion sectors can be found in \citere{Denner:1991kt}.
For a better bookkeeping, renormalization constants for the sine and the cosine of the weak mixing angle are introduced according to
\begin{equation}
s_{\IW,0} = \sw + \de \sw, \quad
c_{\IW,0} = \cw + \de \cw.
\end{equation}
Since~$\MW = \cw \MZ$ is valid both for bare and renormalized quantities, the renormalization constants $\de \sw$ and $\de \cw$ are related to the $\PW$- and $\PZ$-mass renormalization constants by
\begin{equation}
\delta \cw = \frac{\cw}{2} \left( \frac{\delta \MWs}{\MWs} - \frac{\delta \MZs}{\MZs} \right), \quad
\delta \sw = - \frac{\cw}{\sw} \delta \cw.
\end{equation}
Likewise~$v_2 = 2 \MW/g_2$ implies
\begin{equation}
\frac{\de v_2}{v_2} = \frac{\de \MWs}{2 \MWs} - \de Z_e + \frac{\de \sw}{\sw}.
\end{equation}

In the Yukawa Lagrangian, the Higgs field $\Fh_2$ appearing in the doublet~$\Phi$ is consistently rotated using \cref{eq:higgsFieldsRot} and renormalized by the transformations~\eqref{eq:renormTrasfParam} and~\eqref{eq:renormTrasfHiggs}.
In the transition from the \ac{SM} to the \ac{SESM}, the renormalization of the \ac{CKM} matrix does not change; we refer to \citeres{Denner:1991kt,Denner:2004bm,Kniehl:2009kk} for different formulations.

The fermion fields are renormalized using the simple transformations (in the absence of flavour mixing)
\begin{equation}
\Pf_{i,0}^\si = \left( 1 + \frac{1}{2} \de Z_i^{\Pf,\si} \right) \Pf_i^\si,
\end{equation}
where $\Pf = \Pnu, \Pl, \Pu, \Pd$ identify the fermion 
{type,} $i = 1,2,3$ is the generation index, and $\si = L, R$ identify left- and right-handed fermion fields.
%
\section{Renormalization conditions}
\label{sec:renorm}
To fix the renormalization constants introduced in the previous section we adopt, as far as possible, \ac{OS} renormalization conditions.
On-shell renormalization can be performed for all the parameters that are directly accessible by experiments, such as the masses $\Mh$ and $\MH$ in the Higgs sector, but there is no obvious prescription for the mixing angle $\al$ and the coupling constant $\lsd$.
Our choice is to fix these parameters using $\MSbar$ renormalization conditions, so that the corresponding renormalization constants are proportional to the \ac{UV} divergence
\begin{equation}
 \DeUV
 = \frac{2}{4-D} - \gamma_\text{E} + \ln 4 \pi
 = \frac{1}{\epsilon} - \gamma_\text{E} + \ln 4 \pi,
\end{equation}
where $\gamma_\text{E}$ is the Euler$-$Mascheroni constant and the loop integrals are regularized in $D = 4 - 2 \epsilon$ dimensions.
In this way, we have a mixed \ac{OS}$-\MSbar$ renormalization scheme, and particular care has to be taken in the renormalization of the tadpole constants.

In the following, we describe two renormalization schemes used in our analysis, which differ in the treatment of the tadpoles.
The two schemes have the same \ac{OS} renormalization conditions in common, which are presented in \cref{ssec:OSrenCond}. However, because of the different tadpole handling, there are differences between the two schemes in the renormalization of the parameters~$\al$ and~$\lsd$, which are renormalized with $\MSbar$ conditions.
In \cref{sssec:MRrenCond} we present the $\MSbar$ renormalization conditions for a scheme in which the renormalized tadpole constants $t_\Ph$ and $t_\PH$ are set to zero.
In this scheme, which is referred to as \MSb{} scheme in the following, gauge-dependent tadpole terms enter the relations among bare parameters.
While such gauge-dependent terms drop out in the parametrization of observables in terms of input quantities if the latter are defined by \ac{OS} conditions, gauge dependences can remain in the parametrization of observables if some input parameters are defined by \MSb{} conditions.
Note that this does not automatically ruin the consistency of this scheme.
This shortcoming simply demands that all calculations should be carried out in the same gauge; we always use the 't~Hooft$-$Feynman gauge in this scheme.

\MSb{} conditions can be used without introducing gauge dependences in the parametrization of observables if bare (rather than renormalized) tadpoles are set to zero.
Following the renormalization of the \ac{THDM} described in \citere{Altenkamp:2017ldc}, in \cref{sssec:FJrenCond} we describe the changes in the renormalization constants to obtain this gauge-independent scheme and refer to it as FJ scheme, since this variant was proposed by Fleischer and Jegerlehner in \citere{Fleischer:1980ub}.%
\footnote{A similar approach for the tadpole treatment in the \ac{SM} has been discussed in \citere{Actis:2006ra}.}
Variants of this scheme, which are technically different, but phenomenologically equivalent, were also described in \citeres{Krause:2016oke,Denner:2016etu}.
Note that we use the terminology ``\MSb{}'' in two different contexts: as a name for a {\it renormalization scheme}
and for a type of {\it renormalization condition}. Both our ``\MSb{} scheme'' and our ``FJ scheme'' are based
on a mixture of OS and \MSb{} renormalization conditions.

\subsection{On-shell renormalization conditions}
\label{ssec:OSrenCond}

\subsubsection{Higgs sector}

\begin{figure}
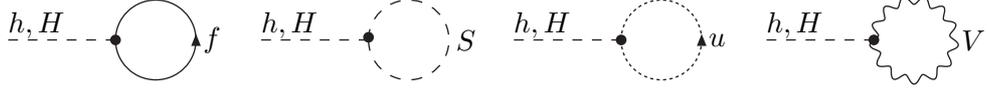

\centering
\include{diagrams/tadpoles}
\vspace*{-1em}
\caption{Tadpole diagrams for the light Higgs boson $\Ph$ and the heavy Higgs boson $\PH$, where $\Ff, \FS, u, \FV$ stand for generic fermion, scalar, ghost, and gauge fields.}
\label{fig:TadpoleDiagrams}
\end{figure}

\paragraph{Tadpoles:}
The renormalized tadpoles $\hat T_\Ph$ and $\hat T_\PH$, defined by the irreducible renormalized one-point vertex functions
\begin{equation}
\label{eq:tadpole}
\hat \Gamma^{\Ph, \PH} \, = \, \im \hat T_{\Ph, \PH} \, =\quad
\parbox{65pt}{\centering
\begin{picture}(65,40)(0,0)
\DashLine(0,20)(35,20){4}
\Text(0,23)[lb]{$\scriptstyle{\Fh, \FH}$}
\CCirc(50,20){15}{Gray}{Gray}
\end{picture}
}
\,
\end{equation}
are demanded to be zero in our {\MSb{} scheme.}%
\footnote{Equation~\eqref{eq:tadpole} in fact is an
on-shell condition; the name ``\MSb{} scheme'' refers to the
use of \MSb{} conditions imposed on the parameters $\alpha$
and $\lambda_{12}$.}
At \ac{NLO}, these contain contributions from the diagrams illustrated in \cref{fig:TadpoleDiagrams} and from the counterterms $\de t_{\Ph, \PH}$. 
The renormalized tadpoles vanish if the two contributions cancel each other,
\begin{equation}
\label{eq:tadpoleRenCondMR}
\hat T_\Ph = \de t_\Ph + T_\Ph = 0,
\qquad
\hat T_\PH = \de t_\PH + T_\PH = 0.
\end{equation}
With this choice, in any process, explicit tadpole diagrams are canceled by tadpole counterterms, so that both contributions can be omitted.
Note, however, that the tadpole constants~$\de t_\Ph$ and $\de t_\PH$ enter the expressions of some coupling counterterms and, thus, need to be computed.
The tadpole treatment in the FJ scheme is described in \cref{ssec:MSbarren}.

\begin{figure}
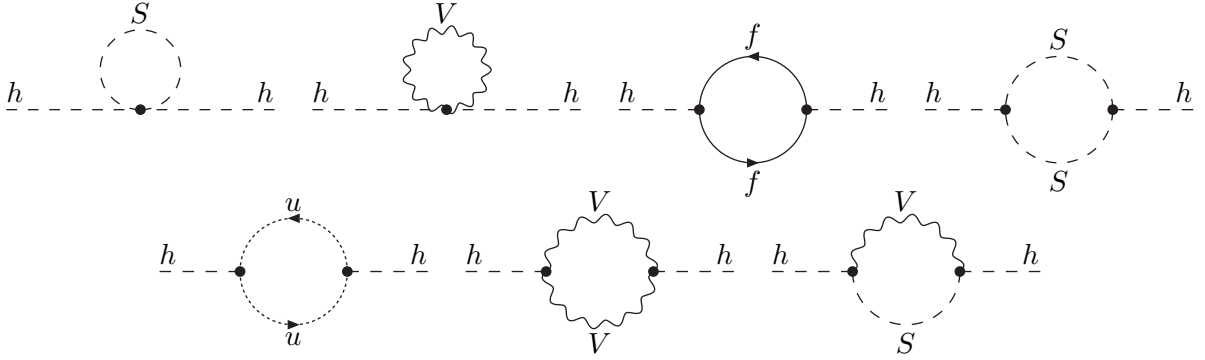

\centering
\include{diagrams/selfenergies}
\vspace*{-1em}
\caption{Generic self-energy diagrams for the light Higgs boson $\Ph$. Analogous diagrams contribute to the self-energy of the heavy Higgs boson $\PH$ and to the mixing self-energy between the two scalars.}
\label{fig:SEDiagrams}
\end{figure}

\paragraph{Higgs self-energies:}
The scalar sector of the \ac{SESM} is characterized by the presence of two Higgs bosons, $\Ph$ and $\PH$, and loop corrections lead to a mixing between the two scalars. Therefore, the renormalized one-particle irreducible two-point function for two external scalar fields is not diagonal and can be split into a diagonal \ac{LO} term plus a non-diagonal \ac{NLO} contribution,
\begin{equation}
\hat \Gamma^{ab}(k^2) \, = \quad
\parbox{90pt}{\centering
\begin{picture}(90,40)(0,0)
\DashLine(0,20)(30,20){4}
\Text(0,23)[lb]{$\scriptstyle{a}$}
\Text(12,15)[lt]{$\scriptstyle{\overset{\rightarrow}{k}}$}
\CCirc(45,20){15}{Gray}{Gray}
\DashLine(60,20)(90,20){4}
\Text(90,23)[rb]{$\scriptstyle{b}$}
\end{picture}
} \quad
= \,
\im \delta_{ab} (k^2 - M_a^2) + \im \hat \Sigma^{ab}(k^2),
\qquad
a,b = \Fh, \FH.
\end{equation}
The functions $\hat \Sigma^{ab}$ are the renormalized self-energies (containing loop and counterterm contributions) with the fields $a$ and $b$ on the external legs. These can be cast in the form
\begin{equation}
\label{eq:higgsRenSelfEnergies}
\begin{split}
\hat \Sigma^{\Ph \Ph}(k^2) &= \Sigma^{\Ph \Ph}(k^2) + \left( k^2 - \Mhs \right) \de Z_{\Ph \Ph} - \de \Mhs,
\\
\hat \Sigma^{\PH \PH}(k^2) &= \Sigma^{\PH \PH}(k^2) + \left( k^2 - \MHs \right) \de Z_{\PH \PH} - \de \MHs,
\\
\hat \Sigma^{\Ph \PH}(k^2) &= \Sigma^{\Ph \PH}(k^2) + \frac{1}{2} \left( k^2 - \Mhs \right) \de Z_{\Ph \PH} + \frac{1}{2} \left( k^2 - \MHs \right) \de Z_{\PH \Ph},
\end{split}
\end{equation}
where the unrenormalized self-energies $\Sigma^{a b}$, for $a,b = \Ph, \PH$, contain loop contributions of the types shown in \cref{fig:SEDiagrams}.
In the \ac{OS} scheme, the renormalized masses $\Mhs$ and $\MHs$ are determined by the zeroes of the (real parts of the) diagonal two-point functions,
\begin{equation}
\label{eq:massOScond}
\Re \hat \Sigma^{\Ph\Ph} \left( \Mhs \right) = 0,
\qquad
\Re \hat \Sigma^{\PH\PH} \left( \MHs \right) = 0.
\end{equation}
Since the matrix-valued two-point function $\hat \Gamma^{ab}$ is (up to sign factors) given by the inverse propagator of the $\Ph, \PH$ system, the renormalized \ac{OS} masses $\Mh, \, \MH$ are directly tied to the propagator poles.
Using the expressions for the renormalized self-energies given in \cref{eq:higgsRenSelfEnergies}, the conditions~\eqref{eq:massOScond} fix the mass renormalization constants $\de \Mhs$ and $\de \MHs$ to
\begin{equation}
\de \Mhs = \Re \Sigma^{\Ph \Ph} \left( \Mhs \right),
\qquad
\de \MHs = \Re \Sigma^{\PH \PH} \left( \MHs \right).
\end{equation}
The diagonal field renormalization constants are fixed by requiring that the (real parts of the) residues of the propagators at their respective poles are not changed by \ac{NLO} corrections, \ie
\begin{equation}
\lim_{k^2 \rightarrow \Mhs} \Re \frac{\im \, \hat \Gamma^{\Ph \Ph}(k^2)}{k^2 - \Mhs} = -1,
 \qquad
\lim_{k^2 \rightarrow \MHs} \Re \frac{\im \, \hat \Gamma^{\PH \PH}(k^2)}{k^2 - \MHs} = -1.
\end{equation}
The renormalization constants $\de Z_{\Ph \Ph}$ and $\de Z_{\PH \PH}$ are then given by
\begin{equation}
\delta Z_{\Ph\Ph} = - \Re \Sigma^{\prime \Ph\Ph}(\Mhs),
\qquad
\delta Z_{\PH\PH} = - \Re \Sigma^{\prime \PH\PH}(\MHs),
\end{equation}
where $\Sigma^\prime(k^2)$ is the derivative of the unrenormalized self-energy with respect to the argument~$k^2$.
Finally, to fix the mixing renormalization constants, we enforce the conditions that fields on their mass shells do not mix, \ie
\begin{equation}
\label{eq:renCondMixing}
\Re \hat \Sigma^{\Ph \PH} \left( \Mhs \right) = 0,
\qquad
\Re \hat \Sigma^{\Ph \PH} \left( \MHs \right) = 0.
\end{equation}
Using the expressions for the renormalized self-energies given in \cref{eq:higgsRenSelfEnergies}, the conditions \eqref{eq:renCondMixing} lead to
\begin{equation}
\label{eq:renConstMixing}
\de Z_{\Ph \PH} = 2 \Re \frac{\Sigma^{\Ph \PH} \left( \MHs \right)}{\Mhs - \MHs},
\qquad
\de Z_{\PH \Ph} = 2 \Re \frac{\Sigma^{\Ph \PH} \left( \Mhs \right)}{\MHs - \Mhs}.
\end{equation}
{In summary, the use of these on-shell conditions
to fix $\de Z_{ij}$ for $i,j=\Ph,\PH$ ensures that no wave function
renormalization or Higgs mixing corrections for external Higgs
states needs to be taken into account (these corrections
are shifted to self-energy and vertex counterterms).
Note also that this matrix field renormalization does not fix
the mixing angle counterterm $\delta\alpha$, although its
UV~divergences are connected to the ones in $\de Z_{ij}$ via
Eq.~\eqref{eq:renConstRelations}. In fact these relations will
be used below to determine $\delta\alpha$ in the $\MSbar$ scheme, where
$\delta\alpha$ only receives contributions from UV~divergences.}

\subsubsection{Gauge-boson sector}
\label{se:gaugefieldren}
The renormalization conditions for the gauge-boson sector of the \ac{SESM} are identical to the ones used in the \ac{SM}. The mass renormalization constants are fixed by imposing \ac{OS} conditions on the $\PW$- and $\PZ$-boson masses $\MW$ and $\MZ$, so that the renormalized squared masses correspond to the real parts of the locations of the propagator poles.%
\footnote{This statement holds at \ac{NLO}, i.e.\ in $\O(\alem)$. The relation of the real \ac{OS} masses to the complex location of the poles, including $\O(\alem^2)$ contributions is described below.}
The field renormalization constants are fixed by requiring that the residues of \ac{OS} propagators are not changed by NLO corrections; mixing renormalization constants are fixed in such a way that \ac{OS} gauge bosons do not mix. The renormalization constants for the \ac{EW} sector are then given by~\cite{Denner:1991kt}
\begin{align}
\label{eq:renCondGauge}
\delta \MWs &= \Re \Sigma_\text{T}^{\PW\PW}(\MWs),		&	 \delta Z_{\PW} &= - \Re \Sigma^{\prime \PW\PW}_\text{T}(\MWs),	\notag \\
\delta \MZs &= \Re \Sigma_\text{T}^{\PZ\PZ}(\MZs),		&		&						\notag \\
\delta Z_{\PZ\PZ} &= - \Re \Sigma^{\prime \PZ\PZ}_\text{T}(\MZs),	&	\delta Z_{\PA\PA} &= - \Re \Sigma^{\prime \PA\PA}_\text{T}(0),		\\
\delta Z_{\PA\PZ} &= - 2 \Re \frac{\Sigma^{\PA\PZ}_\text{T}(\MZs)}{\MZs},	&	\delta Z_{\PZ\PA} &= 2 \Re \frac{\Sigma^{\PZ\PA}_\text{T}(0)}{\MZs},	\notag
\end{align}
where $\Sigma^{\PV\PV^\prime}_\text{T}$ are the transverse parts of the self-energies for generic gauge bosons $\PV, \PV^\prime$.
The renormalized electric charge $e$ is defined as the electron$-$photon coupling in the Thomson limit of \ac{OS} electrons with a photon at zero momentum transfer. The resulting charge renormalization constant reads~\cite{Denner:1991kt}
\begin{equation}
\label{eq:electricChargeRenconst}
 \delta Z_e = - \frac{1}{2} \left( \delta Z_{\PA\PA} + \frac{\sw}{\cw} \delta Z_{\PZ\PA} \right).
\end{equation}

\subsubsection{Fermion sector}
To fix the renormalization constants in the fermion sector, the procedure is identical to the \ac{SM} case, described in detail in~\citere{Denner:1991kt}.
We require that the real parts of the locations of the poles of the fermion propagators correspond to the squares of the renormalized fermion masses, and that the residues of the fermion propagators do not receive loop corrections. 
Setting the \ac{CKM} matrix to the unit matrix, the renormalization constants are given by
\begin{equation}
\begin{split}
\delta m_{\Pf,i} &= \frac{m_{\Pf,i}}{2} \Re \left[
      \Sigma^{\Pf,\text{L}}_i \left( m_{\Pf,i}^2 \right)
      + \Sigma^{\Pf,\text{R}}_i \left( m_{\Pf,i}^2 \right)
      + 2 \Sigma^{\Pf,\text{S}}_i \left( m_{\Pf,i}^2 \right)
      \right],
\\
\delta Z^{\Pf,\si}_i &= - \Re \Sigma^{\Pf,\si}_i \left(m_{\Pf,i}^2 \right) - m_{\Pf,i}^2 \Re \left[ \Sigma^{\prime \Pf,\text{L}}_i \left( m_{\Pf,i}^2 \right)
									    + \Sigma^{\prime \Pf,\text{R}}_i \left( m_{\Pf,i}^2 \right)
									    + 2 \Sigma^{\prime \Pf,\text{S}}_i \left( m_{\Pf,i}^2 \right)  \right],
\quad
\si = \text{L}, \text{R},
\end{split}
\end{equation}
where $\Sigma^{\Pf,\text{L}}$, $\Sigma^{\Pf,\text{R}}$, $\Sigma^{\Pf,\text{S}}$ are, respectively, the left-handed, right-handed, and scalar parts of the fermion self-energy, as defined in~\citere{Denner:1991kt}.
The generalization to a non-trivial \ac{CKM} matrix can be found in \citeres{Denner:1991kt,Denner:2004bm,Kniehl:2009kk}.

\subsection{$\MSbar$ renormalization conditions}
\label{ssec:MSbarren}
The mixing angle $\al$ and the coupling constant $\lsd$ still
need to be fixed, but there is no obvious formulation of \ac{OS} conditions,
{which are based on physical S-matrix elements, thereby
avoiding any problems with gauge dependences.%
\footnote{The use of physical S-matrix elements is crucial here to avoid gauge dependences.
If instead renormalization conditions are imposed on off-shell Green functions or parts thereof
(such as mixing self-energies, Green functions involving unphysical fields, etc.) at some
momentum transfer, in general gauge dependences will result.
Employing, for instance, the last two equations of Eq.~\eqref{eq:renConstRelations} to
derive $\delta\alpha$ from $\de Z_{\Ph\PH}$ and $\de Z_{\PH\Ph}$ 
as given in Eq.~\eqref{eq:renConstMixing} including UV-finite terms,
leads to a gauge-dependent result, since 
$\Sigma^{\Ph\PH}(\Mhs)$ and $\Sigma^{\Ph\PH}(\MHs)$ are not
directly derived from S-matrix elements.}}

In principle, these parameters could be extracted from the Higgs couplings to other particles and Higgs self-couplings, but we are far from having the precision required for such measurements.
Also, requiring vanishing \ac{NLO} contributions to a specific process could lead to artificially large contributions when computing other observables, as pointed out for other \ac{SM} extensions~\cite{Freitas:2002um,Krause:2016oke}.

Here, we present the $\MSbar$ renormalization conditions for $\al$ and $\lsd$ 
adopted in the two schemes considered in this work. 
Each scheme employs the same \ac{OS} conditions 
{for the other parameters and for the
fields as described in \cref{ssec:OSrenCond}.
Imposing \MSb{} conditions (or conditions involving off-shell quantities)
on mixing angles or couplings in spontaneously broken
gauge theories is prone to introduce gauge dependences in the relations between
physical observables and input parameters. Detailed discussions of this issue,
which is intrinsically linked to the treatment of tadpole contributions,
can, e.g., be found in 
\citeres{Fleischer:1980ub,Actis:2006ra,Krause:2016oke,Denner:2016etu,%
Altenkamp:2017ldc,Denner:2017vms}.
In the following, we describe two different schemes, called ``\MSb{} scheme'' and
``FJ scheme'', which both renormalize $\al$ and $\lsd$ with \MSb{} conditions, but
differ in the treatment of tadpole contributions. The former involves gauge
dependences, while the latter does not.

In our \MSb{} scheme,}
the renormalization conditions for $\al$ and $\lsd$ are fixed using \MSb{} conditions, requiring \ac{UV} finiteness for certain loop vertex functions, and demanding vanishing renormalized tadpoles.
In spite of the issue of involving gauge dependences, this scheme is known to produce results that are rather stable with respect to variations of the renormalization scale.
This was, e.g., observed in the 
\ac{THDM}~\cite{Krause:2016oke,Altenkamp:2017ldc} and supersymmetric models~\cite{Freitas:2002um}.
In the second renormalization scheme, called here FJ scheme, we keep \MSb{} conditions for $\al$ and $\lsd$, but change the tadpole treatment a la Fleischer and Jegerlehner~\cite{Fleischer:1980ub} by setting bare tadpoles to zero consistently, which eliminates the gauge dependences.
Technically, we follow the procedure described in \citere{Altenkamp:2017ldc} in detail for the \ac{THDM}, i.e.\ we implement the FJ scheme by including appropriate finite terms in the renormalization constants obtained for $\al$ and $\lsd$ in the \MSb{} scheme.
In applications to the \ac{THDM}~\cite{Krause:2016oke,Denner:2016etu,Altenkamp:2017ldc,Altenkamp:2017kxk}, it was observed that this scheme is prone to introduce large corrections that may also spoil the stability of predictions with respect to renormalization scale variations.
To distinguish the two schemes we mark the renormalized parameters $\al$ and $\lsd$ and the corresponding renormalization constants with the superscripts \MSb{} and FJ if it is not clear from the context.
The parameters in the two schemes are related by the coincidence of the respective bare parameters, which define the original Lagrangian,
\begin{equation}
\label{eq:bareMSParamDef}
\begin{split}
\al_0 &= \al^\MSbar + \de \al^\MSbar = \al^\text{FJ} + \de \al^\text{FJ},
\\
\la_{12,0} &= \lsd^\MSbar + \de \lsd^\MSbar = \lsd^\text{FJ} + \de \lsd^\text{FJ},
\end{split}
\end{equation}
where we left implicit the dependence of the renormalization constants on the renormalized parameters. We will address the conversion between the two schemes in \cref{sssec:schemeConv}.

\subsubsection{\MSb{} scheme (with vanishing renormalized tadpoles)}
\label{sssec:MRrenCond}

\paragraph{Mixing angle {\boldmath$\al$}:}
The renormalization constant for the mixing angle $\al$ can be determined from Higgs-boson self-energies using the relations given in \cref{eq:renConstRelations}. The last two relations yield
\begin{equation}
\left. \de \al^\MSbar \right|_\text{UV} = \frac{1}{4} \left. \left( \de Z_{\PH \Ph} - \de Z_{\Ph \PH} \right) \right|_\text{UV}.
\end{equation}
Using the explicit expressions of \cref{eq:renConstMixing} for the mixing renormalization constant, and recalling that $\MSbar$ renormalization constants contain only \ac{UV}-divergent terms proportional to $\DeUV$, the counterterm $\de \al$ is given by
\begin{equation}
\label{eq:renConstMixingRes}
\de \al^\MSbar = \left. \Re \frac{\Sigma^{\Ph \PH} \left( \Mhs \right) + \Sigma^{\Ph \PH} \left( \MHs \right)}{2 \left( \MHs - \Mhs \right)} \right|_\text{UV}.
\end{equation}
Contributions induced by closed fermion loops are given by
\begin{equation}
\label{eq:renConstMixingMRferm}
\left. \de \al^\MSbar \right|_\text{ferm} = \, \DeUV \frac{e^2 \sa \ca}{64 \pi^2 \MWs \sws \left( \MHs - \Mhs \right)} \sum_\Pf c_\Pf \, \Mfs \left( \Mhs + \MHs - 12 \Mfs \right),
\end{equation}
with $c_\text{quark} = 3, c_\text{lepton} = 1$, and the remaining bosonic contributions are
\begin{equation}
\label{eq:renConstMixingMRbos}
\begin{split}
\left. \de \al^\MSbar \right|_\text{bos} &=
  \DeUV \frac{\sws \MWs \lsd^2}{2 \pi^2 \ca \sa e^2 \left( \Mhs-\MHs \right)^3} \, \mathcal{F}_2 \left( \Mhs, \MHs, \al \right)
  \\& \quad
  + \DeUV \frac{\ca \sa \lsd}{16 \pi^2 \cws \left( \Mhs - \MHs \right)} \, \mathcal{F}_1 \left( \Mhs, \MHs, \al, \MWs, \thetaw \right)
  \\& \quad
  + \DeUV \frac{\ca \sa e^2}{128 \pi^2 \cw^4 \sws \MWs \left( \Mhs - \MHs \right)} \, \mathcal{F}_0 \left( \Mhs, \MHs, \al, \MWs, \thetaw \right),
\end{split}
\end{equation}
where, to keep the notation compact, we introduced the functions $\mathcal{F}_i$, given by
\begin{equation}
\begin{split}
\mathcal{F}_2 &=
    \Mhs \MHs \left( 3 + 10 \cas - 10 \ca^4 \right)
    + \cas \MH^4 \left( 4 + 5 \cas \right)
    + \sas \Mh^4 \left( 9 - 5 \cas \right),
\\
\mathcal{F}_1 &=
    5 \cws \left( 2 \cas - 1 \right) \left( \Mhs - \MHs  \right)
    + \MWs \left( 2 \cws +1 \right),
\\
\mathcal{F}_0 &=
    - \cas \cw^4 \Mh^4 \left( 5 \cas + 4 \right)
    - \sas \cw^4 \MH^4 \left( 9 - 5 \cas \right)
    + \cws \MHs \MWs \left( 2 \cws + 1 \right) \left( \cas + 1 \right)
    \\& \quad
    - 18 \MW^4 \left( 2 \cw^4 + 1 \right)
    - 2 \cw^4 \Mhs \MHs \left( 3 + 5 \cas - 5 \ca^4  \right)
    - \cws \MWs \Mhs \left( \cas - 2 \right) \left( 2 \cws + 1 \right).
\end{split}
\end{equation}
This counterterm has been computed in the 't~Hooft$-$Feynman gauge and should be entirely used in this gauge.

\begin{figure}
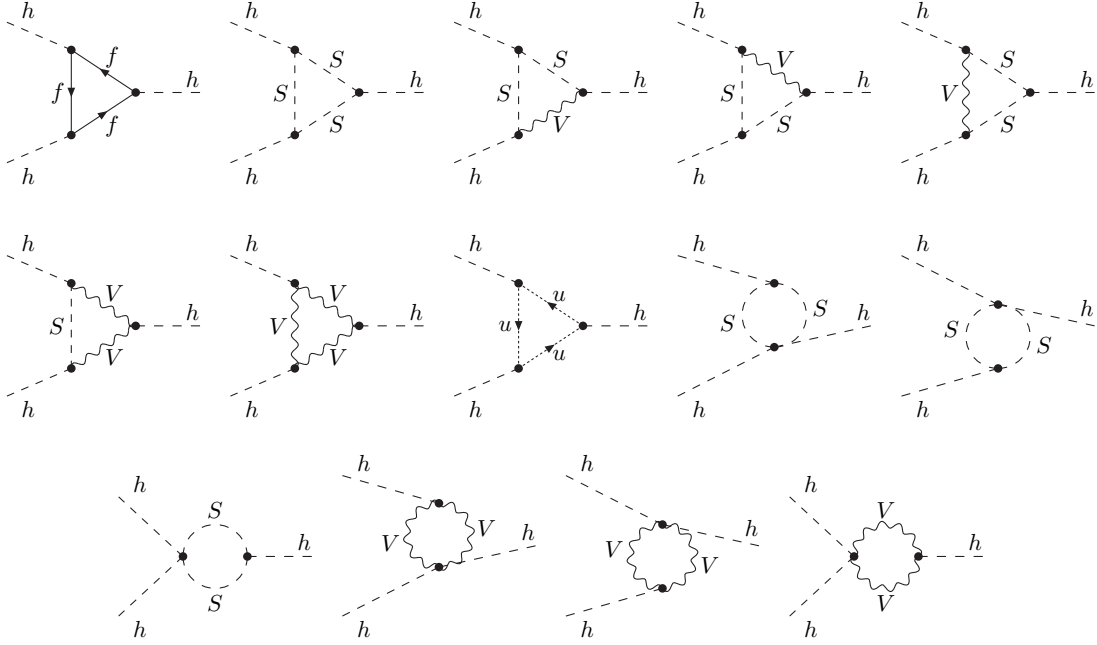

\centering
\include{diagrams/triangles}
\vspace*{-1em}
\caption{Diagram types contributing to the $\Fh \Fh \Fh$ vertex used for the renormalization of $\lsd$.}
\label{fig:TriangleDiagrams}
\end{figure}

\paragraph{Higgs self-coupling {\boldmath$\lsd$}:}
We fix the renormalization constant $\de \lsd^\MSbar$ considering the loop corrections to the vertex function with three external light Higgs bosons,
{similar to the procedure pursued in \citere{Kanemura:2016lkz}.}
Typical diagrams contributing to this vertex function are illustrated in \cref{fig:TriangleDiagrams}.
We require the one-loop renormalized vertex function $\hat \Gamma^{\Ph \Ph \Ph}$ to be \ac{UV} finite,
\begin{equation}
\left. \hat \Gamma^{\Ph \Ph \Ph} \right|_\text{UV} =
\,
\left.
\parbox{60pt}{\centering
\begin{picture}(60,60)(0,0)
\DashLine(60,30)(40,30){4}
\Text(60,33)[br]{$\scriptstyle{\Fh}$}
\DashLine(25,30)(0,60){4}
\Text(6,55)[bl]{$\scriptstyle{\Fh}$}
\DashLine(25,30)(0,0){4}
\Text(6,5)[tl]{$\scriptstyle{\Fh}$}
\CCirc(25,30){15}{Gray}{Gray}
\end{picture}
} \quad
\right|_\text{UV}
= 0.
\end{equation}
This automatically renders all scalar three- and four-point vertex functions \ac{UV} finite, and completes the set of renormalization conditions for the \ac{SESM}.
An explicit calculation of the counterterm yields
\begin{equation}
\label{eq:renConstlsdMRferm}
\left. \de \lsd^\MSbar \right|_\text{ferm} = \DeUV \frac{e^2 \lsd}{32 \pi^2 \sws \MWs} \sum_\Pf c_\Pf \Mfs
\end{equation}
for the contribution from closed fermion loops and
\begin{equation}
\label{eq:renConstlsdMRbos}
\begin{split}
\left. \de \lsd^\MSbar \right|_\text{bos} &= \,
    \DeUV \frac{3 \sws \MWs \lsd^3}{2 \pi^2 \cas \sas e^2 \left( \Mhs - \MHs \right)^2} \left( \cas \MHs + \sas \Mhs \right)
    + \DeUV \frac{\lsd^2}{4 \pi^2}
    \\& \quad
    + \DeUV \frac{3 e^2 \lsd }{64 \pi^2 \cws \sws \MWs} \left[ \cws \left( \cas \Mhs + \sas \MHs \right) - \MWs \left( 2 \cws + 1 \right) \right]
\end{split}
\end{equation}
for the bosonic contribution.
Since $\lsd$ is a fundamental parameter of the original Lagrangian, the $\MSbar$ definition given above leads to a gauge-independent counterterm $\de \lsd^\MSbar$.

\begin{figure}
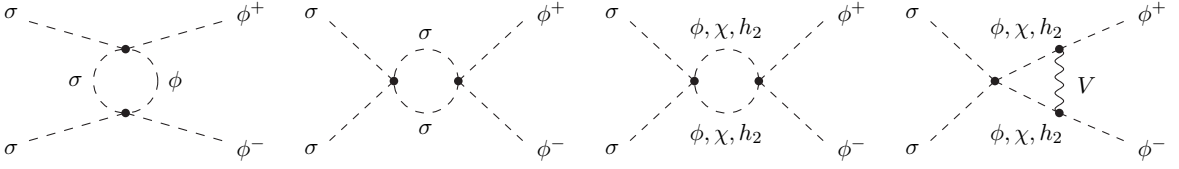

\centering
\include{diagrams/ssppdiagrams}
\vspace*{-1em}
\caption{UV-divergent diagrams contributing to the $\sigma\sigma\phi^+\phi^-$
vertex correction in the unbroken phase of the SESM, with $V$ denoting the
EW gauge bosons. For the first diagram, a crossed version exists as well.}
\label{fig:sspp-loops}
\end{figure}

We have checked our results on $\de \lsd^\MSbar$ against a simpler derivation,
which makes use of the fact that UV divergences in the CTs of dimensionless
couplings are the same in the broken and unbroken phase of the theory.
In the SESM, we can, thus, deduce $\de \lsd^\MSbar$ in the unbroken phase where 
$v_1=v_2=0$. 
In this phase, $h_1\equiv\sigma$, and the coupling $\lsd$ only appears in 
the quartic couplings $\sigma\sigma h_2 h_2$, $\sigma\sigma \chi\chi$, 
and $\sigma\sigma \phi^+\phi^-$.
At tree level, the $\sigma\sigma \phi^+\phi^-$ vertex function is given by
\begin{equation}
\Gamma^{\sigma\sigma\phi^+\phi^-}_0 = -2\im\lsd,
\end{equation}
and its $\MSbar$ CT reads
\begin{equation}
\de\Gamma^{\sigma\sigma\phi^+\phi^-} = -2\im\left(\de\lsd^\MSbar+\lsd \,\de Z_\Phi^\MSbar\right).
\end{equation}
Here we have used the fact that the $\sigma$ field renormalization constant 
$\de Z_\sigma=0$ in the unbroken phase, because $\sigma$ appears only in quartic
couplings, so that the $\sigma$ self-energy is momentum independent.
The $\MSbar$ field renormalization constant $\de Z_\Phi^\MSbar$ can be easily
determined from the UV divergences in any of the Higgs- or Goldstone-boson
self-energies, using \cref{eq:renConstRelations}. Only graphs with intermediate
Goldstone--gauge-boson pairs or fermion--antifermion pairs contribute, yielding
\begin{equation}
\de Z_\Phi^\MSbar = 
-\DeUV \frac{e^2}{32 \pi^2 \sws \MWs} \sum_\Pf c_\Pf \Mfs
+\DeUV \frac{e^2}{32\pi^2\cw^2\sw^2} \left(2\cw^2+1\right). 
\end{equation}
The UV-divergent diagrams contributing to the unrenormalized vertex
function at one loop, $\Gamma^{\sigma\sigma\phi^+\phi^-}_1$, are depicted in
\cref{fig:sspp-loops}. 
The corresponding divergences are easily calculated to
\begin{equation}
\left.\Gamma^{\sigma\sigma\phi^+\phi^-}_1\right|_{\mathrm{UV}} = 
\DeUV\frac{\im\lsd}{16\pi^2} \left(8\lsd+24\la_1+3\la_2\right) 
-\DeUV\frac{\im e^2\lsd}{32\pi^2\cw^2\sw^2} \left(2\cw^2+1\right),
\end{equation}
where the order of terms follows the order of diagrams in \cref{fig:sspp-loops}. 
Demanding that the renormalized vertex function 
$\hat\Gamma^{\sigma\sigma\phi^+\phi^-}$ is UV finite,
\begin{equation}
0=\left.\hat\Gamma^{\sigma\sigma\phi^+\phi^-}\right|_{\mathrm{UV}} = 
\left.\Gamma^{\sigma\sigma\phi^+\phi^-}_1\right|_{\mathrm{UV}} 
-2\im\left(\de\lsd^\MSbar+\lsd \,\de Z_\Phi^\MSbar\right),
\end{equation}
directly leads to
\begin{equation}
\de\lsd^\MSbar = 
\DeUV \frac{e^2\lsd}{32 \pi^2 \sws \MWs} \sum_\Pf c_\Pf \Mfs
+\DeUV\frac{\lsd}{32\pi^2} \left(8\lsd+24\la_1+3\la_2\right)
-\DeUV \frac{3e^2\lsd}{64\pi^2\cw^2\sw^2} \left(2\cw^2+1\right). 
\end{equation}
This is in agreement with the results \eqref{eq:renConstlsdMRferm} and
\eqref{eq:renConstlsdMRbos} for $\de\lsd^\MSbar$ 
given above, as can be checked by
trading the couplings $\la_1$ and $\la_2$ for the chosen independent
input parameters of the SESM with the help of Eq.~\eqref{eq:inversionRel}.

\subsubsection{FJ scheme (with vanishing bare tadpoles)}
\label{sssec:FJrenCond}
To obtain gauge-independent relations between observables and renormalized input parameters, we make use of the FJ scheme, proposed in \citeres{Fleischer:1980ub,Actis:2006ra} for the \ac{SM} and applied to the \ac{THDM} in \citeres{Krause:2016oke,Denner:2016etu,Altenkamp:2017ldc}.
In this scheme, the bare tadpoles $t_{\Ph,0}$ and $t_{\PH,0}$ are set to zero, and any kind of reshuffling of tadpole terms is a mere question of taste, which does not change the results for observables.
In principle, it is even possible to include explicit tadpole diagrams wherever they appear.

We have performed the FJ renormalization in two independent, but equivalent ways: 
Firstly, following the strategy proposed in \citeres{Krause:2016oke,Denner:2016etu}, 
we have set the bare tadpole terms $t_{\Ph,0}$ and $t_{\PH,0}$ to zero and omitted the introduction of tadpole 
renormalization constants $\de t_{\Ph}$ and $\de t_{\PH}$. Instead we have
reintroduced the tadpole counterterms by shifting the Higgs fields according to
\begin{equation}
\label{eq:higgsShiftFJ}
\Fh \to \Fh + \De v_\Ph,
\qquad
\FH \to \FH + \De v_\PH,
\end{equation}
with constants $\De v_\Ph, \De v_\PH$, which can be interpreted as shifts in the integration variables in the path integral.
The constants $\De v_\Ph, \De v_\PH$ can be chosen freely and are usually introduced to cancel the explicit tadpole loops $T_\Ph, T_\PH$, leading to vanishing one-point functions $\hat T_\Ph, \hat T_\PH$.
The shifts~\eqref{eq:higgsShiftFJ} spreads $\De v_{\Ph,\PH}$ contributions to all Feynman rules for vertices that result from setting $\Fh$ and $\FH$ lines to the corresponding constant $\De v_{\Ph, \PH}$.

The second method, described for the \ac{THDM} in \citere{Altenkamp:2017ldc}, takes advantage of the fact that,
if all counterterms of independent parameters are determined by the same physical conditions, different choices for the tadpole renormalization lead to the same physical results.
In this approach we keep the tadpole 
renormalization constants $\de t_{\Ph,\PH}$ with the 
conditions \eqref{eq:tadpoleRenCondMR}, so that the renormalization constants given in \cref{ssec:OSrenCond} remain unchanged.
The renormalization constant $\de \al^\text{FJ}$, which reproduces the results in the FJ scheme, is related to the renormalization constant $\de \al^\MSbar$ of \cref{eq:renConstMixingRes} by
\begin{equation}
\de \al^\text{FJ} = \de \al^\MSbar + \text{finite terms},
\end{equation}
where the additional finite terms depend on the tadpole contributions $T_\Ph$ and $T_\PH$.
To compute these terms, we consider a variant in which the bare tadpole constants vanish, and the tadpole contributions are explicitly included in Green functions. Denoting the quantities defined in this scheme with a superscript 
``$t$'', the same physical results are obtained using the counterterm
\begin{equation}
\label{eq:tvariantRen}
\de \al^t = \de \al^\MSbar + \De \al^t \left( T_\Ph, T_\PH \right),
\end{equation}
where $\de \al^t$ can be obtained by setting to zero the tadpole renormalization constants~$\de t_{\Ph,\PH}$ in~$\de \al^\MSbar$ and including tadpole diagrams in the related Green functions.
For consistency, $\De \al^t$ contains also the finite terms coming from tadpole diagrams, otherwise the new terms could not be compensated by the tadpole contributions occurring elsewhere, leading to different renormalized amplitudes. The FJ renormalization scheme in the ``$t$-variant'' is obtained by reducing relation \eqref{eq:tvariantRen} to \ac{UV} divergences only, since $\de \al^{t,\text{FJ}}$ has to be proportional to $\DeUV$.
Since $\de \al^\MSbar$ is proportional to $\DeUV$ as well, we can write
\begin{equation}
\label{eq:tvariantFJRen1}
\de \al^{t,\text{FJ}} = \de \al^\MSbar + \left. \De \al^t \left( T_\Ph, T_\PH \right) \right|_\text{UV}.
\end{equation}
Taking the FJ version of \cref{eq:tvariantRen} leads to
\begin{equation}
\label{eq:tvariantFJRen2}
\de \al^{t,\text{FJ}} = \de \al^\text{FJ} + \De \al^t \left( T_\Ph, T_\PH \right),
\end{equation}
where $\de \al^\text{FJ}$ is the counterterm we have to use in our counterterm Lagrangian to compute renormalized amplitudes in the FJ scheme. Finally, combining \cref{eq:tvariantFJRen1,eq:tvariantFJRen2}, the relation between the $\al$ renormalization constant in the two schemes is given by
\begin{equation}
\de \al^\text{FJ} = \de \al^\MSbar - \left. \De \al^t \left( T_\Ph, T_\PH \right) \right|_\text{finite}.
\end{equation}
The term $\De \al^t$, according to \cref{eq:tvariantRen}, is the difference between $\de \al^t$ and $\de \al^\MSbar$, and is given by the tadpole contributions (that must be included in the 
``$t$-variant'') to the self-energies used to define $\de \al^\MSbar$ in \cref{eq:renConstMixingRes}, leading to
\begin{equation}
\label{eq:renConstMixingFJ}
\De \al^t \left( T_\Ph, T_\PH \right) = \de \al^t - \de \al = \Re \left. \frac{ \Sigma^{t,\Ph \PH} \left( \Mhs \right) + \Sigma^{t,\Ph \PH} \left( \MHs \right)}{2 \left( \MHs - \Mhs \right)} \right|_{T_\Ph, T_\PH},
\end{equation}
where the superscript 
``$t$'' in the self-energy~$\Sigma^{t,\Ph\PH}$ indicates that it is computed in the ``$t$-variant'', \ie includes explicit tadpoles.
Representing the unrenormalized tadpoles with black blobs, \cref{eq:renConstMixingFJ} leads to the expression
\begin{equation}
\label{eq:DeltaAlfaFJ}
\begin{split}
\De \al^t \left( T_\Ph, T_\PH \right)
\, &= \, \frac{1}{\MHs - \Mhs} \left(
\, \, \parbox{80pt}{\centering
\begin{picture}(80,40)(0,0)
\DashLine(5,5)(75,5){4}
\Text(0,5)[]{$\scriptstyle{\Fh}$}
\Text(80,5)[]{$\scriptstyle{\FH}$}
\DashLine(40,5)(40,25){4}
\CCirc(40,30){10}{Black}{Black}
\Vertex(40,5){2}
\Text(45,12)[]{$\scriptstyle{\Fh}$}
\end{picture}
} \,
+
\, \parbox{80pt}{\centering
\begin{picture}(80,40)(0,0)
\DashLine(5,5)(75,5){4}
\Text(0,5)[]{$\scriptstyle{\Fh}$}
\Text(80,5)[]{$\scriptstyle{\FH}$}
\DashLine(40,5)(40,25){4}
\CCirc(40,30){10}{Black}{Black}
\Vertex(40,5){2}
\Text(45,12)[]{$\scriptstyle{\FH}$}
\end{picture}
} \, \,
\right)
\\
\,&
\\
\, &= \, \frac{e}{\MHs - \Mhs} \left( T_\Ph \, \frac{C_{\Ph\Ph\PH}}{\Mhs} + T_\PH \, \frac{C_{\Ph\PH\PH}}{\MHs} \right),
\end{split}
\end{equation}
with the factors for the $\Fh\Fh\FH$ and $\Fh\FH\FH$ tree-level couplings given by
\begin{equation}
\label{eq:couplingFactorsFJ}
\begin{split}
C_{\Ph\Ph\PH} &= \sa \left( 2 \Mhs + \MHs \right) \left[ \frac{4 \sw \MW \lsd}{e^2 \left( \Mhs - \MHs \right)} - \frac{\cas}{2 \sw \MW} \right],
\\
C_{\Ph\PH\PH} &= \ca \left( \Mhs + 2 \MHs \right) \left[ \frac{4 \sw \MW \lsd}{e^2 \left( \MHs - \Mhs \right)} - \frac{\sas}{2 \sw \MW} \right],
\end{split}
\end{equation}
which are related to the couplings of Eq.~\eqref{eq:scalarCoupl} by
{$c_{\Ph\Ph\PH}=-eC_{\Ph\Ph\PH}/2$ and $c_{\Ph\PH\PH}=-eC_{\Ph\PH\PH}/2$.}
Therefore, in order to reproduce the result in the FJ scheme in the framework of our \MSb{} scheme, where we use vanishing renormalized tadpoles, we use the counterterm
\begin{equation}
\de \al^\text{FJ} = \de \al^\MSbar - \left. \De \al^t \left( T_\Ph, T_\PH \right) \right|_\text{finite}
= \de \al^\MSbar + \frac{e}{\Mhs - \MHs} \left. \left( T_\Ph \, \frac{C_{\Ph\Ph\PH}}{\Mhs} + T_\PH \, \frac{C_{\Ph\PH\PH}}{\MHs} \right) \right|_\text{finite},
\end{equation}
where the finite part of the last term is obtained by dropping the contributions proportional to $\DeUV$ from the expression \eqref{eq:DeltaAlfaFJ}.
When computing a physical observable the use of this renormalization constant ensures a gauge-independent result.

\section{Predictions for {\boldmath$\Ph \to \PW \PW / \PZ \PZ \to 4 \Pf$} in the SESM with the Monte Carlo program \prophecy}
\label{sec:impl}
%
\subsection{Features of \prophecy}
The program \prophecy{} (\textbf{Prop}er description of the \textbf{H}iggs d\textbf{ec}a\textbf{y} into \textbf{4} \textbf{f}ermions)~\cite{Bredenstein:2006ha,Bredenstein:2006nk,Bredenstein:2006rh} is a Monte Carlo generator for the computation of any partial width for the decay of the Higgs boson into four light fermions at \ac{NLO}, including both \ac{EW} and \ac{QCD} corrections. The generator can be used to produce differential distributions for any leptonic and semi-leptonic final state, as well as unweighted events for the leptonic final states. The first versions of \prophecy{} dealt with the decay of a \ac{SM} Higgs boson and supported the presence of a fourth generation of massive fermions~\cite{Denner:2011vt}. Recently, the program has been extended to allow for the same calculations in \acp{THDM}~\cite{Altenkamp:2017ldc,Altenkamp:2017kxk}.

In the implementation, the final-state fermions are considered to be massless, but the physical mass values are kept in closed fermion loops which contribute to the virtual corrections. In the considered massless limit, the results are the same for final-state fermions of different generations (given that the same diagrams contribute), so that only the 19 independent final states reported in \cref{tab:finalStates} 
need to be considered. 
\begin{table}
  \centering
   \renewcommand{\arraystretch}{1.1}
\begin{tabular}{|c|ccc|}\hline
Final states & leptonic & semi-leptonic & hadronic\\\hline
\multirow{4}{*}{neutral current}& $\Pnu_\Pe \bar{\Pnu}_\Pe \Pnu_\Pmu \bar{\Pnu}_\Pmu$~(3) &$\Pnu_\Pe \bar{\Pnu}_\Pe \Pu \bar{\Pu}$~(6)&$\Pu \bar{\Pu} \Pc \bar{\Pc}$~(1)\\
& $ \Pe^- \Pe^+ \Pmu^- \Pmu^+$~(3) &$\Pnu_\Pe \bar{\Pnu}_\Pe \Pd \bar{\Pd}$~(9) & $\Pd \bar{\Pd} \Ps \bar{\Ps}$~(3)\\
& $\Pnu_\Pe \bar{\Pnu}_\Pe  \Pmu^-\Pmu^+$~(6) & $ \Pe^-\Pe^+ \Pu \bar{\Pu}$~(6) & $\Pu \bar{\Pu} \Ps \bar{\Ps}$~(4)\\
&&  $\Pe^- \Pe^+  \Pd \bar{\Pd}$~(9)&\\\hline
\multirow{2}{*}{neutral current with interference}& $ \Pe^- \Pe^+ \Pe^-\Pe^+$~(3) &&$\Pu \bar{\Pu} \Pu \bar{\Pu}$~(2)\\
& $\Pnu_\Pe \bar{\Pnu}_\Pe \Pnu_\Pe \bar{\Pnu}_\Pe$~(3) &&$\Pd \bar{\Pd} \Pd \bar{\Pd}$~(3)\\
\hline
charged current & $ \Pnu_\Pe \Pe^+ \Pmu^- \bar{\Pnu}_\Pmu$~(6) & $  \Pnu_\Pe \Pe^+ \Pd \bar{\Pu}$~(12)& $\Pu \bar{\Pd} \Ps \bar{\Pc}$~(2)\\\hline
charged and neutral current & $\Pnu_\Pe \Pe^+  \Pe^- \bar{\Pnu}_\Pe$~(3) &&  $\Pu \bar{\Pd} \Pd \bar{\Pu}$~(2)\\\hline
\end{tabular}
  \caption{Classification of the possible final states for the decays $\Ph \to \PW \PW/ \PZ \PZ \to 4 \Pf$. Final states that differ only by generation indices, but have the same diagrams, are only stated once. The numbers in parentheses are the numbers of inequivalent final states that are represented by the given state.}
\label{tab:finalStates}
\end{table}
In the table, these are classified by the intermediate gauge bosons appearing in the \ac{LO} matrix element of the corresponding decay and by the number of lepton pairs in the final state.
The $\PW$- and $\PZ$-boson resonances are treated in the complex-mass scheme~\cite{Denner:1999gp,Denner:2005es,Denner:2005fg}, and the vector bosons are kept off-shell, so that the results have NLO accuracy both in resonant and non-resonant phase-space regions. The proper inclusion of off-shell effects is of fundamental importance, since the discovered Higgs boson at $125 \, \GeV$ is below the $\PW\PW$ and $\PZ\PZ$ thresholds.
To calculate the virtual corrections, the loop integrals are computed using the \fortran{} library \collier{}~\cite{Denner:2016kdg}, which makes use of dimensional regularization to handle the \ac{UV} divergences. \ac{IR} divergences are regulated using small masses for the final-state fermions as well as for the emitted photon or gluon, and the divergences are canceled between virtual and real corrections using some slicing or the dipole-subtraction method~\cite{Catani:1996vz,Dittmaier:1999mb,Dittmaier:2008md}.

The phase-space integral is performed by the adaptive algorithm implemented in the original \prophecy{} version, which evaluates the integrand at pseudo-random phase-space points, adapting iteratively the selection of channels in order to provide a better convergence.

Computing the widths for the decays of the Higgs boson into all the possible final states listed in \cref{tab:finalStates} allows to get the total width for the inclusive decay of the Higgs boson into four fermions,~$\Gamma_{\Ph \to 4 \Pf}$. The width $\Gamma_{\Ph \to 4 \Pf}$ is the sum over the decay widths for the $19$ independent final states, each of them weighted with the corresponding multiplicity given in \cref{tab:finalStates}.

In order to define a width for the decay of the Higgs boson into a pair of $\PW$ or $\PZ$ bosons, it is possible to separate contributions to $\Gamma_{\Ph \to 4 \Pf}$ for which, in the \ac{LO} matrix element, the intermediate vector bosons are two $\PW$ or $\PZ$ bosons.
If both $\PW\PW$ and $\PZ\PZ$ are possible intermediate final states at LO, the $\PW\PW$ and $\PZ\PZ$ decay parts are defined by formally taking the two respective fermion$-$antifermion pairs of the $\PW$- or $\PZ$-boson decays from different generations.
This procedure attributes all contributions to $\PW\PW$ or $\PZ\PZ$ channels except for terms that are interferences of $\PW\PW$- and $\PZ\PZ$-mediated contributions or corrections thereof.
The sum of these interferences is denoted by $\Gamma_{\PW \PW / \PZ \PZ - \text{int}}$ (see also \citeres{Dittmaier:2011ti,Dittmaier:2012vm}),
\begin{equation}
\label{eq:totalDecayWidth}
 \Gamma_{\Ph \to 4 \Pf} =
    \Gamma_{\Ph \to \PW \PW \to 4 \Pf}
    + \Gamma_{\Ph \to \PZ \PZ \to 4 \Pf}
    + \Gamma_{\PW \PW / \PZ \PZ - \text{int}}.
\end{equation}
Note that interference contributions between~$\PZ\PZ$ channels with different fermion-number flow are included in~$\Gamma_{\Ph \to \PZ\PZ \to 4 \Pf}$.
As a trivial example, consider the decay into $\Pnue \Pep \Pmum \Pnumub$, for which the LO process is entirely mediated by two $\PW$ bosons,
\begin{equation}
\label{eq:partialDecayWidthsWW}
\Gamma_{\Ph \to \PW \PW \to \Pnue \Pep \Pmum \Pnumub} = \Gamma_{\Ph \to \Pnu_\Pe \Pe^+ \Pmu^- \bar{\Pnu}_\Pmu}.
\end{equation}
On the other hand, the leptonic final state~$\Pnu_\Pe \Pe^+ \Pe^- \bar{\Pnu}_\Pe$ contributes to all three parts of \cref{eq:totalDecayWidth},
\begin{equation}
\label{eq:partialDecayWidthsVV}
\begin{split}
 \Gamma_{\Ph \to \PW \PW \to \Pnu_\Pe \Pe^+ \Pe^- \bar{\Pnu}_\Pe} &= \Gamma_{\Ph \to \Pnu_\Pe \Pe^+ \Pmu^- \bar{\Pnu}_\Pmu},
 \\
 \Gamma_{\Ph \to \PZ \PZ \to \Pnu_\Pe \Pe^+ \Pe^- \bar{\Pnu}_\Pe} &= \Gamma_{\Ph \to \Pnu_\Pe \bar{\Pnu}_\Pe \Pmu^- \Pmu^+},
 \\
 \Gamma_{\PW \PW / \PZ \PZ - \text{int}, \Pnu_\Pe \Pe^+ \Pe^- \bar{\Pnu}_\Pe} &=
    \Gamma_{\Ph \to \Pnu_\Pe \Pe^+ \Pe^- \bar{\Pnu}_\Pe}
    - \Gamma_{\Ph \to \Pnu_\Pe \Pe^+ \Pmu^- \bar{\Pnu}_\Pmu}
    - \Gamma_{\Ph \to \Pnu_\Pe \bar{\Pnu}_\Pe \Pmu^- \Pmu^+}.
\end{split}
\end{equation}
Following this procedure for all four-fermion final states leads to the definition of $\Gamma_{\Ph \to 4 \Pf}$ into $\PW\PW$- and $\PZ\PZ$-mediated parts and corresponding interference,
\begin{equation}
\label{eq:partialDecayWidthVV1}
\begin{split}
\Gamma_{\Ph \to \PW\PW \to 4 \Pf}
  &= 9 \, \Gamma_{\Ph \to \Pnue \Pep \Pmum \Pnumub}
  + 12 \, \Gamma_{\Ph \to \Pnue \Pep \Pd \Pub}
  + 4 \, \Gamma_{\Ph \to \Pu \Pdb \Ps \Pcb},
\\
\Gamma_{\Ph \to \PZ\PZ\to4f}
  &= 3 \, \Gamma_{\Ph \to \nu_\Pe \bar{\nu}_\Pe \nu_\mu \bar{\nu}_\mu}
  + 3 \, \Gamma_{\Ph \to \Pe^+ \Pe^- \mu^+ \mu^-}
  + 9 \, \Gamma_{\Ph \to \nu_\Pe \bar{\nu}_\Pe \mu^+ \mu^-}
  + 3 \, \Gamma_{\Ph \to \Pe^+ \Pe^- \Pe^+ \Pe^-}
  \\& \quad
  + 3 \, \Gamma_{\Ph \to \nu_\Pe \bar{\nu}_\Pe \nu_\Pe \bar{\nu}_\Pe}
  + 6 \, \Gamma_{\Ph \to \nu_\Pe \bar{\nu}_\Pe \Pu \bar{\Pu}}
  + 9 \, \Gamma_{\Ph \to \nu_\Pe \bar{\nu}_\Pe \Pd \bar{\Pd}}
  + 6 \, \Gamma_{\Ph \to \Pe^+ \Pe^- \Pu \bar{\Pu}}
  + 9 \, \Gamma_{\Ph \to \Pe^+ \Pe^- \Pd \bar{\Pd}}
  \\& \quad
  + \Gamma_{\Ph \to \Pu \bar{\Pu} \Pc \bar{\Pc}}
  + 3 \, \Gamma_{\Ph \to \Pd \bar{\Pd} \Ps \bar{\Ps}}
  + 6 \, \Gamma_{\Ph \to \Pu \bar{\Pu} \Ps \bar{\Ps}}
  + 2 \, \Gamma_{\Ph \to \Pu \bar{\Pu} \Pu \bar{\Pu}}
  + 3 \, \Gamma_{\Ph \to \Pd \bar{\Pd} \Pd \bar{\Pd}},
\\
\Gamma_{\PW\PW/\PZ\PZ-\text{int}}
  &= 3 \, \Gamma_{\Ph \to \nu_\Pe \Pe^+  \Pe^- \bar{\nu}_\Pe}
  - 3 \, \Gamma_{\Ph \to \nu_\Pe \bar{\nu}_\Pe \mu^+ \mu^-}
  -3  \, \Gamma_{\Ph \to \Pe^+ \nu_\Pe \bar{\nu}_\mu \mu^-}
  \\& \quad
  + 2 \, \Gamma_{\Ph \to \Pu \bar{\Pd} \Pd \bar{\Pu}}
  - 2 \, \Gamma_{\Ph \to \Pu \bar{\Pu} \Ps \bar{\Ps}}
  - 2 \, \Gamma_{\Ph \to \Pu \bar{\Pd} \Ps \bar{\Pc}}.
\end{split}
\end{equation}

\subsection{Details on the calculation of the process {\boldmath$\Ph \to \PW\PW/\PZ\PZ \to 4 \Pf$} at NLO}
\label{sec:NLOdecay}

\subsubsection{Leading order}
\begin{figure}
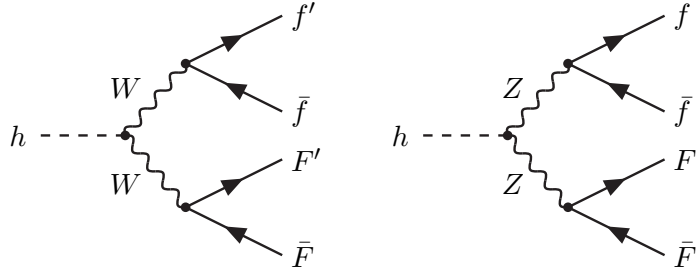

\centering
\include{diagrams/diagramsDecay_LO}
\vspace*{-.5em}
\caption{Charged- and neutral-current LO diagrams contributing to the process $\Ph \to 4 \Pf$. The primed fermions $\Pf^\prime$ and $\PF^\prime$ stand for the isospin partners of $\Pf$ and $\PF$, respectively. For $\PF = \Pf^\prime$ in the final state, both contributions must be taken into account. For $\PF = \Pf$ the neutral-current diagram on the right-hand side appears in a second version with the $\Pf$ and $\PF$ lines (or the $\Pfb$ and $\PFb$ lines) interchanged.}
\label{fig:diagramsLO}
\end{figure}

At the Born level, in the massless limit for the final-state fermions, the decay $\Ph \to 4 \Pf$ is mediated by a pair of (off-shell) gauge bosons, each of them decaying into two fermions.
The contributions to the matrix element for the generic process are given by the Feynman diagrams reported in \cref{fig:diagramsLO}.
Compared to the \ac{SM} case, there are no additional diagrams, and the matrix element for the \ac{LO} process is simply rescaled by a $\ca$ factor,
\begin{equation}
\label{eq:matrixElementLO}
\ME_\text{SESM}^\text{LO} = \ca \ME_\text{SM}^\text{LO},
\end{equation}
so that \ac{LO} predictions for the decay widths in the \ac{SESM} can be easily obtained rescaling the \ac{SM} results by a factor $\cas$.

Depending on the fermions in the final state, the tree-level process either involves only the first diagram of \cref{fig:diagramsLO} (``charged current''), only the second diagram (``neutral current''), both diagrams (``charged and neutral current'', for $\Pf = \PF^\prime$), or two diagrams of the second kind (``neutral current with interference'', for $\Pf = \PF$).

\subsubsection{Virtual corrections}
Moving beyond the LO computation, loop and real-emission contributions must be taken into account. The one-loop virtual corrections to the decay process $\Ph \to 4 \Pf$ in the \ac{SESM} receive contributions from self-energy, vertex, box, and pentagon diagrams, as well as from counterterms in the self-energy and vertex corrections. These are very similar to the contributions arising in the \ac{SM} case, which are described in detail in \citeres{Bredenstein:2006ha,Bredenstein:2006nk,Bredenstein:2006rh}. Indeed, the set of Feynman diagrams that contribute to the decay in the \ac{SESM} is given by all the \ac{SM} diagrams, supplemented by additional diagrams involving the heavy Higgs boson~$\PH$.
The computation of the corresponding matrix element can be performed using the same technology used in the \ac{SM} case, keeping in mind that the ``\ac{SM}-like'' diagrams, \ie the diagrams which do not involve the heavy Higgs, may have different expressions with respect to the \ac{SM}, since the coupling factors are different in the \ac{SESM}.
Note that diagrams without internal Higgs-boson lines are simply copies of the SM counterparts, rescaled by a factor $\ca$; this class of diagrams, however, is neither forming a gauge-invariant nor a UV-finite subset.

\paragraph{QCD loops:}
The \ac{QCD} corrections, relevant for the semi-leptonic and hadronic decays, can be obtained easily from the \ac{SM} case, since no additional diagrams involving the strong interaction are changed by the presence of the heavy Higgs boson, and the only modification is the multiplicative factor $\ca$ in the $\Fh\FV\FV$ and $\Fh\Pf\Pf$ couplings. Consequently, as in the LO result, the matrix element for the one-loop \ac{QCD} matrix element is given by
\begin{equation}
\label{eq:matrixElementQCD}
\ME_\text{SESM,QCD}^\text{NLO,virt} = \ca \ME_\text{SM,QCD}^\text{NLO,virt}.
\end{equation}
A survey of the generic diagrams contributing to the \ac{QCD} matrix element of \cref{eq:matrixElementQCD} is reported in \citere{Bredenstein:2006ha}.

\begin{figure}
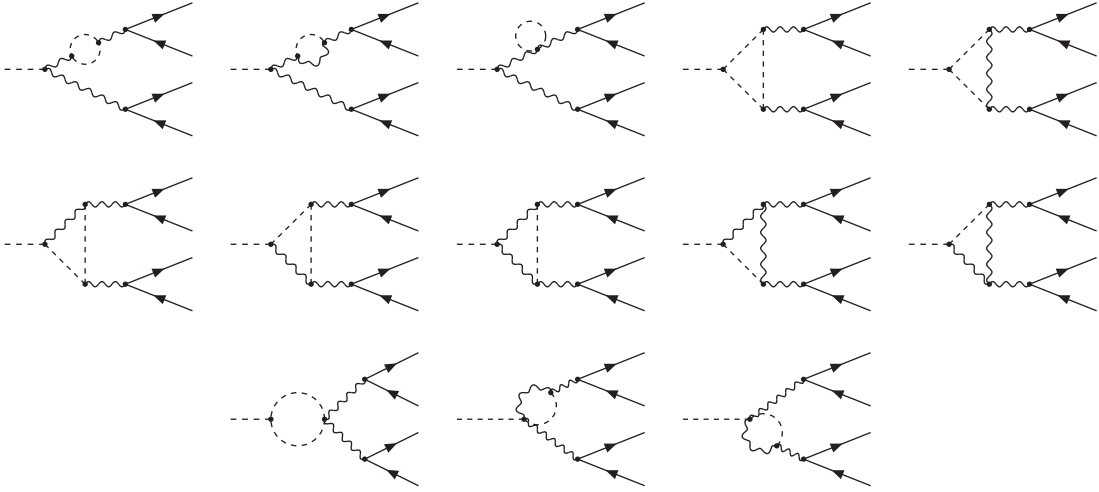

\centering
\include{diagrams/diagramsDecay_NLO}
\vspace*{-1em}
\caption{Loop diagrams involving internal scalar particles. Each internal scalar line can be either a light or a heavy Higgs field; self-energy topologies are shown only for one of the intermediate gauge bosons.}
\label{fig:diagramsNLO}
\end{figure}

\paragraph{EW loops:}
Comparing the \ac{SESM} to the \ac{SM}, the presence of the singlet has an impact on the \ac{EW} corrections, giving rise to a higher number of loop diagrams and changing the analytic expressions of the \ac{SM}-like contributions. For a list of the generic diagrams contributing to the \ac{EW} matrix element, see \citere{Bredenstein:2006rh}. Since, in these diagrams, no internal scalar lines appear in box and pentagon graphs, the heavy Higgs yields only additional self-energy and vertex diagrams of the type reported in \cref{fig:diagramsNLO}. The computation of the \ac{EW} loops can be performed with the standard machinery.  

\subsubsection{Real-emission corrections}
\begin{figure}
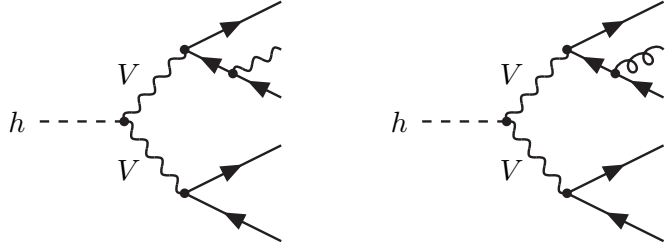

\centering
\include{diagrams/diagramsDecay_NLOreal}
\vspace*{-1em}
\caption{Exemplary diagrams for the real-emission processes~$\Ph \to 4 \Pf + \gamma$ and~$\Ph \to 4 \Pf + \Pg$ included in the \ac{NLO} computation.}
\label{fig:diagramsEmission}
\end{figure}

At \ac{NLO}, the final-state fermions can emit a photon or a gluon, so that it is necessary to include diagrams as depicted in \cref{fig:diagramsEmission}.
The gluon-emission diagram can be obtained straightforwardly attaching a gluon line to the \ac{LO} diagram and, since the gluon$-$quark couplings in the \ac{SESM} are the same as in the \ac{SM}, the matrix element for the process will be the \ac{SM} matrix element, rescaled by the prefactor $\ca$ arising from the $\Fh \FV \FV$ coupling,
\begin{equation}
\label{eq:matrixElementQCDreal}
\ME_\text{SESM,QCD}^\text{NLO,real} = \ca \ME_\text{SM,QCD}^\text{NLO,real}.
\end{equation}
In the same way, since the photon$-$fermion couplings in the \ac{SESM} are equal to the \ac{SM} couplings, the photon-emission matrix element can be easily obtained from the corresponding matrix element for the \ac{SM},
\begin{equation}
\label{eq:matrixElementEWreal}
\ME_\text{SESM,EW}^\text{NLO,real} = \ca \ME_\text{SM,EW}^\text{NLO,real}.
\end{equation}
In the real-emission contributions the \ac{IR} structure is the same as in the \ac{SM}, and the extraction of the soft and collinear divergences appearing in the phase-space integration of the squared matrix elements of \cref{eq:matrixElementQCDreal,eq:matrixElementEWreal} can be performed with the same methods used in the standard case~\cite{Bredenstein:2006rh}.
Note that all \ac{LO} and (real and virtual) \ac{QCD} amplitudes are related to the corresponding \ac{SM} counterparts by the factor~$\ca$, so that the relative \ac{QCD} corrections to the partial widths (normalized to \ac{LO}) are the same in the \ac{SESM} and \ac{SM}.

\subsubsection{Complex-mass scheme}
\label{sssec:complexMassScheme}

Within the complex-mass scheme~\cite{Denner:1999gp,Denner:2005es,Denner:2005fg}, the renormalized masses of the $\PW$ and $\PZ$ bosons are replaced by the complex masses $\mu_\PW$ and $\mu_\PZ$, defined via the real pole masses $\MW$, $\MZ$, and the decay widths $\GammaW$, $\GammaZ$, of the gauge bosons by
\begin{equation}
\mu^2_\PW = \MWs - \im \MW \GammaW,
\qquad
\mu^2_\PZ = \MZs - \im \MZ \GammaZ.
\end{equation}
The cosine of the weak mixing angle, which is defined by the ratio of the $\PW$- and $\PZ$-boson masses, is replaced by the complex quantity $\cw = \mu_\PW/\mu_\PZ$.
This relation ensures the gauge independence of NLO matrix elements in spite of the use of 
complex W and Z masses,
whose imaginary parts result from a partial resummation of self-energy contributions.
Even though the Higgs particles are unstable, we treat only the $\PW$ and the $\PZ$ bosons in the complex-mass scheme. Indeed, effects induced by a complex Higgs-boson mass are of the order $\Gammah/\Mh$ and, assuming that the light Higgs boson of the \ac{SESM} has a small width (as it happens for the \ac{SM} Higgs), these are negligible compared to the \ac{NLO} contributions considered in this work. The heavy Higgs enters only loop diagrams, so that corrections from a complex mass are negligible, as long as $\GammaH \ll \MH$.

In the complex-mass scheme, the renormalization constants for the $\PW$- and $\PZ$-boson masses are complex to guarantee that~$\mu^2_\PW$ and~$\mu^2_\PZ$ correspond to the complex locations of the~$\PW$ and~$\PZ$-propagator poles.
This implies that the renormalization constant for the weak mixing angle becomes complex as well.
The $\FW$- and $\FZ$-field renormalization constants are defined in the complex-mass scheme by self-energies (and not only by their real parts) which depend on the complex parameters, so that the field renormalization constants are complex. The electric charge renormalization constant, which depends on the (complex) field renormalization constants, 
becomes complex as well.
Explicit definitions for the renormalization constants in the complex-mass scheme are given in \citere{Denner:2005fg} for the SM.
In the SESM, the definitions of the additional renormalization constants $\de \al$ and $\de \lsd$ are not changed in the complex-mass scheme, but the constants are treated as complex quantities, since they are defined using two- and three-point loop functions which contain the complex $\PW$- and $\PZ$-boson masses and complex couplings.

\subsubsection{{\boldmath$\Gf$} scheme}
Adopting the so-called ``$\Gf$ scheme'', we use the Fermi constant $\Gf$ as input parameter and compute the electromagnetic coupling constant $\alem = e^2/(4 \pi)$ according to
\begin{equation}
\label{eq:fermiConstant}
  \alem = \frac{\sqrt{2} \Gf \MW^2}{\pi} \left( 1 - \frac{\MW^2}{\MZ^2} \right).
\end{equation}
In this way, a large universal part of the $\O(\alem)$ corrections is absorbed into the \ac{LO} prediction.
More precisely, this choice absorbs the running of $\alem$ from zero-momentum transfer to the weak scale and the universal corrections to the $\rho$ parameter into the lowest-order coupling $\alem/\sws$. Following this procedure, to avoid double-counting, we have to subtract the \ac{EW} corrections to muon decay from the explicit \ac{NLO} contributions to the electric charge renormalization constant~$\de Z_e$ (see also \citere{Bredenstein:2006rh}),
\begin{equation}
\left. \de Z_e \right|_{\Gf} = \de Z_e - \frac{1}{2} \left( \De r \right)_\text{1-loop},
\end{equation}
where the renormalization constant $\de Z_e$ is given by \cref{eq:electricChargeRenconst}, 
and $\left( \De r \right)_\text{1-loop}$ is the one-loop weak correction to the muon 
decay $\De r$~\cite{Sirlin:1980nh,Denner:1991kt}, but now calculated in the \ac{SESM},
as, e.g., done in \citere{Lopez-Val:2014jva}. 
For consistency, both contributions are computed in the complex-mass scheme.
Using nevertheless the real value for $\alem$ defined in \cref{eq:fermiConstant} is consistent at \ac{NLO}.

\subsection{Implementation into \prophecy{}}
To take advantage of the capabilities of the original \prophecy{} version, we have modified the code in order to include the expressions for the \ac{SESM} matrix elements described in \cref{sec:NLOdecay}. For the \ac{LO} contributions, the \ac{QCD} corrections, and the photonic real-emission contributions to the decay process $\Ph \to 4 \Pf$, this can be easily achieved by rescaling the \ac{SM} Higgs couplings to vector bosons by the appropriate prefactor $\ca$, according to Eqs.\ \eqref{eq:matrixElementLO}, \eqref{eq:matrixElementQCD}, \eqref{eq:matrixElementQCDreal}, and \eqref{eq:matrixElementEWreal}.
For the \ac{EW} virtual corrections, we computed the matrix elements in two independent ways.

In the first computation of the \ac{NLO} matrix elements contributing to the decay $\Ph \to 4 \Pf$, we constructed a model file for the \ac{SESM}, including all the one-loop counterterm vertices and the definitions for the renormalization constants, for the amplitude generator \feynarts~\cite{Hahn:2000kx}.
To produce the model file we used the \mathematica{} package \feynrules~\cite{Christensen:2008py,Alloul:2013bka}.
\feynrules{} allowed us to get the Feynman rules from the \ac{SESM} Lagrangian (including the vertices from the counterterm Lagrangian described in \cref{sec:loopLag}) and to generate the \feynarts{} model file in an automated way.
Afterwards, we have added the definitions of the renormalization constants to the \feynarts{} model file as they are reported in \cref{sec:renorm}, using the \formcalc{} format~\cite{Hahn:1998yk}, both for the \MSb{} and the FJ renormalization schemes.
The model file can be used to generate, to compute, and to simplify one-loop matrix elements with the packages \feynarts{} and \formcalc{} for (in principle) any process within the \ac{SESM}.
The model file has been tested by checking 
\ac{UV} finiteness for many processes, both analytically and numerically, devoting special attention to the multi-scalar vertex functions, which involve the renormalization constants $\de \alpha$ and $\de \lsd$.
We adapted the model file to the demands of \prophecy{}, using complex masses for the gauge bosons and keeping the full mass dependence in the closed fermion loops, and used it to generate the \fortran{} routines for the computation of the virtual matrix elements contributing to the decay $\Ph \to 4 \Pf$. Finally, we have incorporated the \fortran{} code in \prophecy{}.

In the second calculation, we generated the amplitudes using a tree-level \feynarts{} 1~\cite{Kublbeck:1990xc} model file, and we inserted the counterterms by hand and processed them further with in-house \mathematica{} routines.
The results from both calculations are UV- and IR-finite and in good mutual numerical agreement.

\section{Input parameters and benchmark scenarios}
\label{sec:scenarios}
In this section we fix the input parameters used to derive our numerical results.
In \cref{ssec:SMIPS} we present the \ac{SM} input parameter set and in \cref{ssec:BSMconstr} we discuss how the parameter space of the theory is constrained by the requirements of vacuum stability and perturbativity of the couplings. In \cref{ssec:BSMBP} we define the benchmark scenarios 
used for the numerical evaluations.

\subsection{SM parameters}
\label{ssec:SMIPS}

We identify the light Higgs boson $\Ph$ with the known Higgs particle and set
\begin{equation}
 \Mh = 125.1 \, \GeV,
\end{equation}
in agreement with the mass value measured by ATLAS and CMS~\cite{Aad:2015zhl}.
The numerical values of the other parameters are fixed according to the recommendations of the \ac{HXSWG}~\cite{deFlorian:2016spz}, mostly based on \citere{Agashe:2014kda}. The Fermi and the strong coupling constants are
\begin{equation}
\Gf = 1.1663787 \cdot 10^{-5} \, \GeV^{-2},
\qquad
\alphas = 0.118.
\end{equation}
We simply take $\alphas$ at the scale of the Z-boson mass, i.e.\ we do not change
the QCD renormalization scale in the scale variations discussed below, because 
it merely leads to changes at next-to-next-to-leading order, which are part of the
residual theoretical uncertainty from missing higher orders.
The \ac{OS} gauge-bosons masses and widths and the fermion masses are
\begin{align}
 \label{eq:inputMasses}
  \MW^\text{OS} &= 80.385 \, \GeV, & \GammaW^\text{OS} &= 2.085 \, \GeV, && \notag
  \\ \notag
  \MZ^\text{OS} &= 91.1876 \, \GeV,& \GammaZ^\text{OS} &= 2.4952 \, \GeV, &&
  \\
  \me   &= 0.510998928 \, \MeV, & \mmu  &= 105.6583715 \, \MeV, & \mtau &= 1776.82 \, \MeV,
  \\ \notag
  \muq &= 0.1 \, \GeV, & \mcq &= 1.51 \, \GeV, & \mtq  &= 172.5 \, \GeV,
  \\ \notag
  \mdq &= 0.1 \, \GeV, & \msq &= 0.1 \, \GeV,  & \mbq  &= 4.92 \, \GeV.
\end{align}

For a consistent use of the complex-mass scheme~\cite{Denner:1999gp,Denner:2005es,Denner:2005fg}, we convert the experimental values of the \ac{OS} masses and widths of the vector bosons reported in \cref{eq:inputMasses} to the related pole quantities by
\begin{equation}
\label{eq:convOSPole}
  M_V = \frac{M_V^\text{OS}}{\sqrt{1 + \left( \Gamma_V^\text{OS}/M_V^\text{OS} \right)^2 }},
  \quad
  \Gamma_V = \frac{\Gamma_V^\text{OS}}{\sqrt{1 + \left( \Gamma_V^\text{OS}/M_V^\text{OS} \right)^2 }},
  \quad
  V = \PW,\PZ.
\end{equation}
In the numerical analysis, we use the $\PW$- and $\PZ$-boson masses obtained from \cref{eq:convOSPole}. The decay widths $\GammaW$ and $\GammaZ$ are calculated from the given experimental input, taking into account $\O(\alem)$ corrections and using real masses. We do not use the pole widths of \cref{eq:convOSPole}, but we compute $\Gamma_\PV$ at \ac{NLO}, in order to ensure that the effective~$\PW/\PZ$ branching ratios add up to one in the sum over all decay channels. In this step, we neglect effects due to the presence of the singlet; in principle, it contributes to the \ac{NLO} corrections to the~$\PW$ and~$\PZ$ widths, but for the small $\alpha$ value we consider, the effect is negligible.

As in the original \ac{SM} version of \prophecy{} the full dependence on the fermion masses given in \cref{eq:inputMasses} is kept in corrections induced by closed fermion loops, while external (light) fermions are treated in the massless limit.
Since quark mixing to the third generation as well as the differences in the (internal) light-quark
masses are negligible, the CKM matrix drops out in the calculation of the inclusive (flavour-summed) width
$\Gamma_{\Ph\to4f}$. We, thus, set the CKM matrix to the unit matrix in the following.

\subsection{Constraints on BSM parameters}
\label{ssec:BSMconstr}

Even without taking into account the data collected from the experiments, the parameter space of 
the \ac{SESM} is limited by theoretical 
constraints~\cite{Pruna:2013bma,%
Robens:2015gla,%
Kanemura:2015fra,%
Bojarski:2015kra,%
Costa:2015llh,%
Robens:2016xkb}. 
Before choosing the input values for the free parameters of the theory, it is worth recalling these constraints and how the free parameters can fulfill such conditions. 

\paragraph{Perturbativity of the couplings:}
The scalar couplings of the \ac{SESM} must not exceed a certain value, so that the perturbative approach used in the calculations remains valid.
Thus, we require that the contributions from the coupling constants $\la_i$ to the coefficients of the quartic coupling terms in the Higgs potential,
\begin{equation}
V_4 = \frac{\la_2}{16} \Fh_2^4 + \la_1 \Fh_1^4 + \frac{\la_{12}}{2} \Fh_2^2 \Fh_1^2,
\end{equation}
respect some limit $\O( |\la_i| / \pi ) \lesssim 1$.
The following choice is made in order to replicate the results of \citere{Robens:2016xkb}, where a similar analysis was performed using a different input parameter set and different conventions. The conditions from there translate into the bounds
\begin{equation}
\label{eq:BSMconstrPert}
 |\lambda_1| < \pi, \quad
 |\lambda_2| < 16 \pi, \quad
 |\lambda_{12}| < 2 \pi.
\end{equation}
These values are meant to be rough estimates that are used to show where perturbativity problems can arise, rather than sharp boundaries on the allowed values. 

\paragraph{Vacuum stability:}
As discussed in \cref{ssec:higgsLag}, vacuum stability at \ac{LO} is guaranteed by the conditions given in \cref{eq:vacuumStabCond2}.

\begin{figure}
\centering
\includegraphics[scale=0.9]{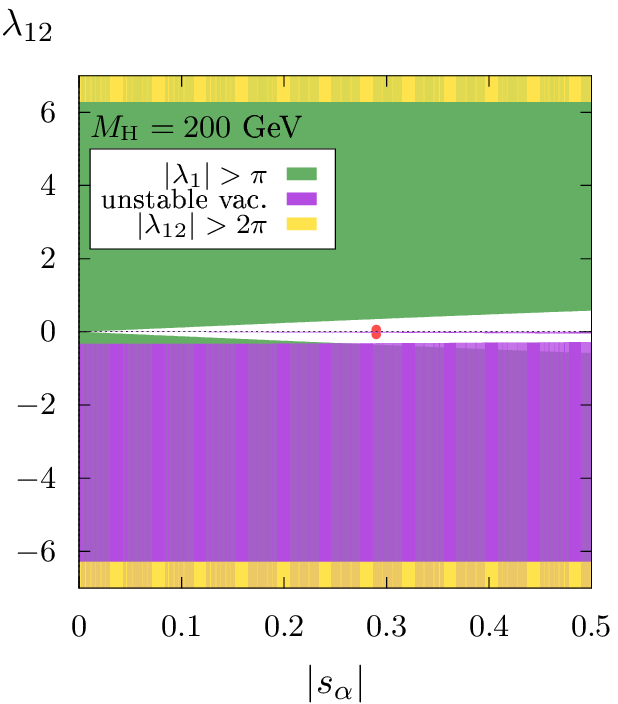}
\qquad
\includegraphics[scale=0.9]{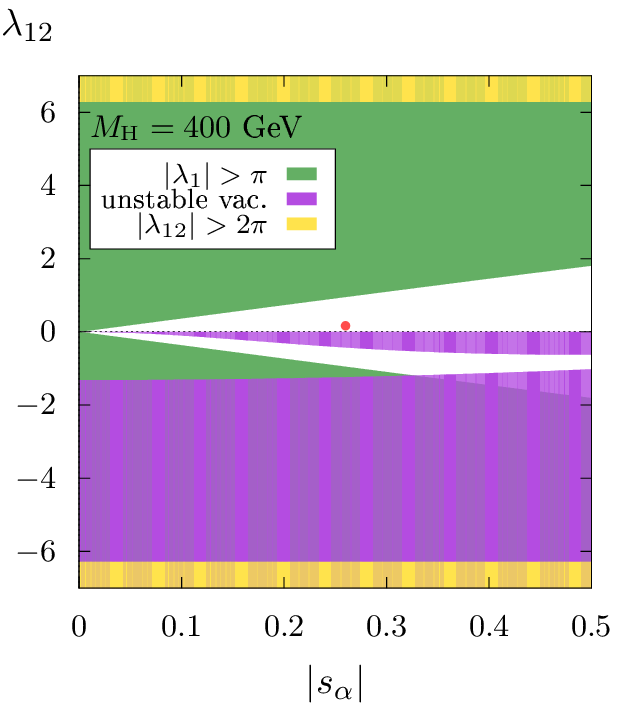}
\\ \vspace{10pt}
\includegraphics[scale=0.9]{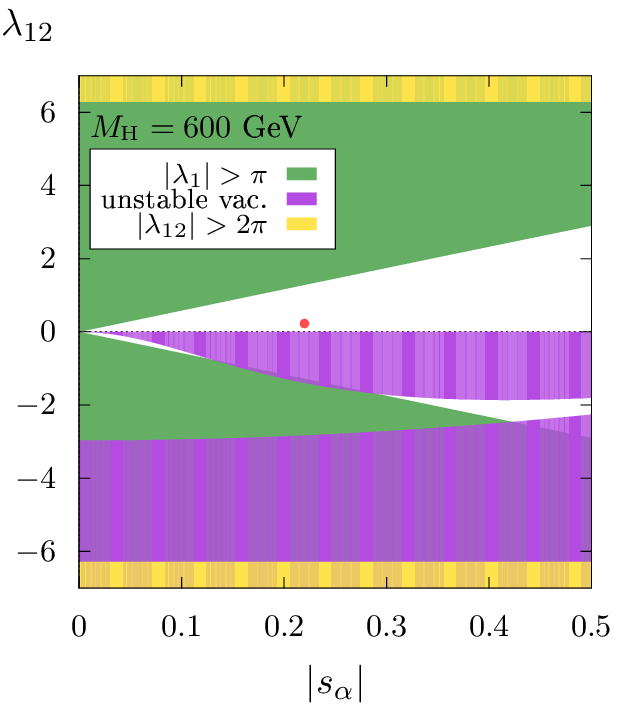}
\qquad
\includegraphics[scale=0.9]{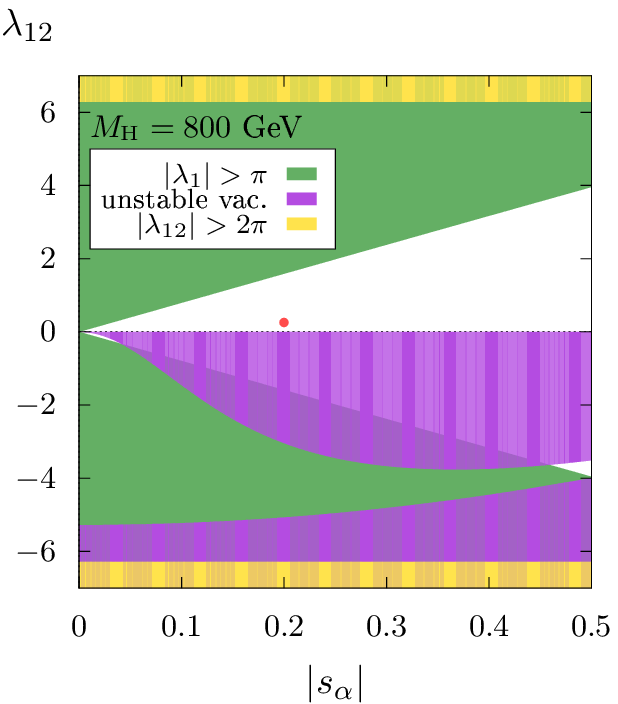}
\caption{Theory constraints on the parameters $\lsd$ and $\sa$, for different $\MH$ values. Note that $\lsd > 0 \, (<0)$ corresponds to $\sa > 0 \, (<0)$. The green and the yellow regions are excluded, respectively, by perturbativity constraints on the couplings $\lambda_1$ and $\lsd$. The purple region is excluded by the vacuum stability constraint. The white regions correspond to the allowed parameter space, and the red dots correspond to the chosen benchmark scenarios.}
\label{fig:inputparamconstr}
\end{figure}

\paragraph{}
In \cref{fig:inputparamconstr} we show the effects of the requirements of perturbativity~\eqref{eq:BSMconstrPert} and vacuum stability~\eqref{eq:vacuumStabCond2} on the input parameter space for different heavy Higgs masses in the range $\MH = 200{-}800 \, \GeV$. 
The allowed region is restricted to the white area, and it is possible to see how the condition on $\lambda_1$ presented in \cref{eq:BSMconstrPert} plays an important role both for negative and positive $\lsd$ values. The vacuum stability condition~\eqref{eq:vacuumStabCond2} has an impact only on negative $\lsd$ values, ruling out a large part of space that is not excluded by perturbativity requirements. Green and yellow areas indicate regions where one can expect perturbativity problems, but should not be intended as sharp-cut regions. For the $\MH$ values considered here ($200, 400, 600$ and $800 \, \GeV$), the perturbativity of $\lsd$ does not affect the parameter space, but it becomes relevant for higher $\MH$ values. The perturbativity constraint on the coupling~$\lambda_2$ is irrelevant in the considered regions.

\subsection{Benchmark scenarios}
\label{ssec:BSMBP}
For the numerical analysis we consider some of the benchmark scenarios proposed in \citere{deFlorian:2016spz}, which were originally suggested in \citere{Robens:2016xkb}, adapting the input values to our needs. In \citere{Robens:2016xkb}, different values for the mass $\MH$ are considered (both lighter and heavier than $\Mh$), and for each mass the mixing angle~$\al$ is fixed to the maximal allowed value. Moreover, for mass values $\MH \ge 2 \Mh$ (\ie when the $\PH \to \Ph \Ph$ decay is kinematically allowed), two values are proposed for $\tan \beta \equiv v_2/v_1$, corresponding to the maximal and the minimal branching ratios for the $\PH \to \Ph \Ph$ decay.

Among these possibilities, we only consider scenarios in which $\MH > \Mh$ (since the other possibility is phenomenologically disfavoured) and vary the heavy Higgs mass in the interval $200{-}800 \, \GeV$ with $200 \, \GeV$ steps. When two different $\tan \beta$ values are proposed, we consider the average. Since $\tan \beta$ enters in our calculation only at \ac{NLO} and the two proposed values are always quite close, we expect negligible differences due to this choice. 

We have to convert the numerical values of the input parameters given in \citere{Robens:2016xkb} to our conventions. The \ac{SESM} Higgs Lagrangian used here, as given in \cref{eq:scalarLag}, is equivalent to the one given in \citere{Robens:2016xkb} using the following substitutions,
\begin{equation}
\label{eq:convParamRef}
\begin{split}
 \mu^\text{ref} & \to \sqrt{2} \mu_1, \quad
 \lambda_1^\text{ref} \to \frac{\lambda_2}{4}, \quad
 \lambda_2^\text{ref} \to 4 \lambda_1, \quad
 \lambda_3^\text{ref} \to 2 \lsd,\\&
 \tan \beta^\text{ref} \to \frac{v_2}{v_1}, \quad
 v^\text{ref} \to v_2, \quad
 x^\text{ref} \to v_1,
\end{split}
\end{equation}
where the label ``ref'' indicates the parameters used in \citere{Robens:2016xkb}, in which the numerical input is given in terms of $\MH$, $\al$,  and $\tan \beta^\text{ref}$. The heavy mass $\MH$ and the mixing angle $\al$
can be taken over directly. 
The scalar coupling $\lsd$, which is a free parameter in our conventions, can be obtained from $\tan \beta^\text{ref}$ using the relations
\begin{equation}
\label{eq:convParamlsd}
 \lsd = \frac{\ca \sa}{2 v_2 v_1} \left( \MH^2 - \Mh^2 \right),
 \quad
 \tan \beta^\text{ref} = \frac{v_2}{v_1}.
\end{equation}
We convert the benchmark points using \cref{eq:convParamRef,eq:convParamlsd}, rounding the $\lsd$ values to two decimal digits. The input values, in our convention, for the scenarios considered in our analysis are reported in \cref{tab:benchmarkPoints} (together with the corresponding $\tan\beta^\text{ref}$ values).
\begin{table}
 \[
 \begin{array}{lccc}
  \toprule
  \text{Scenario} & \MH [\GeV] & \sin \alpha & \lsd \, (\tan \beta^\text{ref}) \\
  \midrule
  \text{\BHMapm} & 200 &  \pm 0.29 &  \pm 0.07 \, (1.19) \\
  \text{\BHMb}   & 400 &      0.26 &      0.17 \, (0.585) \\
  \text{\BHMc}   & 600 &      0.22 &      0.23 \, (0.375) \\
  \text{\BHMd}   & 800 &      0.20 &      0.26 \, (0.260) \\
  \bottomrule
 \end{array}
 \]
 \caption{Input values for the \ac{SESM} for a selection of benchmark scenarios for the \ac{SESM} proposed in \citeres{deFlorian:2016spz,Robens:2016xkb}, converted to the notation used in this work. In brackets, the $\tan \beta^\text{ref}$ values used in \cref{eq:convParamlsd} to compute the corresponding $\lsd$ values are given.}
 \label{tab:benchmarkPoints}
\end{table}
For $\MH = 200 \, \GeV$ we discuss both signs of~$\sa$ with~$|\sa| = 0.29$; for higher~$\MH$ values we consider only positive~$\sa$ values, since the corresponding negative values are ruled out by the vacuum stability constraint.
In the following we will make use of these scenarios in each of the
renormalization schemes proposed in this paper.

\section{Numerical analysis}
\label{sec:num}
%
In the following, we present the numerical results relevant for the decay $\Ph \to 4 \Pf$ of the light Higgs boson of the \ac{SESM}. Starting from 
benchmark scenario \BHMap, we show the effects of the conversion of the input variables between the two renormalization schemes presented in \cref{sec:renorm}.
Then we investigate the scale dependence of the parameters $\al$ and $\lsd$, which are defined by $\MSbar$ renormalization conditions, by solving numerically the corresponding \acp{RGE}. Afterwards, we present the results for the decay width $\Gamma_{\Ph \to 4 \Pf}$ computed at different renormalization scales and show the deviations from the \ac{SM} results as a function of the mixing angle. The same analysis is presented for 
benchmark scenario \BHMc{}, while results for the scenarios \BHMam{} and \BHMb{} are reported, respectively, in \cref{app:resBHMb,app:resBHMam}.
Finally, we show some differential distributions, comparing the results in the \ac{SM} with the ones in the benchmark scenarios of \cref{tab:benchmarkPoints}.

\subsection{\BHMap}
\label{ssec:resultsBHM200}

\subsubsection{Scheme conversion}
\label{sssec:schemeConv}
When computing a physical observable at \ac{NLO} accuracy, starting from a set of input parameters, it is crucial to realize that the input values correspond to a specific renormalization scheme adopted in the calculation.
This becomes even more important when comparing \ac{NLO} results for the same observable obtained using different renormalization schemes. In different schemes, the same numerical values for the input parameters represent different physical scenarios and, in order to have a sensible comparison of predictions for an observable in a given scenario, a proper conversion of the input parameters between the schemes is required.

In general, defining $N$ renormalized parameters $p_i$ in two renormalization schemes, denoted, respectively, by $p_i^{(1)}$ and $p_i^{(2)}$, the relation between them is given by the solution of the following system of equations,
\begin{equation}
\label{eq:schemeConv}
p_{i,0} = p_i^{(1)} + \de p_i^{(1)} \bigl( p_1^{(1)}, \dots, p_N^{(1)} \bigr) = p_i^{(2)} + \de p_i^{(2)} \bigl( p_1^{(2)}, \dots, p_N^{(2)}  \bigr),
\end{equation}
where the connection between the parameters in the two schemes is given by the bare parameters $p_{i,0}$, which are independent of the renormalization scheme.
In our particular case, converting the input values from the \MSb{} to the FJ scheme is quite simple, since, apart from the mixing angle $\al$, all the other input parameters of the \ac{SESM} have the same definition in the two schemes.
Ignoring effects beyond \ac{NLO}, the input parameters $p_i \neq \al$ are defined by identical renormalization conditions in the two schemes, \ie
\begin{equation}
p_{i,0} = p_i^\MSbar + \de p_i \bigl( \al^\MSbar, \bigl\{ p_i^\MSbar \bigr\} \bigr)
        = p_i^\text{FJ} + \de p_i \bigl( \al^\text{FJ}, \bigl\{ p_i^\text{FJ} \bigr\} \bigr),
\end{equation}
with identical counterterm functions $\de p_i$ at \ac{NLO}.
This implies $p_i^\MSbar = p_i^\text{FJ} + \O \left( \alem^2 \right)$, and we do not distinguish between $p_i^\MSbar$ and $p_i^\text{FJ}$ for parameters other than $\al$.
\Cref{eq:schemeConv} reduces to
\begin{equation}
\label{eq:schemeConvMixing}
\al_0 = \al^\MSbar + \de \al^\MSbar\bigl(\al^\MSbar\bigr) = \al^\text{FJ} + \de \al^\text{FJ} \bigl(\al^\text{FJ} \bigr).
\end{equation}
To solve the equation and find the relation between $\al^\MSbar$ and $\al^\text{FJ}$, we adopt two strategies.
In the first approach we linearize \cref{eq:schemeConvMixing}
and obtain
\begin{equation}
\label{eq:schemeConvLin}
\al^\text{FJ} = \al^\MSbar + \de \al^\MSbar\bigl(\al^\MSbar\bigr) - \de \al^\text{FJ}\bigl( \al^\MSbar \bigr) + \O(\alem^2).
\end{equation}
Since our computations are performed at \ac{NLO}, the $\O(\alem^2)$ term in \cref{eq:schemeConvLin} can be neglected. An analogous procedure can be applied to determine $\al^\MSbar$ when $\al^\text{FJ}$ is given as input. Using this method, converting an input value for the mixing angle from one scheme to the other and repeating the procedure to go back to the initial scheme, the final numerical result for $\al$ will change by contributions that are formally beyond \ac{NLO}.

In the second approach, we solve \cref{eq:schemeConvMixing} numerically, in order to keep the contributions of 
$\O(\alem^2)$, which can become relevant for large counterterms or small tree-level values.
Using this method, converting $\al$ to the other scheme and back, does not change the value of $\al$.
In the following results we use, as much as possible, the second method, \ie we include the $\O(\alem^2)$ terms.

\begin{figure}
\centering
\includegraphics[scale=1.]{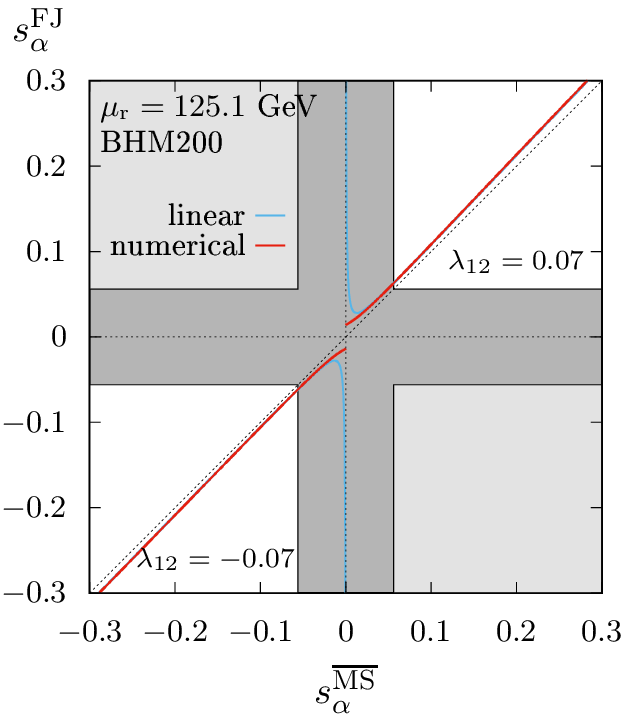}
\qquad
\includegraphics[scale=1.]{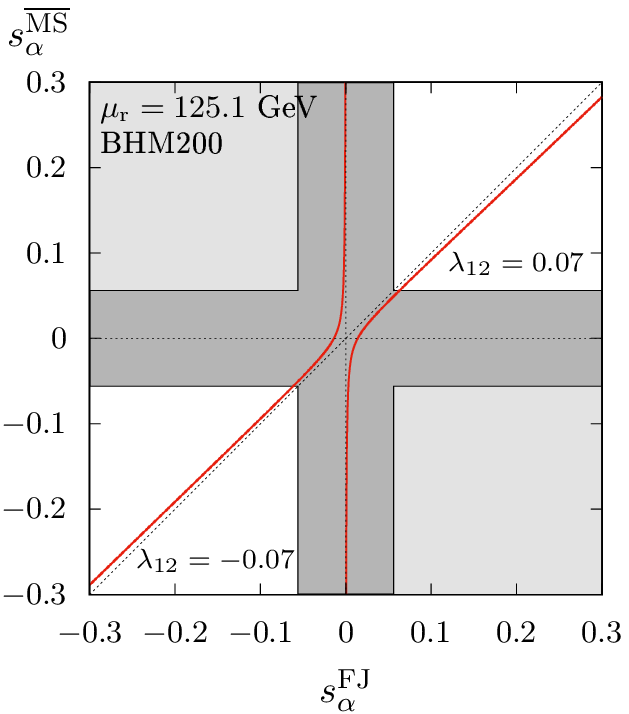}
\caption{Conversion of the input parameter $\sa$ between the \MSb{} and the FJ schemes. The heavy Higgs-boson mass $\MH$ is fixed according to the benchmark scenarios \BHMapm. The dark-gray area denotes~$\sa$ values for which~$\la_1$ becomes non-perturbative. In the light-gray regions the sign of~$\sa$ is flipped by the conversion and becomes inconsistent with the sign of~$\lsd$.}
\label{fig:schemeconversion2BHM200}
\end{figure}

In {\cref{ssec:MSbarren}}
we have derived the counterterm $\de \al$ in the two schemes, which differs by finite contributions,
\begin{equation}
\begin{split}
\left. \de \al^\MSbar \right|_\text{finite} &= 0,
\\
\left. \de \al^\text{FJ} \right|_\text{finite} &= - \left. \De \al^t \left( T_\Ph, T_\PH \right) \right|_\text{finite} = \frac{e}{\Mhs - \MHs} \left. \left( T_\Ph \, \frac{C_{\Ph\Ph\PH}}{\Mhs} + T_\PH \, \frac{C_{\Ph\PH\PH}}{\MHs} \right) \right|_\text{finite},
\end{split}
\end{equation}
where the tadpoles $T_\Ph$, $T_\PH$, and the coupling factors $C_{\Ph\Ph\PH}$, $C_{\Ph\PH\PH}$ in the last term depend on $\al^\text{FJ}$. 
Using these expressions, it is straightforward to get the conversion from the FJ to the \MSb{} scheme,
\begin{equation}
\label{eq:FJtoMRconversion}
\al^\MSbar = \al^\text{FJ} - \left. \De \al^t \left( T_\Ph, T_\PH \right) \right|_\text{finite},
\end{equation}
while the conversion from \MSb{} to FJ requires a numerical solution of \cref{eq:FJtoMRconversion} for $\al^\text{FJ}$, which appears also in the $\De \al^t$ term.
In \cref{fig:schemeconversion2BHM200}, we show the results for the conversion of the sine of the mixing angle, $\sa$, 
between the two schemes,
{both for the full solution of Eq.~\eqref{eq:schemeConvMixing} 
and using the linearized solution \eqref{eq:schemeConvLin}.%
\footnote{If Eq.~\eqref{eq:schemeConvMixing} is used for the conversion,
the corresponding curves in the two plots are related by a simple reflection about the
diagonal $\sa^\MSbar \equiv \sa^\text{FJ}$ (apart from the different truncation of the curves 
in the non-perturbative region).
The reflection symmetry is not there in the linearized version 
\eqref{eq:schemeConvLin}, but broken by effects beyond NLO. Note also that both
versions coincide on the r.h.s., because $\de\alpha^\MSbar$ does not contain finite contributions
(UV divergent terms are canceled analytically).}
The curves on the right sides inside the plots}
are obtained fixing the mass $\MH$ of the heavy Higgs boson and the coupling $\lsd$ according to their values in the scenario \BHMap{}, reported in \cref{tab:benchmarkPoints}.
On the left sides,
similar curves show the conversion effects for negative~$\sa$ values. For consistency, we adjust the sign of~$\lsd$ so that $\sgn(\sa) = \sgn(\lsd)$ (for the input~$\sa$) and \cref{eq:mixingAngle} is not violated.
The renormalization scale is fixed to the mass of the light Higgs boson, $\mur = \Mh$; the motivation for this choice will become clear in \cref{sssec:scaleDepBHM200}.
The dark-gray shaded areas in the plots mark the values of $\sa$ for which the perturbativity constraint 
\eqref{eq:BSMconstrPert} on $\la_1$ is violated; 
{from the last line of Eq.~\eqref{eq:inversionRel} it is easily seen that $\la_1$ necessarily violates its
perturbativity bound for $\sa\to0$, since we keep $\Mh,\MH,v_2,\lambda_{12}$ fixed.}
The light-gray shaded areas denote regions where the sign of~$\sa$ is flipped by the conversion and becomes inconsistent with the sign of the considered~$\lsd$.
The conversion effects 
{in the perturbative regions}
are small: The red line is, in general, very close to the dashed diagonal line, which corresponds to the absence of any conversion effect (i.e.~$\sa^\MSbar = \sa^\text{FJ}$),
{and the linearized solution reproduces the full conversion very well.
Large effects (and deviations between full and linearized solutions)}
are only observed when approaching the non-perturbative regime, corresponding to small values of the mixing angle.
In both plots of \cref{fig:schemeconversion2BHM200}, a slight asymmetry can be observed between positive and negative~$\sa$ values, due to the different \ac{NLO} contributions obtained 
by changing the sign of the input values for~$\sa$ and~$\lsd$.

\subsubsection{Running of {\boldmath$\sa$} and {\boldmath$\lsd$}}
\label{sssec:runningParam}
Since we have defined the parameters $\al$ and $\lsd$ by $\MSbar$ renormalization conditions, they depend on an unphysical renormalization scale $\mur$. The dependence on this scale is governed by the \acp{RGE}
\begin{equation}
\label{eq:runningCoupl}
\frac{\partial}{\partial \ln \murs} \,  \al \left( \murs \right) = \beta_\al \left( \murs \right),
\qquad
\frac{\partial}{\partial \ln \murs} \, \lsd \left( \murs \right) = \beta_{\lsd} \left( \murs \right),
\end{equation}
where the $\beta_\al$ and $\beta_{\lsd}$ functions can be extracted from the expressions of the counterterms $\de \al$ and $\de \lsd$, taking the coefficients of the \ac{UV} divergence $\DeUV$. These functions are different for the two considered renormalization schemes: For the \MSb{} scheme, the $\be$ functions can be obtained considering the following derivatives with respect to the \ac{UV} divergence,
\begin{equation}
\be_{\al^\MSbar} = \frac{\partial}{\partial \DeUV} \de \al^\MSbar,
\qquad
\be_{\lsd^\MSbar} = \frac{\partial}{\partial \DeUV} \de \lsd^\MSbar,
\end{equation}
where the counterterm $\de \al^\MSbar$ is given in \cref{eq:renConstMixingMRferm,eq:renConstMixingMRbos}, and $\de \lsd^\MSbar$ in \cref{eq:renConstlsdMRferm,eq:renConstlsdMRbos}. For the FJ renormalization scheme, also the \ac{UV} contributions due to the tadpoles must be taken into account, leading to the $\be$ functions
\begin{equation}
\be_{\al^\text{FJ}} = \be_{\al^\MSbar} + \frac{\partial}{\partial \DeUV} \De \al^t \left( T_\Ph, T_\PH \right),
\qquad
\be_{\lsd^\text{FJ}} = \be_{\lsd^\MSbar},
\end{equation}
where $\be_{\lsd}$ is not changed due to the fact that $\de \lsd$ is the same in the two schemes.

\begin{figure}
\centering
\includegraphics[scale=1.]{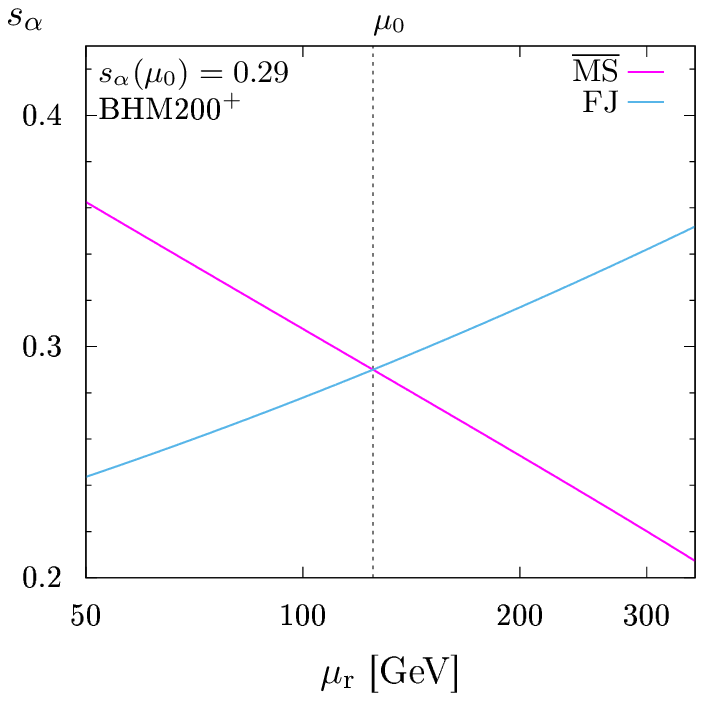}
\qquad
\includegraphics[scale=1.]{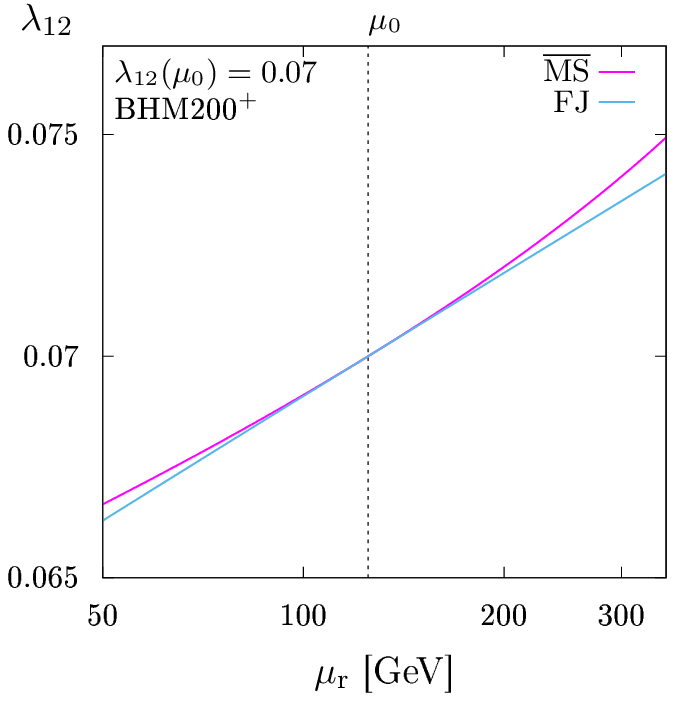}
\caption{The running of the input parameters $\sa$ and $\lsd$ in the \MSb{} and FJ renormalization schemes for benchmark scenario \BHMap.}
\label{fig:murscanparam2BHM200}
\end{figure}

The \acp{RGE} are coupled differential equations for which, in general, an analytical solution is not possible. 
We solve the equations numerically, using a Runge$-$Kutta algorithm, obtaining the scale dependence for the sine of the mixing angle, $\sa$,
and the coupling $\lsd$, as shown in \cref{fig:murscanparam2BHM200}.
These results are
obtained fixing the parameters $\MH$, $\al$, and $\lsd$ according to the values reported in \cref{tab:benchmarkPoints} for 
benchmark scenario \BHMap{} at the scale $\mu_0 = \Mh$, and changing the scale $\mur$ in the range $50{-}350\, \GeV$, using the $\be$ functions for the two schemes. Since the purpose here is the assessment of the scale dependence of the $\MSbar$ parameters, no conversion between schemes is
applied on the input values.
In the two schemes, the scale dependence of $\sa$ shows a completely different behaviour, while the running of $\lsd$ displays the same trend in the two schemes. As we will discuss below, the scale dependence of the mixing angle has a big impact on the scale variation of the decay width $\Gamma_{\Ph \to 4 \Pf}$.

\subsubsection{Scale dependence of the inclusive decay width {\boldmath$\Gamma_{\Ph \to 4 \Pf}$}}
\label{sssec:scaleDepBHM200}
\begin{figure}
\centering
\includegraphics[scale=1.]{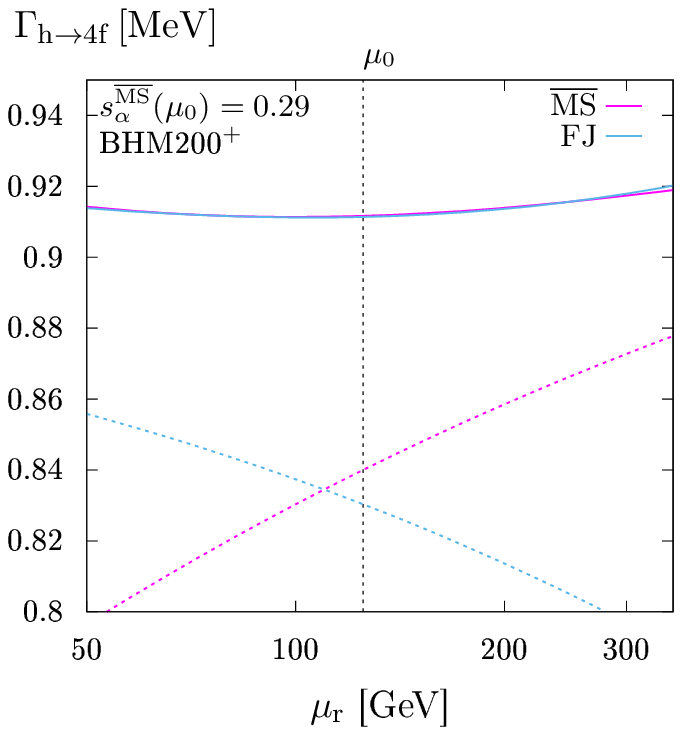}
\qquad
\includegraphics[scale=1.]{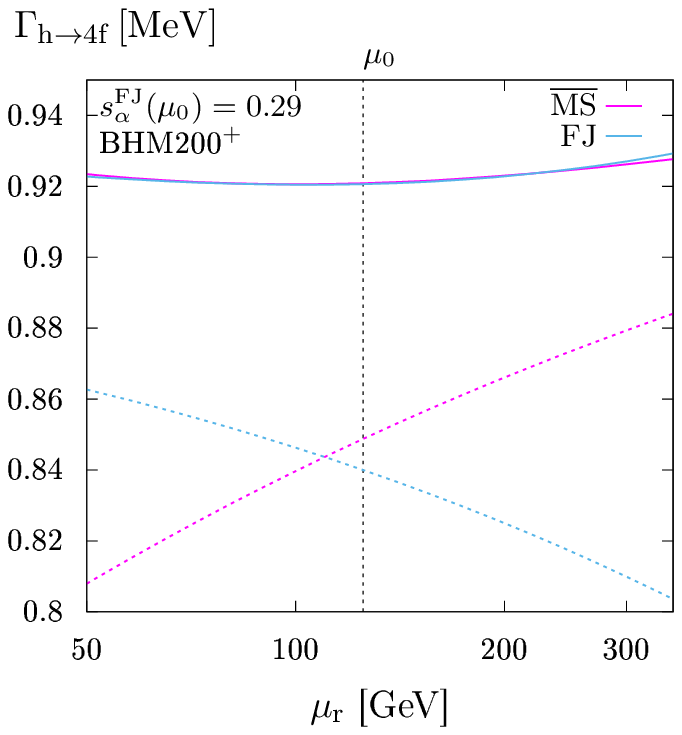}
\caption{Scale dependence of the decay width $\Gamma_{\Ph \to 4 \Pf}$ at \ac{LO} (dashed curves) and NLO EW+QCD (solid curves) for benchmark scenario \BHMap, using the central scale $\mu_0 = \Mh$. On the left (right) the input parameters are defined in the \MSb{} (FJ) scheme and converted to the FJ (\MSb{}) scheme at the scale $\mu_0$ (both for LO and NLO predictions).}
\label{fig:murscan2BHM200}
\end{figure}

In order to assign a sensible value to the renormalization scale $\mur$, we study the impact of the scale choice on the results for $\Gamma_{\Ph \to 4 \Pf}$, the inclusive decay width of the light Higgs into four fermions.
Taking the light Higgs mass as central renormalization scale, \ie $\mu_0 = \Mh$, we have computed the decay width at the scale $\mur$ in the range $50{-}350 \, \GeV$. 
The results are obtained fixing the input values for $\MH$, $\al$, and $\lsd$ at the central scale according to the scenario \BHMap{} of \cref{tab:benchmarkPoints} and shown in \cref{fig:murscan2BHM200}. In the figure, dashed lines correspond to the results for the \ac{LO} decay width, and solid lines include NLO EW and QCD corrections.
Magenta and blue lines represent, respectively, the results obtained in the \MSb{} and in the FJ renormalization schemes.
On the left (right) panel the input parameters are defined in the \MSb{} (FJ) scheme and converted to the FJ (\MSb{}) scheme at the scale $\mu_0$. 
The conversion 
is applied both for LO and NLO predictions.
Note that the difference in the \ac{LO} width observed at the central scale $\mur = \mu_0 = \Mh$ reflects the effect of
the scheme conversion of~$\sa$.
Recalling that the expression of the \ac{LO} matrix element is proportional to $\ca$, the behaviour of the \ac{LO} results as a function of the renormalization scale is explained by the running of the mixing angle (shown in \cref{fig:murscanparam2BHM200}).
%

Comparing \ac{LO} and \ac{NLO} results, it is evident how the inclusion of loop contributions drastically reduces the scale dependence, as well as the scheme dependence: The solid lines are not only much flatter than the dashed lines, proving a reduced scale dependence, but solid curves of different colours are much closer to each other than the corresponding dashed curves. This is expected, since results obtained within different renormalization schemes should be equal up to higher-order contributions, provided that the input parameters are properly converted.
The \ac{NLO} predictions, in particular, perfectly coincide at the central scale $\mur = \mu_0 = \Mh$.
Quantifying the scale dependence by the change in $\Gamma_{\Ph \to 4 \Pf}$ obtained by varying the
scale $\mur$ by factors of two up and down ($\mur=2\mu_0$ and $\mur=\mu_0/2$), we observe
a reduction from $\sim3{-}4\%$ at LO to $\lsim0.5\%$ at NLO.
The scheme dependence, defined by the relative difference between the results in the $\MSbar$ and FJ schemes
at the central scale, on the other hand, reduces from $\sim1\%$ at LO to $\lsim0.1\%$ at NLO.

As discussed in \citeres{Altenkamp:2017ldc,Altenkamp:2017kxk} for \ac{NLO} predictions of $\Gamma_{\Ph \to 4 \Pf}$ in the \ac{THDM}, when the computation involves multiple mass scales in the loops, it is not clear a priori that the central scale $\mu_0 = \Mh$ is an appropriate choice. In principle, this applies also to the \ac{SESM}, where the heavy Higgs boson appears in loop diagrams. However, it is evident from \cref{fig:murscan2BHM200} how the scale dependence is minimized for values around the light Higgs mass.
On the other hand, using the alternative scale $\mu_0 = (\Mh + \MH)/2$ analogous to the scale choice advocated in \citere{Altenkamp:2017ldc} for the \ac{THDM}, would not make a big difference for the scenario \BHMap, due to the relatively small value for the heavy Higgs mass, fixed at $\MH = 200 \, \GeV$. As discussed later in the section, repeating the same analysis on the scale variation with higher $\MH$ values, we will see how the scale $\mu_0 = \Mh$ is better than the alternative scale given by the arithmetic mean of the Higgs-boson masses.
Note that finding optimal scales, to some extent, is empirical, i.e.\ the scale dependence should be investigated whenever qualitatively new scenarios are considered.

\subsubsection{Mixing-angle dependence of the inclusive decay width {\boldmath$\Gamma_{\Ph \to 4 \Pf}$}}
\begin{figure}
\centering
\includegraphics[scale=1.]{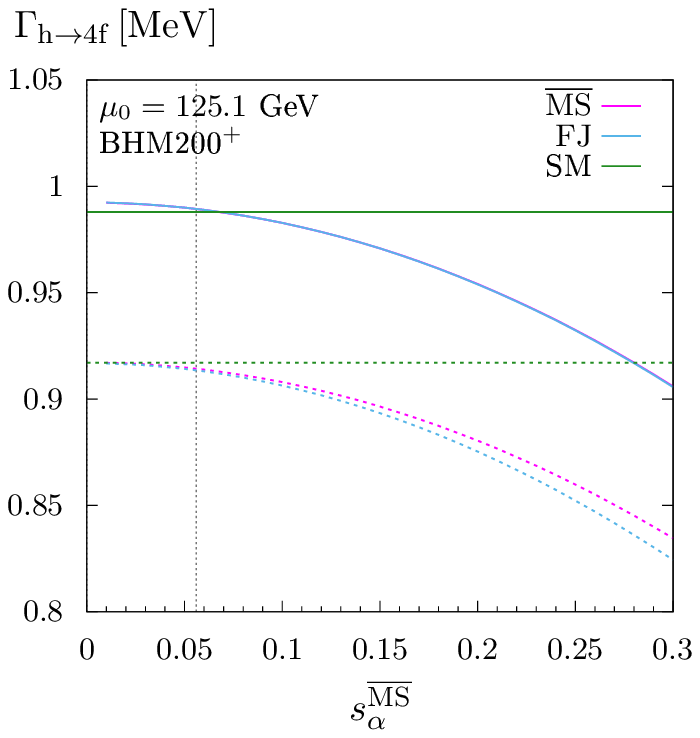}
\quad
\includegraphics[scale=1.]{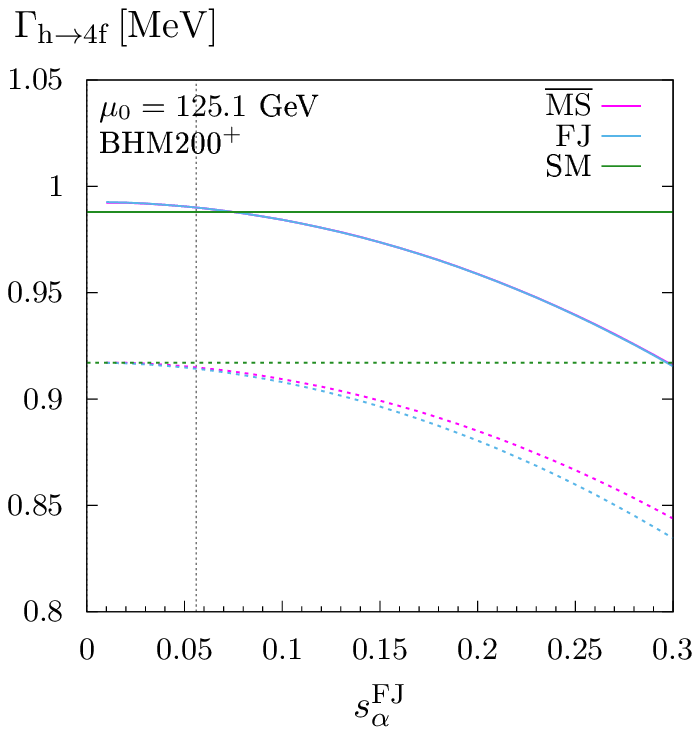}
\caption{Dependence of the decay width $\Gamma_{\Ph \to 4 \Pf}$ at LO and NLO EW+QCD with respect to the variation of $\sa$. Free parameters other than $\sa$ are fixed according to benchmark scenario \BHMap{} (see \cref{tab:benchmarkPoints}).
On the left (right) the input parameters are defined in the \MSb{} (FJ) scheme and converted to the FJ (\MSb{}) scheme (both for LO and NLO predictions).
The vertical dashed line signals that for $\sa \lesssim 0.06$ perturbativity problems might arise.}
\label{fig:sascanBHM200}
\end{figure}

Among the free parameters of the \ac{SESM}, the mixing angle~$\al$ plays the central role in the computation of the decay width $\Gamma_{\Ph \to 4 \Pf}$. Its value affects already the \ac{LO} result, while the heavy Higgs mass $\MH$ and the coupling $\lsd$ enter only the \ac{NLO} decay amplitudes. For this reason, we compute the decay width varying $\sa$ in the range $0.01{-}0.3$, keeping $\MH$ and $\lsd$ fixed according to the values for 
benchmark scenario \BHMap.%
\footnote{Within this analysis, we exclude the value $\sa = 0$, since in this case \cref{eq:inversionRel} would imply $v_1 = 0$, for a given non-vanishing $\lsd$.}
We consider both \MSb{} and FJ as input schemes and compute the decay width in the two schemes, using the renormalization scale $\mur = \Mh$.
The results for $\Gamma_{\Ph \to 4 \Pf}$ are reported in \cref{fig:sascanBHM200}, where we also show the \ac{SM} value of the decay width identifying~$\Ph$ with the \ac{SM} Higgs boson.
Dashed and solid lines denote, respectively, the \ac{LO} and the \ac{NLO} results, where the latter include both EW and QCD corrections. The dashed vertical line indicates the minimal $\sa$ value for which the perturbativity conditions of \cref{eq:BSMconstrPert} are satisfied. Differences between the \ac{LO} results in the two schemes (within the same plot) are due to the scheme conversion, which is done at \ac{NLO}.
Comparing the \ac{LO} results, it is possible to observe the suppression with respect to the \ac{SM} given by the $\cas$ factor, coming from the square of \cref{eq:matrixElementLO}. The proportionality to $\cas = 1 - \sas$ is exact if no conversion of the input is done, \ie on the left (right) side for the LO \MSb{} (FJ) curve.
The \ac{NLO} contributions modify the $\sa$ dependence, so that for $\sa \sim 0.08$ the \ac{NLO} decay width is the same as in the \ac{SM}, and in general the difference between \ac{SESM} and \ac{SM} are smaller for the \ac{NLO} decay width.
Note also the reduction of the scheme dependence at NLO, visible by the fact that \MSb{} and FJ curves practically lie on top of each other.

\begin{figure}
\centering
\includegraphics[scale=0.9]{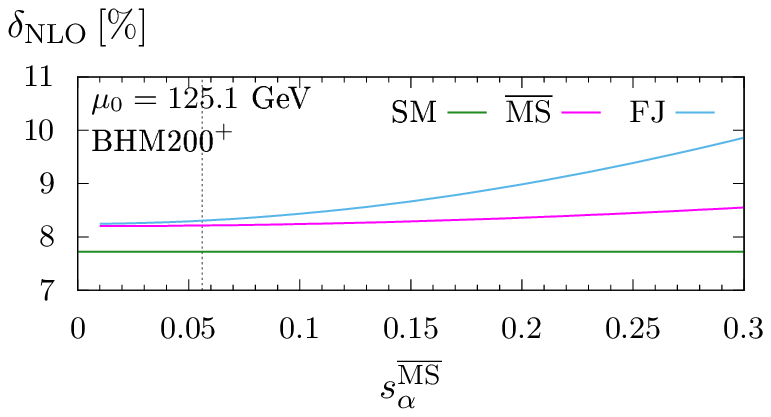}
\quad
\includegraphics[scale=0.9]{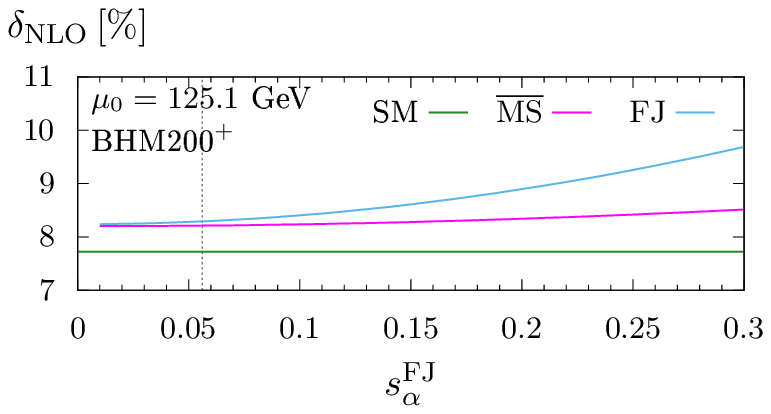}
\caption{$\sa$ dependence of the relative EW+QCD NLO corrections to the decay width $\Gamma_{\Ph \to 4 \Pf}$. Free parameters other than $\sa$ are fixed according to benchmark scenario \BHMap{} (see \cref{tab:benchmarkPoints}). The vertical dashed line signals that for $\sa \lesssim 0.06$ perturbativity problems might arise.}
\label{fig:sascandeltaNLOBHM200}
\end{figure}

In \cref{fig:sascandeltaNLOBHM200} we show the relative corrections~$\de_\text{NLO}$ to the inclusive $\Ph \to 4 \Pf$ decay width, 
defined by
\begin{equation}
\label{eq:deltaNLO}
\de_\text{NLO} = \frac{\Gamma_\text{NLO} - \Gamma_\text{LO}}{\Gamma_\text{LO}}
	       = \de_\text{EW} + \de_\text{QCD},
\end{equation}
where the \ac{NLO} result includes both \ac{EW} ($\de_\text{EW}$) and \ac{QCD} corrections ($\de_\text{QCD}$). The relative 
\ac{QCD} corrections do not depend on the mixing angle and are equal to the \ac{SM} case, providing an offset of about~$5\%$.
The relative corrections in the SESM are, for the scenario considered here, bigger than the relative corrections in the \ac{SM} case, somewhat compensating the \ac{LO} suppression factor $\cas$ mentioned above. For $\sa = 0.29$, as defined for the scenario \BHMap, the relative corrections are $9.6\%$ in the FJ scheme and $8.6\%$ in the \MSb{} scheme.
\Cref{fig:sascandeltaNLOBHM200} illustrates that in the two schemes contributions (related to the tadpole terms) are shared differently between \ac{LO} and \ac{NLO} parts. 
Recall that we have seen in \cref{fig:sascanBHM200} how the NLO decay widths are in good agreement in the two renormalization schemes, independent of the $\sa$ value.

\begin{figure}
\centering
\includegraphics[scale=1.]{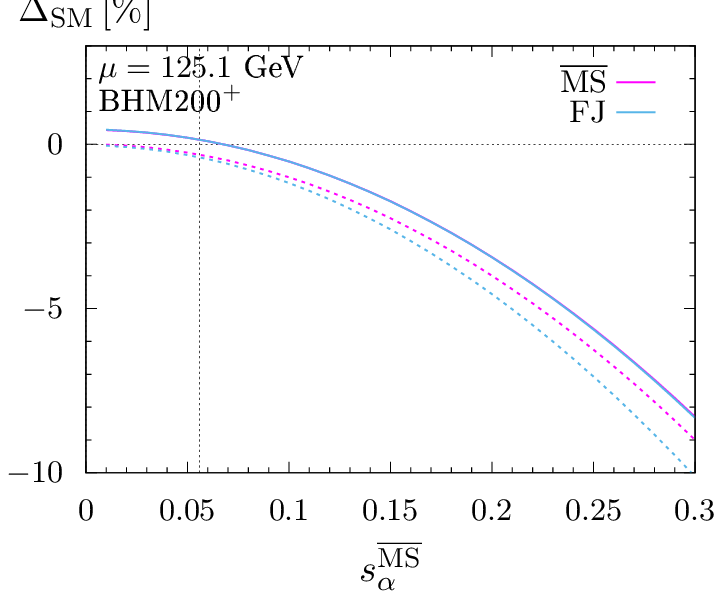}
\quad
\includegraphics[scale=1.]{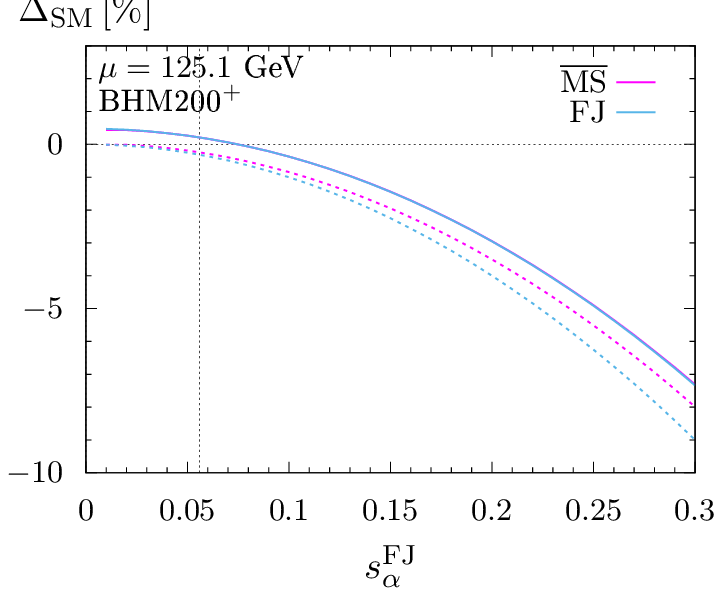}
\caption{Deviation of the LO and NLO decay widths $\Gamma_{\Ph \to 4 \Pf}$ in the SESM from the SM values, as a function of $\sa$. Free parameters other than $\sa$ are fixed according to benchmark scenario \BHMap{} (see \cref{tab:benchmarkPoints}). The vertical dashed line signals that for $\sa \lesssim 0.06$ perturbativity problems might arise.}
\label{fig:sascanDeltaSMBHM200}
\end{figure}

In \cref{fig:sascanDeltaSMBHM200} we compare the \ac{SESM} result with the \ac{SM} prediction, defining the relative deviation by
\begin{equation}
\label{eq:DeltaSM}
\De_\text{SM} = \frac{\Gamma_\text{SESM} - \Gamma_\text{SM}}{\Gamma_\text{SM}}.
\end{equation}
Again, it is possible to see how the negative deviations induced by the $\cas$ factor are somewhat reduced at \ac{NLO} due to the positive loop contributions, which in the \ac{SESM} are bigger than in the \ac{SM}. For $\sa = 0.29$, corresponding to the scenario \BHMap, the deviations from the \ac{SM} are $\De_\text{SM} \sim 7{-}8\%$. The difference between the results shown in the two panels, which correspond to~$\sa$ values in the two different renormalization schemes, is due to the fact that, changing the 
input scheme, the same numerical values correspond to different physical scenarios.

\subsubsection{Decay widths for individual four-fermion final states}
\label{sssec:decayWidthsTab}
In \cref{tab:partialWidthsBHM200} we compile our results on the widths for the decays $\Ph \to 4 \Pf$ into the various different final states for 
benchmark scenario \BHMap, computed in the FJ scheme. 
\begin{table}
\[
\begin{array}{cccccc}
  \toprule
  \text{Final state} & \Gamma^\text{NLO}_{\Ph \to 4 \Pf} & \de_\text{EW} [\%] & \de_\text{QCD} [\%] & \De_\text{SM}^\text{NLO} [\%] & \De_\text{SM}^\text{LO} [\%] \\
  \midrule
 \text{inclusive} \, \Ph \to 4 \Pf &  0.92106(6) & 4.64(0)  &  4.97(0) & -6.82(1) & -8.41(1) \\
 \PZ \PZ                           &  0.101320(5)  & 2.35(0)  &  4.89(0) & -6.78(1) & -8.41(0) \\
 \PW \PW                           &  0.82459(8) & 4.92(0)  &  5.01(1) & -6.82(1) & -8.41(1) \\
 \PW \PW / \PZ \PZ \text{int.}     & -0.00485(5) & 3.3(2) & 11.4(8)& -7(1)    & -8.4(6)  \\
 \nu_\Pe \Pe^+ \mu^- \bar{\nu}_\mu &  0.009719(1)  & 4.95(0)  &  0.00    & -6.75(1) & -8.41(1) \\
 \nu_\Pe \Pe^+ \Pu \bar{\Pd}       &  0.030198(4)  & 4.94(0)  &  3.76(1) & -6.80(2) & -8.41(1) \\
 \Pu \bar{\Pd} \Ps \bar{\Pc}       &  0.09369(2) & 4.89(0)  &  7.52(1) & -6.86(2) & -8.41(1) \\
 \nu_\Pe \Pe^+ \Pe^- \bar{\nu}_\Pe &  0.009716(1)  & 5.05(0)  &  0.00    & -6.75(1) & -8.41(1) \\
 \Pu \bar{\Pd} \Pd \bar{\Pu}       &  0.09562(2) & 4.77(0)  &  7.36(1) & -6.86(2) & -8.41(1) \\
\Pnue \Pnueb \nu_\mu \bar{\nu}_\mu &  0.000906(0)  & 5.02(0)  &  0.00    & -6.75(1) & -8.41(1) \\
 \Pe^- \Pe^+ \mu^- \mu^+           &  0.000228(0)  & 3.31(1)  &  0.00    & -6.72(1) & -8.41(1) \\
 \nu_\Pe \bar{\nu}_\Pe \mu^-\mu^+  &  0.000456(0)  & 4.47(1)  &  0.00    & -6.74(2) & -8.41(1) \\
 \Pnue \Pnueb \Pnue \bar{\nu}_\Pe  &  0.000543(0)  & 4.91(0)  &  0.00    & -6.75(1) & -8.41(1) \\
 \Pe^- \Pe^+ \Pe^-\Pe^+            &  0.000126(0)  & 3.14(1)  &  0.00    & -6.72(1) & -8.41(1) \\
 \Pnue \Pnueb \Pu \bar{\Pu}        &  0.001603(0)  & 2.60(1)  &  3.76(1) & -6.77(1) & -8.41(1) \\
 \Pnue \Pnueb \Pd \bar{\Pd}        &  0.002078(0)  & 3.70(0)  &  3.76(1) & -6.79(2) & -8.41(1) \\
 \Pe^-\Pe^+ \Pu \bar{\Pu}          &  0.000807(0)  & 2.12(1)  &  3.75(1) & -6.76(1) & -8.41(1) \\
 \Pe^- \Pe^+ \Pd \bar{\Pd}         &  0.001039(0)  & 2.48(1)  &  3.76(1) & -6.77(2) & -8.41(1) \\
 \Pu \bar{\Pu} \Pc \bar{\Pc}       &  0.002836(0)  & 0.21(1)  &  7.51(1) & -6.79(2) & -8.41(1) \\
 \Pd \bar{\Pd} \Pd \bar{\Pd}       &  0.002444(1)  & 1.62(0)  &  4.53(2) & -6.76(3) & -8.41(1) \\
 \Pd \bar{\Pd} \Ps \bar{\Ps}       &  0.004729(1)  & 1.65(0)  &  7.51(1) & -6.81(2) & -8.41(1) \\
 \Pu \bar{\Pu} \Ps \bar{\Ps}       &  0.003676(1)  & 1.34(1)  &  7.51(1) & -6.80(2) & -8.41(1) \\
 \Pu \bar{\Pu} \Pu \bar{\Pu}       &  0.001441(0)  & 0.09(1)  &  4.22(2) & -6.73(3) & -8.41(1) \\
  \bottomrule
\end{array}
\]
\caption{Partial widths for scenario \BHMap{} in the FJ renormalization scheme. The integration errors are given in parentheses.}
 \label{tab:partialWidthsBHM200}
\end{table}
The contributions $\Gamma_{\Ph \to \PW \PW \to 4 \Pf}$, $\Gamma_{\Ph \to \PZ \PZ \to 4\Pf}$, and $\Gamma_{\PW\PW/\PZ\PZ-\text{int}}$ are calculated according to \cref{eq:partialDecayWidthVV1}, and the total decay width $\Gamma_{\Ph \to 4 \Pf}$ using \cref{eq:totalDecayWidth}. In \cref{tab:partialWidthsBHM200}, we also show the relative \ac{EW} and \ac{QCD} corrections, $\de_\text{EW}$ and $\de_\text{QCD}$, and in the last two columns the deviation $\De_\text{SM}$ from the \ac{SM} both at \ac{LO} and \ac{NLO}.
For all the quantities, we report the integration uncertainty in parentheses.
To determine the errors of the decay widths to $\PW\PW$, $\PZ\PZ$, $\PW\PW/\PZ\PZ$ interference and of the total width $\Gamma_{\Ph \to 4 \Pf}$, we 
apply the standard error propagation to \cref{eq:partialDecayWidthVV1,eq:totalDecayWidth}, making use of the integration uncertainties for each single final state.

The main contribution to the total $\Ph \to 4 \Pf$ decay width originates from the charge-current final states, while the neutral-current processes have a smaller impact, and the $\PW\PW/\PZ\PZ$ interference gives a very small negative contribution. The \ac{EW} corrections to the $\PW\PW$ contributions are about~$5\%$, and lead to a similar value for the inclusive decay $\Ph \to 4 \Pf$. The \ac{EW} corrections to neutral-current final states range from~$0$ to~$5\%$, depending on the flavour of the final-state fermions.
The \ac{QCD} corrections are mostly due to the corrections to the decays $\PW/\PZ \to \Pq \Pqb$, and amount to $\alphas/\pi$ for each quark pair in the final state. Exceptions are the final states $\Pu \Pub \Pu \Pub$ and $\Pd \Pdb \Pd \Pdb$, where interference contributions from two different topologies of the $\PZ\PZ$ channel occur, and the \ac{QCD} corrections to these final states are only about~$4\%$.
The \ac{SM} deviation, at \ac{LO}, comes from the $\ca$ rescaling factor of the $\Fh\FV\FV$ coupling with respect to the \ac{SM} coupling, and is equal to $\cas - 1 = -0.0841$ in the considered scenario. As already observed in the previous section, at \ac{NLO} the deviations from the \ac{SM} are about~$1.5\%$ smaller.

We have computed the same quantities as in \cref{tab:partialWidthsBHM200} in the \MSb{} scheme, observing somewhat smaller values for the \ac{EW} corrections, since in the two schemes contributions are shared differently between the \ac{LO} and the \ac{NLO} (as observed also in \cref{fig:sascandeltaNLOBHM200}).
Moreover, using the same numerical input in the \MSb{} input scheme leads to \ac{NLO} deviations from the \ac{SM} about $1\%$ higher, since the same numerical input corresponds to a slightly different physical scenario.
In total, the \MSb{} results follow the same qualitative pattern as in the FJ scheme, and are not reported here.

\subsection{\BHMc}
\label{ssec:resultsBHM600}

\subsubsection{Scheme conversion}
\label{sssec:schemeConv600}
\begin{figure}
\centering
\includegraphics[scale=1.]{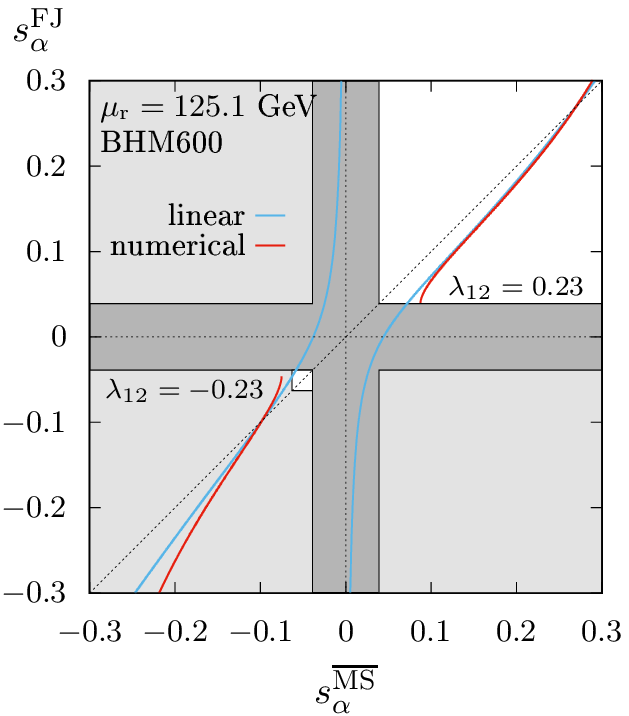}
\qquad
\includegraphics[scale=1.]{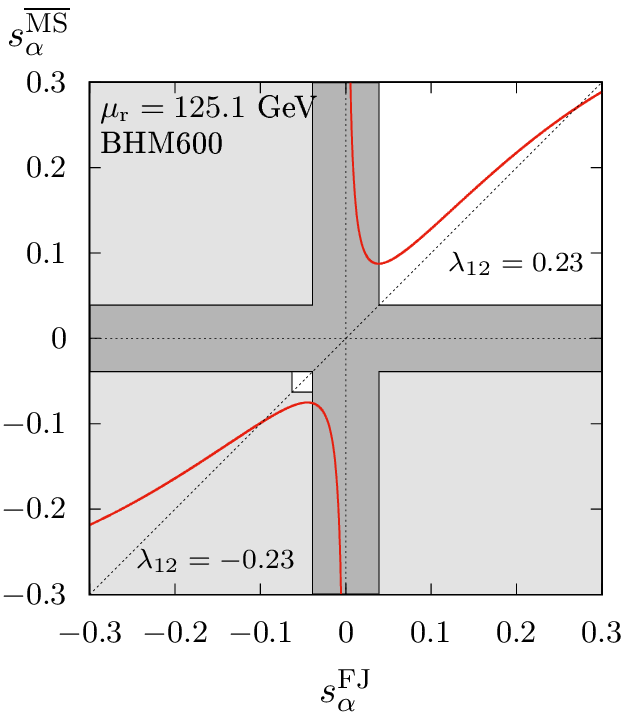}
\caption{Conversion of the input parameter $\sa$ between the \MSb{} and the FJ schemes. The heavy Higgs-boson mass $\MH$ is fixed according to benchmark scenario \BHMc. Red and blue lines correspond, respectively, to the complete and linearized solutions of the matching condition~\eqref{eq:FJtoMRconversion}.}
\label{fig:schemeconversion2BHM600}
\end{figure}

In \cref{fig:schemeconversion2BHM600} we show the conversion of the input value for $\sa$ from the \MSb{} to the FJ scheme (left panel) and vice versa (right panel). 
%
The values of $\MH$ and $|\lsd|$ are given in the scenario \BHMc{} as stated in \cref{tab:benchmarkPoints}. For consistency, for negative $\sa$ input values we consider a negative Higgs self-coupling~$\lsd$.
The renormalization scale $\mur$ is again chosen to be equal to the light Higgs mass~$\Mh$.
For the considered values of the input parameters, the perturbativity constraint on $\la_1$ of \cref{eq:BSMconstrPert} is violated for $|\sa| \lesssim 0.04$. The dark-gray shaded area corresponds to the region where the perturbativity condition breaks down.
The light-gray areas denote regions where the vacuum stability condition \eqref{eq:vacuumStabCond2} is violated and where the sign of~$\sa$ becomes different from the sign of~$\lsd$ by effect of the conversion.
As discussed in \cref{sssec:schemeConv}, the conversion from the FJ to the \MSb{} scheme can be computed straightforwardly from the matching condition~\eqref{eq:schemeConvMixing}, while to compute the inverse conversion the matching condition has to be solved numerically. Alternatively, the linear approximation~\eqref{eq:schemeConvLin} may be used. In the left panel of \cref{fig:schemeconversion2BHM600} we use the linear approximation (shown in blue) in the non-perturbative region, where the numerical 
inversion does not provide a solution.
The red lines correspond to the results obtained from the numerical solution of the non-linearized matching equation \eqref{eq:schemeConvMixing}.
The conversion effects are small for $|\sa|\sim 0.3$ and become large for small angles. The conversion has an important impact in the vicinity of the non-perturbative region, where loop effects exceed the \ac{LO} contributions.
Asymmetries of the plots are due to the different sign used for~$\lsd$, so that $\sgn(\sa) = \sgn(\lsd)$.

\subsubsection{Running of {\boldmath$\sa$} and {\boldmath$\lsd$}}
\begin{figure}
\centering
\includegraphics[scale=1.]{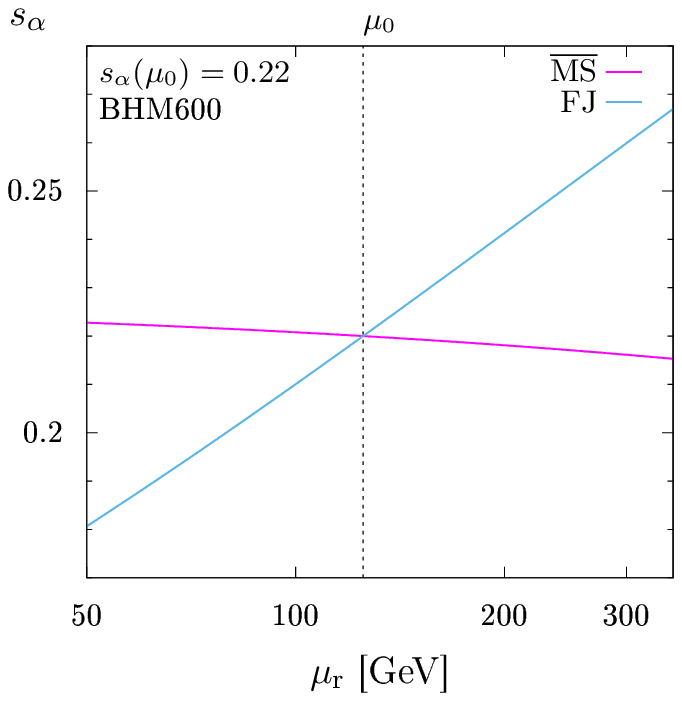}
\qquad
\includegraphics[scale=1.]{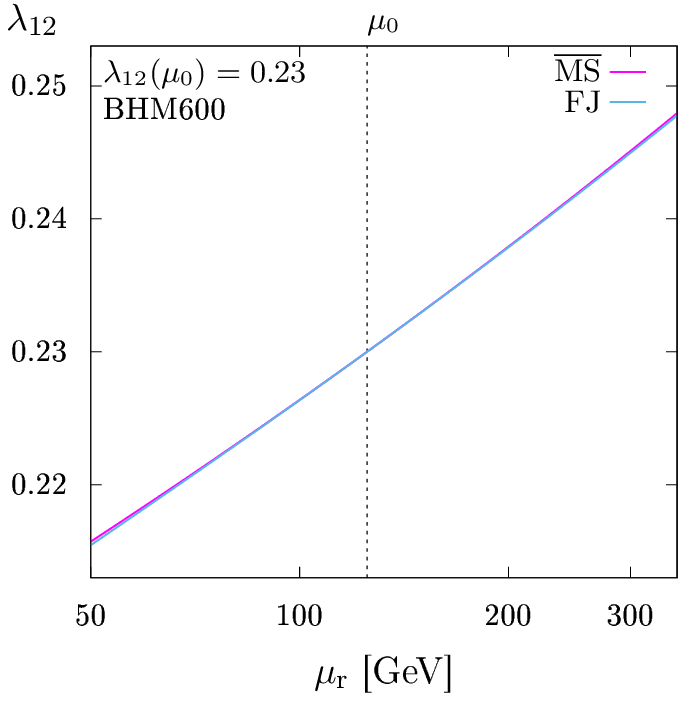}
\caption{The running of the input parameters $\sa$ and $\lsd$ in the \MSb{} and FJ renormalization schemes for benchmark scenario \BHMc.}
\label{fig:murscanparam2BHM600}
\end{figure}

The solution of the \acp{RGE} for the $\MSbar$ parameters $\sa$ and $\lsd$ is shown in \cref{fig:murscanparam2BHM600} for scenario \BHMc{}.
%
The solutions are again obtained numerically by a Runge$-$Kutta algorithm.
The values for the \ac{BSM} parameters $\sa$, $\MH$, and $\lsd$ are fixed according to scenario \BHMc{} at the scale $\mu_0 = \Mh$ and evolved to the renormalization scale $\mur$ in the range $\mur = 50{-}350 \, \GeV$.
We report the results both for the \MSb{} and the FJ schemes, where we can observe a different behaviour in the running of $\sa$: The sine of the mixing angle increases with the scale in the FJ scheme, while it slowly decreases in the \MSb{} scheme. The running of the coupling $\lsd$, in the considered range, is almost identical in the two schemes.
Compared to the scenario \BHMap{}, the running of $\sa$ in the \MSb{} scheme is strongly reduced.

\subsubsection{Scale dependence of the inclusive decay width {\boldmath$\Gamma_{\Ph \to 4 \Pf}$}}
\begin{figure}
\centering
\includegraphics[scale=1.]{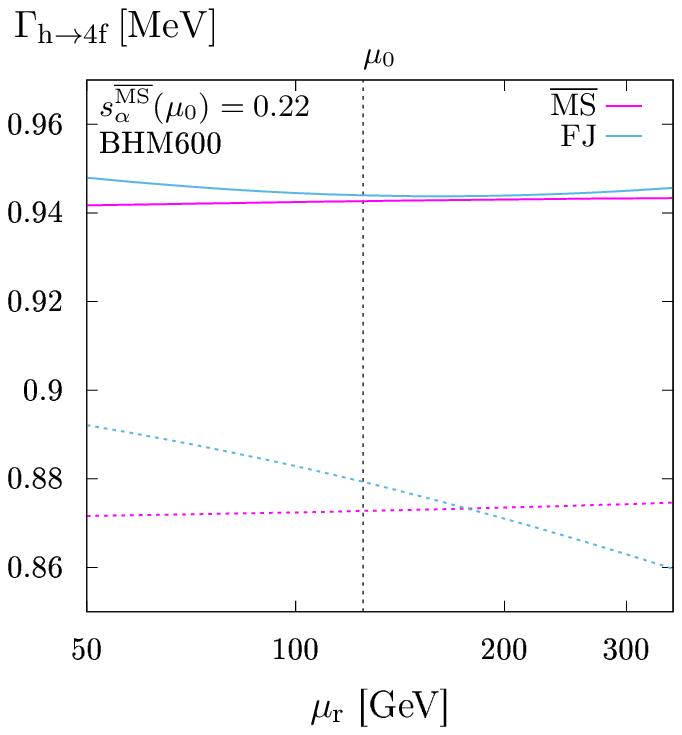}
\qquad
\includegraphics[scale=1.]{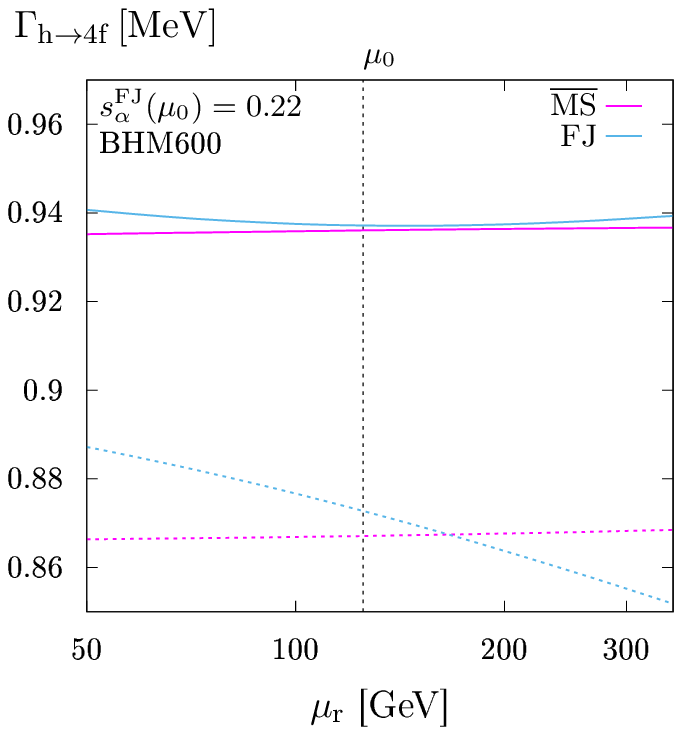}
\caption{Scale dependence of the decay width $\Gamma_{\Ph \to 4 \Pf}$ at \ac{LO} (dashed curves) and NLO EW+QCD (solid curves) for benchmark scenario \BHMc, using the central scale $\mu_0 = \Mh$. On the left (right) the input parameters are defined in the \MSb{} (FJ) scheme and converted to the FJ (\MSb{}) scheme at the scale $\mu_0$ (both for LO and NLO predictions).}
\label{fig:murscan2BHM600}
\end{figure}

In \cref{fig:murscan2BHM600} we present the results for the inclusive decay width $\Gamma_{\Ph \to 4 \Pf}$ for scenario \BHMc{} at scales $\mur$ in the range $50{-}350 \, \GeV$.
%
The results obtained using the \MSb{} (magenta) and the FJ (blue) input schemes are reported, respectively, on the left and the right panels of the figure.
Dashed lines correspond to the \ac{LO} results (with \ac{NLO} conversion of the input parameters), solid lines include \ac{NLO} \ac{EW}+\ac{QCD} corrections.
Similar to the observations made for scenario \BHMap{} above, the \ac{NLO} results show a much milder scale dependence compared to the LO results, and the differences between the two schemes are strongly reduced.
We can see that the scale choice $\mur = \Mh$ seems more suitable than the alternative scale $\mur = (\Mh + \MH)/2 \sim 360 \, \GeV$, where the dependence on the renormalization scale is somewhat stronger.
Quantitatively, the scale dependence (again defined by scale variations of factors $2$ and $1/2$)
reduces from $\sim1{-}2\%$ at LO to $\sim0.3\%$ at NLO in the FJ scheme,
while the scale dependence in the $\MSbar$ scheme is at the $0.1\%$ level both at LO and NLO due
to the suppression of the running of $\alpha$ in the BHM600 scenario.
The scheme dependence at the central scale
reduces from $\sim0.8\%$ at LO to $\sim0.1\%$ at NLO.

\subsubsection{Mixing-angle dependence of the inclusive decay width {\boldmath$\Gamma_{\Ph \to 4 \Pf}$}}
\begin{figure}
\centering
\includegraphics[scale=1.]{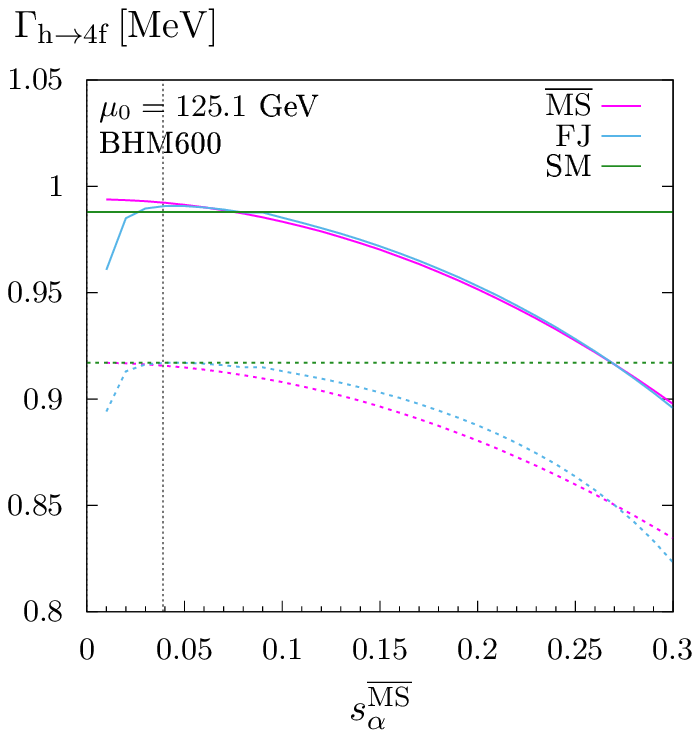}
\quad
\includegraphics[scale=1.]{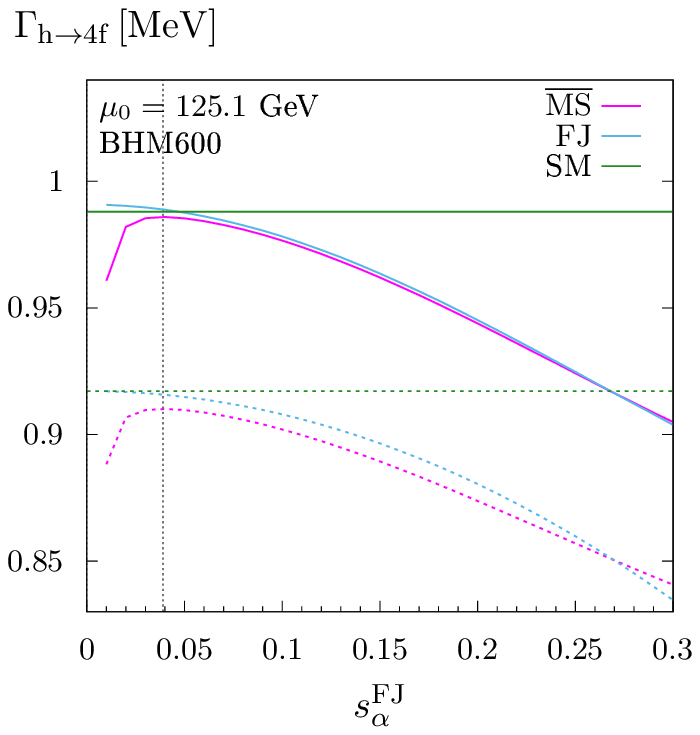}
\caption{Dependence of the decay width $\Gamma_{\Ph \to 4 \Pf}$ at LO and NLO EW+QCD with respect to the variation of $\sa$. Free parameters other than $\sa$ are fixed according to benchmark scenario \BHMc{} (see \cref{tab:benchmarkPoints}).
On the left (right) the input parameters are defined in the \MSb{} (FJ) scheme and converted to the FJ (\MSb{}) scheme (both for LO and NLO predictions).
The vertical dashed line signals that for $\sa \lesssim 0.04$ perturbativity problems might arise.}
\label{fig:sascan2BHM600}
\end{figure}

\begin{figure}
\centering
\includegraphics[scale=0.9]{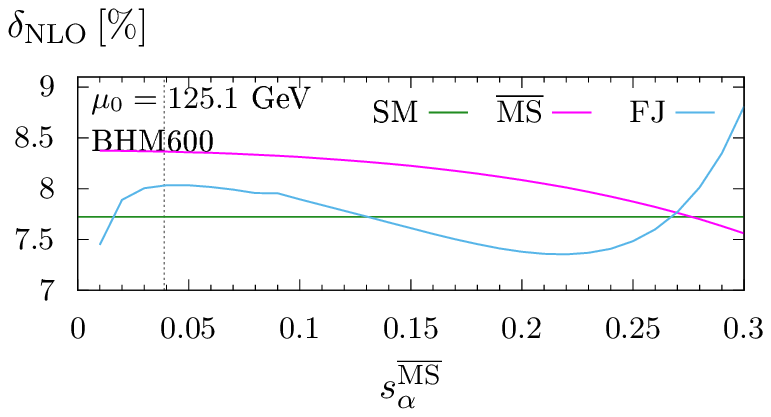}
\quad
\includegraphics[scale=0.9]{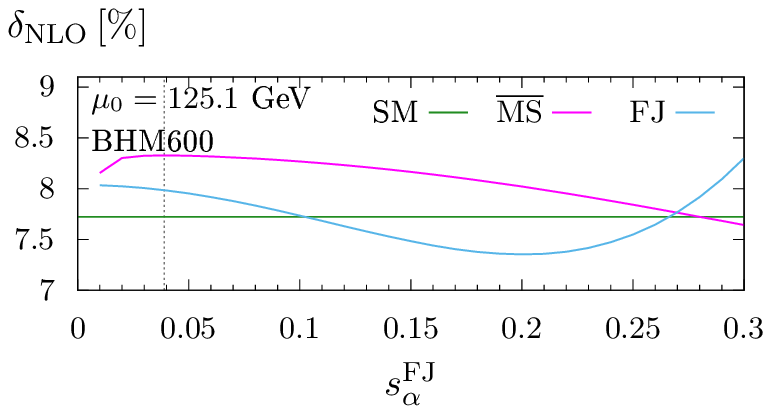}
\caption{$\sa$ dependence of the relative EW+QCD NLO corrections to the decay width $\Gamma_{\Ph \to 4 \Pf}$. Free parameters other than $\sa$ are fixed according to benchmark scenario \BHMc{} (see \cref{tab:benchmarkPoints}). The vertical dashed line signals that for $\sa \lesssim 0.04$ perturbativity problems might arise.}
\label{fig:sascandeltaNLOBHM600}
\end{figure}

\begin{figure}
\centering
\includegraphics[scale=1.]{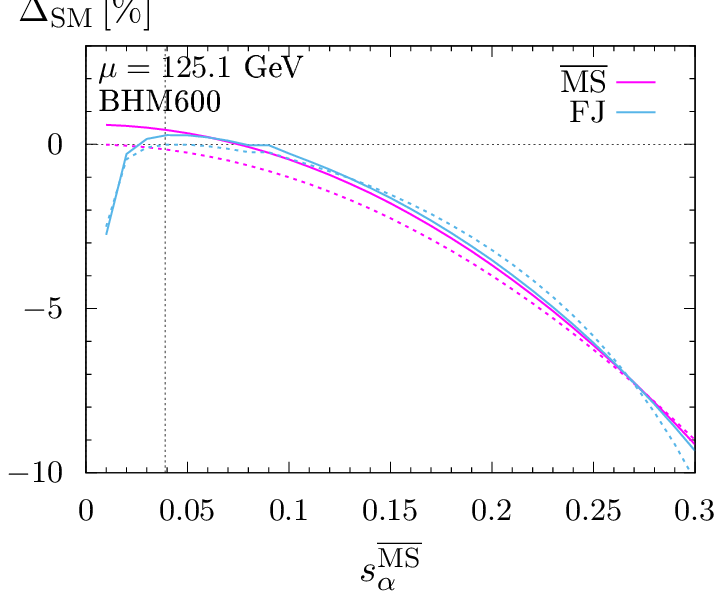}
\quad
\includegraphics[scale=1.]{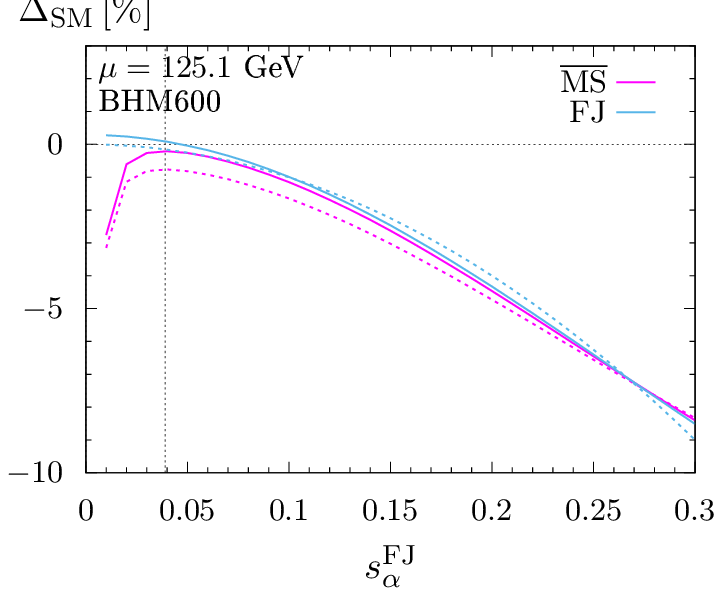}
\caption{Deviation of the LO and NLO decay widths $\Gamma_{\Ph \to 4 \Pf}$ in the SESM from the SM values, as a function of $\sa$. Free parameters other than $\sa$ are fixed according to benchmark scenario \BHMc{} (see \cref{tab:benchmarkPoints}). The vertical dashed line signals that for $\sa \lesssim 0.04$ perturbativity problems might arise.}
\label{fig:sascanDeltaSMBHM600}
\end{figure}

In \cref{fig:sascan2BHM600,fig:sascandeltaNLOBHM600,fig:sascanDeltaSMBHM600} we present, respectively, the decay width $\Gamma_{\Ph \to 4 \Pf}$, the relative corrections to the decay width, and the deviations with respect to the \ac{SM} result as a function of the parameter $\sa$ using our default scale choice $\mur = \mu_0 = \Mh$.
%
The heavy Higgs mass and the coupling $\lsd$ are fixed according to scenario \BHMc{}, as defined in \cref{tab:benchmarkPoints}.
In all the figures, dashed lines correspond to \ac{LO} results and solid lines to \ac{NLO} results, including both \ac{EW} and \ac{QCD} corrections.
We consider values $|\sa|< 0.3$ and mark with a dashed vertical line the minimal $\sa$ value for which the perturbativity constraints of \cref{eq:BSMconstrPert} are fulfilled.
The results reported in the left (right) panel are obtained with input parameters defined in \MSb{} (FJ) scheme with a proper parameter conversion at NLO applied if the input is not directly given in the scheme used in the calculation.

In the \ac{LO} results of \cref{fig:sascan2BHM600}, the $\cas$ dependence of the decay width in the \ac{SESM} is manifest, where the proportionality to~$\cas$ is exact if no scheme conversion is involved. For small mixing angles, where $\ca \to 1$, the deviations from the \ac{SM} are due to the conversion effects.
As observed in the other scenarios, the inclusion of \ac{NLO} corrections slightly compensates the $\cas$ suppression for small $\sa$ values.
Moreover, we observe that the two schemes are in 
much better agreement after the inclusion of \ac{NLO} corrections, i.e.\ the inclusion of NLO corrections reduces the renormalization scheme dependence drastically.

The relative \ac{NLO} corrections, as defined in \cref{eq:deltaNLO}, are shown in \cref{fig:sascandeltaNLOBHM600} including both \ac{EW} and \ac{QCD} contributions (the latter are independent 
of $\sa$ and about $5\%$).
In the considered $\sa$ range the relative corrections amount to $7{-}9\%$ (slightly depending on the input scheme used) and the difference with the SM relative corrections $|\delta^\text{SESM} - \delta^\text{SM}|$ does not exceed the $1\%$ level in the considered region.

In \cref{fig:sascanDeltaSMBHM600} we illustrate the deviations from the \ac{SM}, defined by \cref{eq:DeltaSM}.
It is possible to see how the \ac{NLO} results converge nicely in the two schemes. Note that the large scheme-dependence due to missing higher-orders for small $\sa$ values occurs only in the non-perturbative regime, where calculations generally become unreliable.
The proposed $\sa$ values for the scenario \BHMc{} is $0.22$, yielding a deviation from the \ac{SM} of about $-5\%$.

\subsubsection{Decay widths for individual four-fermion final states}
\label{sssec:decayWidthsTab600}
In \cref{tab:partialWidthsBHM600} we compile the results for the $\Ph \to 4 \Pf$ decay widths for each final state listed in \cref{tab:finalStates}, together with the widths for the decays into two vector bosons and the inclusive decay width $\Gamma_{\Ph \to 4 \Pf}$. 
\begin{table}
\[
\begin{array}{cccccc}
  \toprule
  \text{Final state} & \Gamma^\text{NLO}_{\Ph \to 4 \Pf} & \de_\text{EW} [\%] & \de_\text{QCD} [\%] & \De_\text{SM}^\text{NLO} [\%] & \De_\text{SM}^\text{LO} [\%] \\
  \midrule
 \text{inclusive} \, \Ph \to 4 \Pf &  0.93761(6) &  2.42(0)  &  4.97(0)  & -5.14(1) & -4.84(1) \\
 \PZ \PZ                           &  0.103098(5)  &  0.13(0)  &  4.89(0)  & -5.15(1) & -4.84(0) \\
 \PW \PW                           &  0.83945(8) &  2.71(0)  &  5.01(1)  & -5.14(1) & -4.84(1) \\
 \PW \PW / \PZ \PZ \text{int.}     & -0.00494(5) &  1.1(2) & 11.4(8) & -5(1)    & -4.8(7)  \\
 \nu_\Pe \Pe^+ \mu^- \bar{\nu}_\mu &  0.009884(1)  &  2.73(0)  &  0.00     & -5.16(1) & -4.84(1) \\
 \nu_\Pe \Pe^+ \Pu \bar{\Pd}       &  0.030736(4)  &  2.73(0)  &  3.76(1)  & -5.15(2) & -4.84(1) \\
 \Pu \bar{\Pd} \Ps \bar{\Pc}       &  0.09542(2) &  2.68(0)  &  7.52(1)  & -5.14(2) & -4.84(1) \\
 \nu_\Pe \Pe^+ \Pe^- \bar{\nu}_\Pe &  0.009882(1)  &  2.83(0)  &  0.00     & -5.16(1) & -4.84(1) \\
 \Pu \bar{\Pd} \Pd \bar{\Pu}       &  0.09739(2) &  2.56(0)  &  7.36(1)  & -5.14(2) & -4.84(1) \\
\Pnue \Pnueb \nu_\mu \bar{\nu}_\mu &  0.000921(0)  &  2.80(0)  &  0.00     & -5.16(1) & -4.84(1) \\
 \Pe^- \Pe^+ \mu^- \mu^+           &  0.000232(0)  &  1.10(1)  &  0.00     & -5.16(1) & -4.84(1) \\
 \nu_\Pe \bar{\nu}_\Pe \mu^-\mu^+  &  0.000464(0)  &  2.25(1)  &  0.00     & -5.16(2) & -4.84(1) \\
 \Pnue \Pnueb \Pnue \bar{\nu}_\Pe  &  0.000552(0)  &  2.69(0)  &  0.00     & -5.16(1) & -4.84(1) \\
 \Pe^- \Pe^+ \Pe^-\Pe^+            &  0.000128(0)  &  0.93(1)  &  0.00     & -5.16(1) & -4.84(1) \\
 \Pnue \Pnueb \Pu \bar{\Pu}        &  0.001631(0)  &  0.39(1)  &  3.76(1)  & -5.15(1) & -4.84(1) \\
 \Pnue \Pnueb \Pd \bar{\Pd}        &  0.002114(0)  &  1.48(0)  &  3.76(1)  & -5.15(2) & -4.84(1) \\
 \Pe^-\Pe^+ \Pu \bar{\Pu}          &  0.000821(0)  & -0.09(1)  &  3.75(1)  & -5.15(2) & -4.84(1) \\
 \Pe^- \Pe^+ \Pd \bar{\Pd}         &  0.001057(0)  &  0.27(1)  &  3.76(1)  & -5.15(2) & -4.84(1) \\
 \Pu \bar{\Pu} \Pc \bar{\Pc}       &  0.002886(0)  & -2.00(1)  &  7.51(1)  & -5.14(2) & -4.84(1) \\
 \Pd \bar{\Pd} \Pd \bar{\Pd}       &  0.002486(1)  & -0.59(0)  &  4.53(2)  & -5.15(3) & -4.84(1) \\
 \Pd \bar{\Pd} \Ps \bar{\Ps}       &  0.004814(1)  & -0.56(0)  &  7.51(1)  & -5.14(2) & -4.84(1) \\
 \Pu \bar{\Pu} \Ps \bar{\Ps}       &  0.003742(1)  & -0.87(1)  &  7.51(1)  & -5.14(2) & -4.84(1) \\
 \Pu \bar{\Pu} \Pu \bar{\Pu}       &  0.001465(0)  & -2.11(1)  &  4.22(2)  & -5.14(3) & -4.84(1) \\
  \bottomrule
\end{array}
\]
\caption{Partial widths for scenario \BHMc{} in the FJ renormalization scheme. The integration errors are given in parentheses.}
 \label{tab:partialWidthsBHM600}
\end{table}
For each decay channel, we report the \ac{NLO} result, the \ac{EW} and \ac{QCD} relative corrections, and the deviations from the \ac{SM} result. The latter are reported both for \ac{LO} and \ac{NLO}.
The qualitative picture is basically the same as for the scenarios \BHMap{}, with slightly different values for the \ac{EW} corrections, which are between $-2\%$ and $3\%$, \ie a bit smaller than in the other cases with smaller $\MH$.
The deviation from the \ac{SM} is about $-5\%$, in the FJ scheme, at \ac{LO} and \ac{NLO}. Using the \MSb{} input scheme, the deviations $\Delta_\text{SM}^\text{NLO}$ are only
slightly different, around $-4.5\%$, and display a similar pattern.

\subsection{Differential distributions}
Footprints of potential BSM physics often can be found by looking into the shapes of differential distributions, even if integrated results do not deviate from the SM predictions significantly.
In the following, we discuss some of the distributions produced with \prophecy{} for the \ac{SESM}, reporting results both for charged- and neutral-current final states.
The generator provides invariant-mass and angular distributions for leptonic and semi-leptonic final states, while distributions for fully hadronic final states are not interesting, since they are not experimentally accessible.
A detailed survey of distributions in the \ac{SM} can be found in \citeres{Bredenstein:2006rh,Bredenstein:2006nk,Boselli:2015aha} for fully leptonic final states and in \citere{Bredenstein:2006ha} for the semi-leptonic case. There, the treatment of the final-state radiation and the photon recombination for nearly collinear fermion$-$photon pairs are also discussed in detail.
In the SESM, relative corrections induced by final-state radiation are identical to the SM case and thus not discussed in greater detail here.

\subsubsection{Leptonic final states}
We consider the fully leptonic final states $\Pmum \Pmup \Pem \Pep$ and $\Pnumu \Pmup \Pem \Pnue$, which involve, respectively, intermediate (off-shell) $\PW\PW$ and $\PZ\PZ$ states.
The distributions discussed in the following have been computed in the FJ renormalization scheme, the results obtained in the \MSb{} scheme show the same features and are not reported here.
All the distributions are generated using the renormalization scale $\mur = \Mh$.

\begin{figure}
\centering
\includegraphics[scale=0.85]{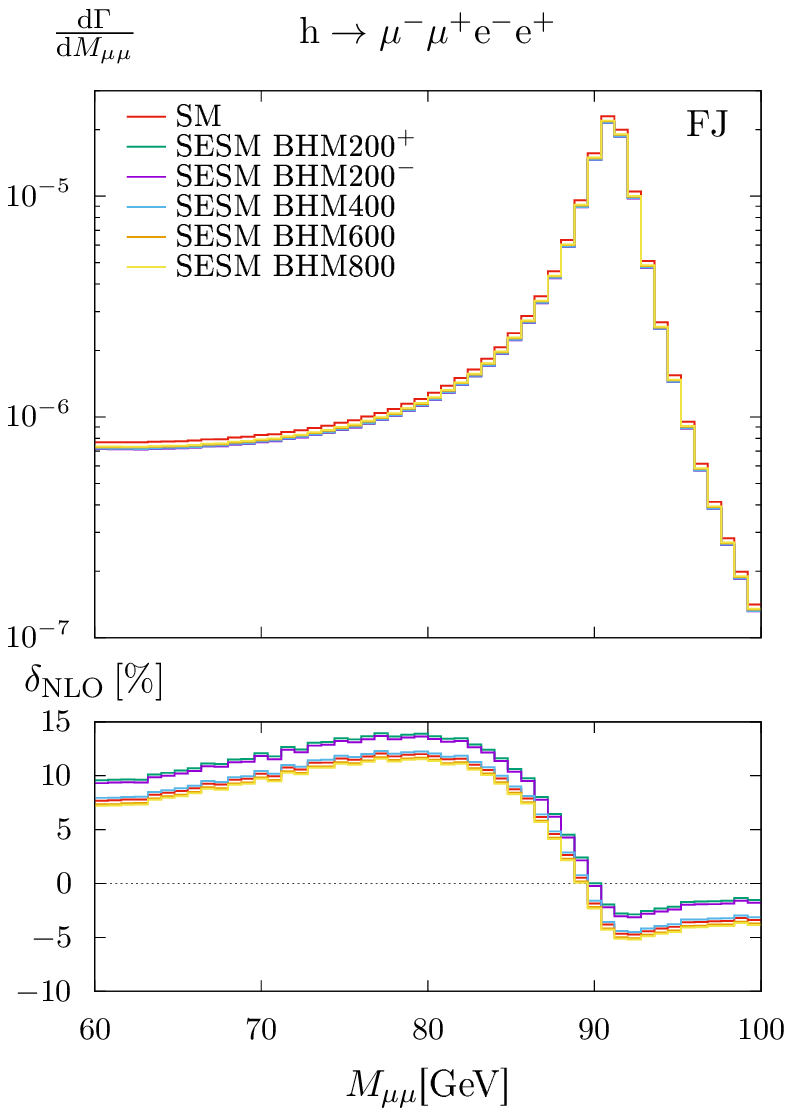}
\quad
\includegraphics[scale=0.85]{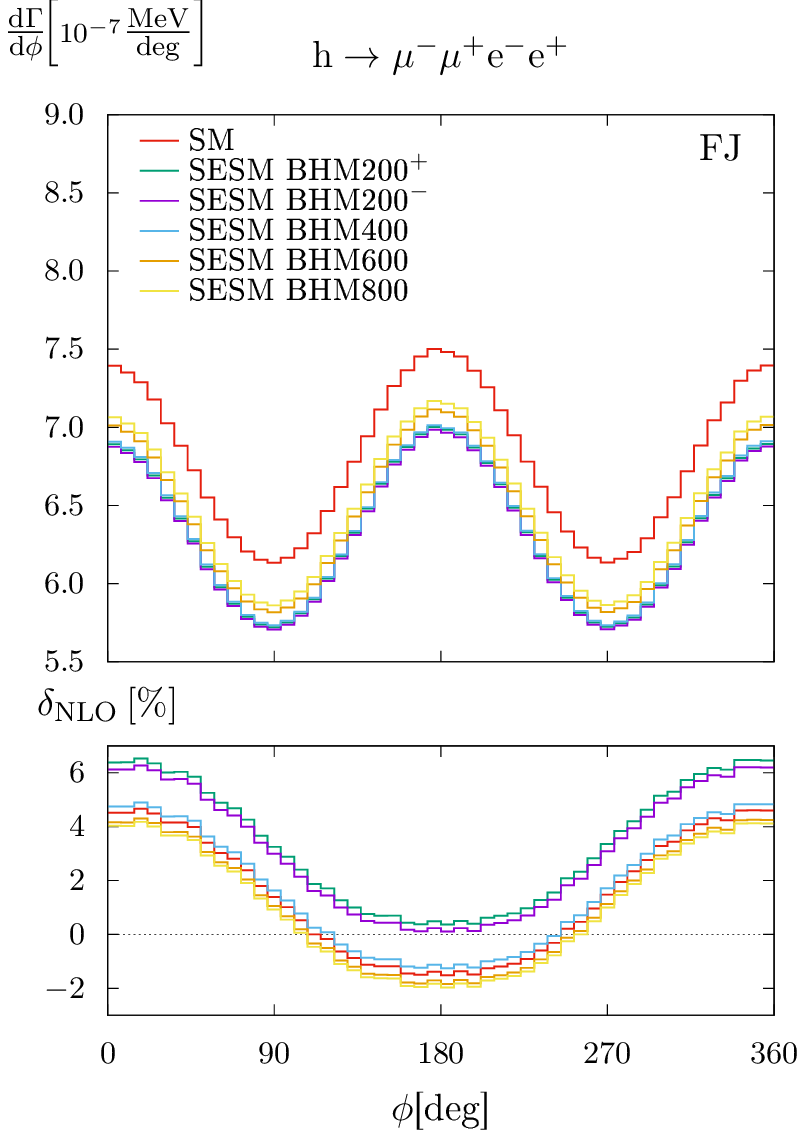}
\caption{Invariant-mass and angular distributions for the neutral-current decay into leptons $\Ph \to \Pmum \Pmup \Pem \Pep$, in the FJ renormalization scheme for the various SESM scenarios.
}
\label{fig:diffdistr_mumuee}
\end{figure}

For the neutral-current final state $\Pmum \Pmup \Pem \Pep$, the left panel of \cref{fig:diffdistr_mumuee} shows the NLO distribution for the invariant mass of the muon pair around the $\PZ$-boson resonance, both for the SM and for the considered SESM scenarios.
In general, the invariant mass of a fermionic pair is defined by
\begin{equation}
M_{\Pf_a \Pfb_b} = \left( k_a + k_b \right)^2,
\end{equation}
where $k_a$ and $k_b$ are the 
four-momenta of the fermion $\Pf_a$ and the anti-fermion $\Pfb_b$ and, in case of photon recombination, the photon momentum is added to the momentum of the fermion or anti-fermion.
The distributions obtained for the SESM display the $\PZ$-boson resonance peak and the ``radiative tail'' observed in the standard case: The real-emission contributions lead to positive corrections for invariant masses below the $\PZ$-boson peak. This is explained by the fact that outgoing leptons lose momentum when radiating a photon, so that these events are shifted towards lower masses.
This can be seen in the lower panel of the plot, where the relative NLO corrections $\de_\text{NLO}$ are shown bin-by-bin.
The shapes of the distributions in the SESM are the same as in the SM, and the only difference is given by an offset, which depends on the SESM scenario.
The difference between the NLO distribution in the SM and in the SESM equals, for each considered scenario, the quantity $\De_\text{SM}^\text{NLO}$ obtained for the corresponding integrated result (see Tables \ref{tab:partialWidthsBHM200} and \ref{tab:partialWidthsBHM600} 
for \BHMap{} and \BHMc{}).
We observe this pattern in all the generated distributions.
The right panel of \cref{fig:diffdistr_mumuee} shows, for the same final state, the differential decay width with respect to the angle $\phi$, which is defined as the angle between the decay planes of the two intermediate~$\PZ$ bosons in the  Higgs rest frame,
\begin{equation}
\cos \phi = \frac{\left(\left(\mathbf{k}_\Pmup+\mathbf{k}_\Pmum\right) \times \mathbf{k}_\Pmup\right)\cdot((\mathbf{k}_\Pmup+\mathbf{k}_\Pmum) \times \mathbf{k}_\Pep) }{|(\mathbf{k}_\Pmup+\mathbf{k}_\Pmum) \times \mathbf{k}_\Pmup|\,|(\mathbf{k}_\Pmup+\mathbf{k}_\Pmum) \times \mathbf{k}_\Pep|},
\end{equation}
with the sign convention
\begin{equation}
 \text{sgn} (\sin \phi)= \text{sgn}\left\{\left(\mathbf{k}_\Pmup+\mathbf{k}_\Pmum\right)
\cdot\left[\left(\left(\mathbf{k}_\Pmup+\mathbf{k}_\Pmum\right) \times \mathbf{k}_\Pmup\right)\times \left(\left(\mathbf{k}_\Pmup+\mathbf{k}_\Pmum\right) \times \mathbf{k}_\Pep\right)\right]\right\}.
\end{equation}
The distribution resembles a $\cos(2 \phi)$ oscillation with some constant offset and can be used to determine the parity of the Higgs boson and to set bounds on BSM couplings to EW gauge bosons of the decaying scalar (see \citeres{Nelson:1986ki,Soni:1993jc,Chang:1993jy,Skjold:1993jd,Buszello:2002uu,Arens:1994wd,Choi:2002jk,Boselli:2017pef}).
As observed for the invariant-mass distributions, the shape of the distributions in the SESM scenarios is the same as in the SM, and the NLO relative corrections differ just by a constant offset which is equivalent to $\De_\text{SM}^\text{NLO}$ observed for the integrated widths.

\begin{figure}
\centering
\includegraphics[scale=0.85]{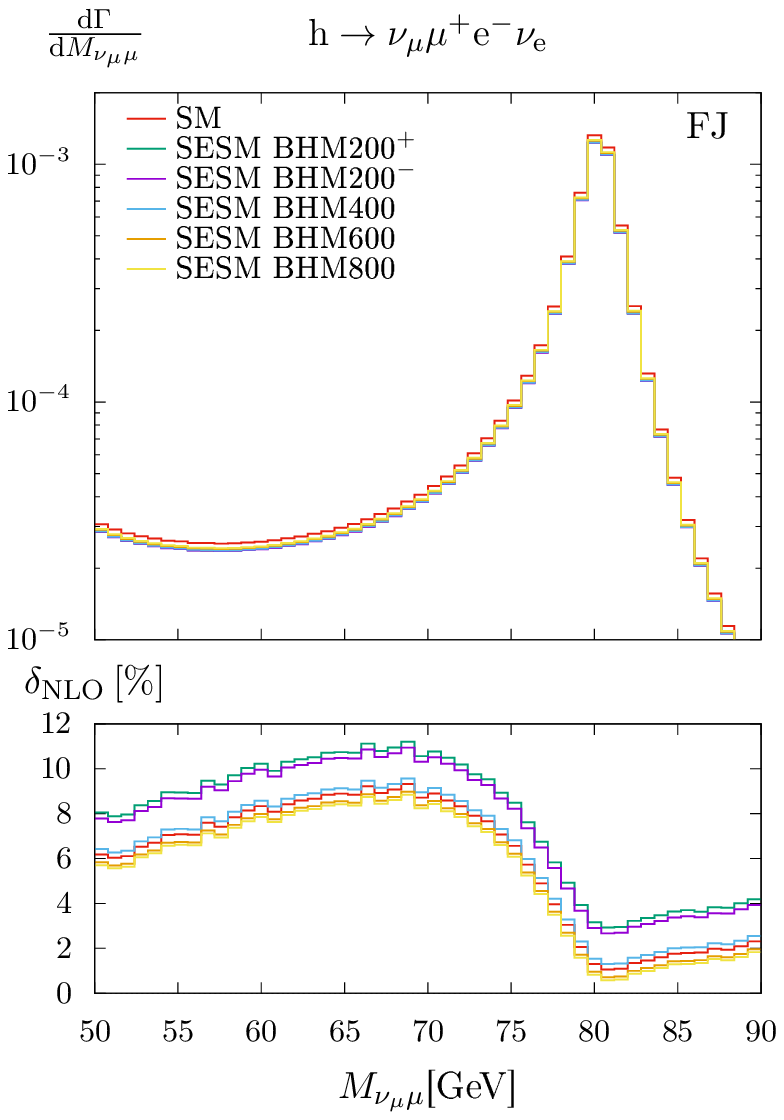}
\quad
\includegraphics[scale=0.85]{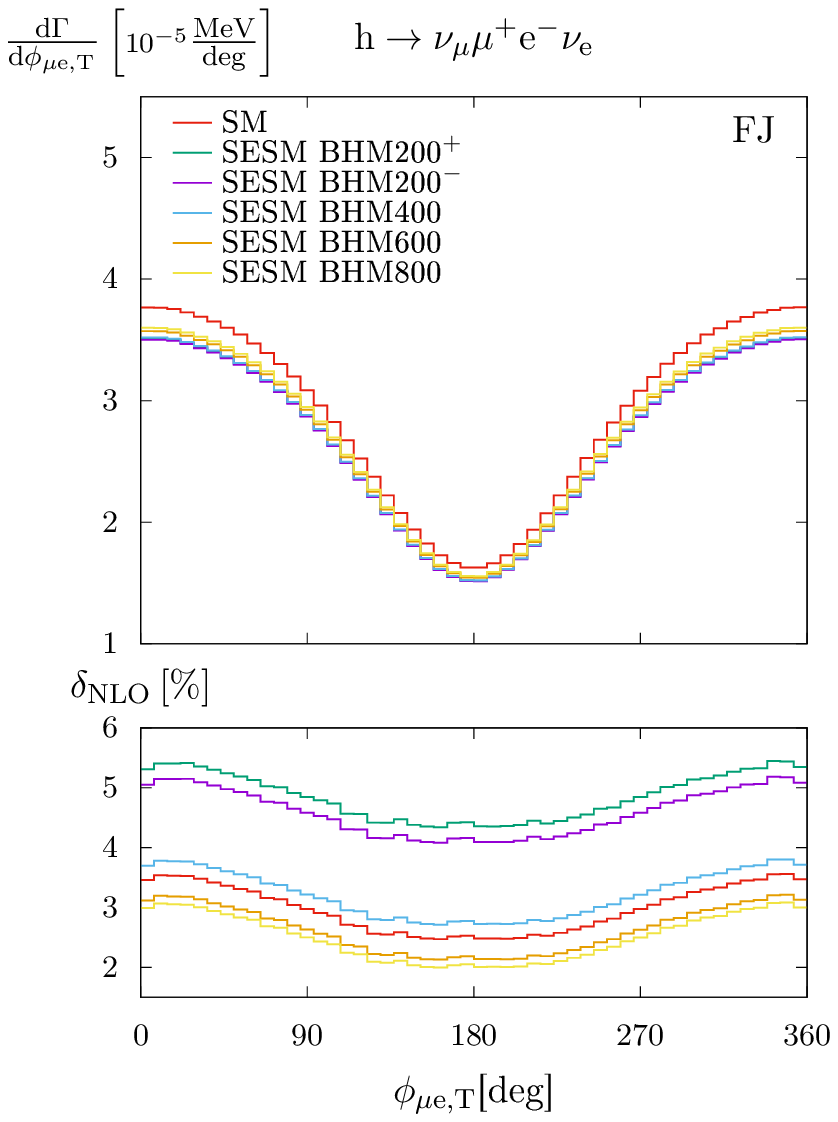}
\caption{Invariant-mass and angular distributions for the charged-current decay into leptons $\Ph \to \Pnumu \Pmup \Pem \Pnue$, in the FJ renormalization scheme for the various SESM scenarios.
}
\label{fig:diffdistr_vmuev}
\end{figure}

\Cref{fig:diffdistr_vmuev} shows differential distributions for the charged-current final state $\Pnumu \Pmup \Pem \Pnue$ for the SM and the SESM.
The invariant-mass distribution for $M_{\Pnumu \Pmu}$, shown in the left panel of the figure, is not experimentally accessible, but is interesting from the theoretical point of view.
The distribution shows the $\PW$-boson resonance and the radiative tail already described for the $\Pmum \Pmup \Pem \Pep$ final state, with an enhancement below the $\PW$-boson peak driven by the real photon emission.
The SESM does not induce any additional distortion on top of the SM shape.
The corrections and the differences between the SM and the SESM scenarios match always the values obtained for the integrated results.
The right panel of the figure shows the distribution for the angle between the transverse momenta of the muon and the electron, in the Higgs rest frame, defined by
\begin{equation}
\cos \phi_{\mu\Pe,\rT}=\frac{\mathbf{k}_{\Pmu,\rT} \cdot \mathbf{k}_{\Pe,\rT}}{|\mathbf{k}_{\Pmu,\rT}|\,| \mathbf{k}_{\Pe,\rT}|},
\qquad
 \text{sgn} (\sin \phi_\rT)=\text{sgn}\{\mathbf{e}_z \cdot (\mathbf{k}_{\Pmu,\rT} \times \mathbf{k}_{\Pe,\rT}) \},
\end{equation}
where $\mathbf{k}_{i,\rT}$ are the projections of the lepton momenta onto the plane orthogonal to the unit vector $\mathbf{e}_z$, which denotes the beam direction of the Higgs production process.
The distributions, in the SESM, have the same shape as in the SM, and the relative NLO corrections, reported in the lower panel, are the same as in the SM up to constant offsets. As observed in the other cases, the differences between the SM result and the ones in the various SESM scenarios can be quantified by the corresponding $\De_\text{SM}^\text{NLO}$ obtained for the integrated decay width.

\subsubsection{Semi-leptonic final states}
In the following, we present differential distributions obtained for the semi-leptonic final states $\Pd \Pdb \Pem \Pep$ and $\Pnue \Pep \Pd \Pub$, computed in the FJ renormalization scheme.
In the $\MSbar$ renormalization scheme, the distributions show similar features and are not reported here.
The renormalization scale used to generate the distributions is again $\mur = \Mh$.

\begin{figure}
\centering
\includegraphics[scale=0.85]{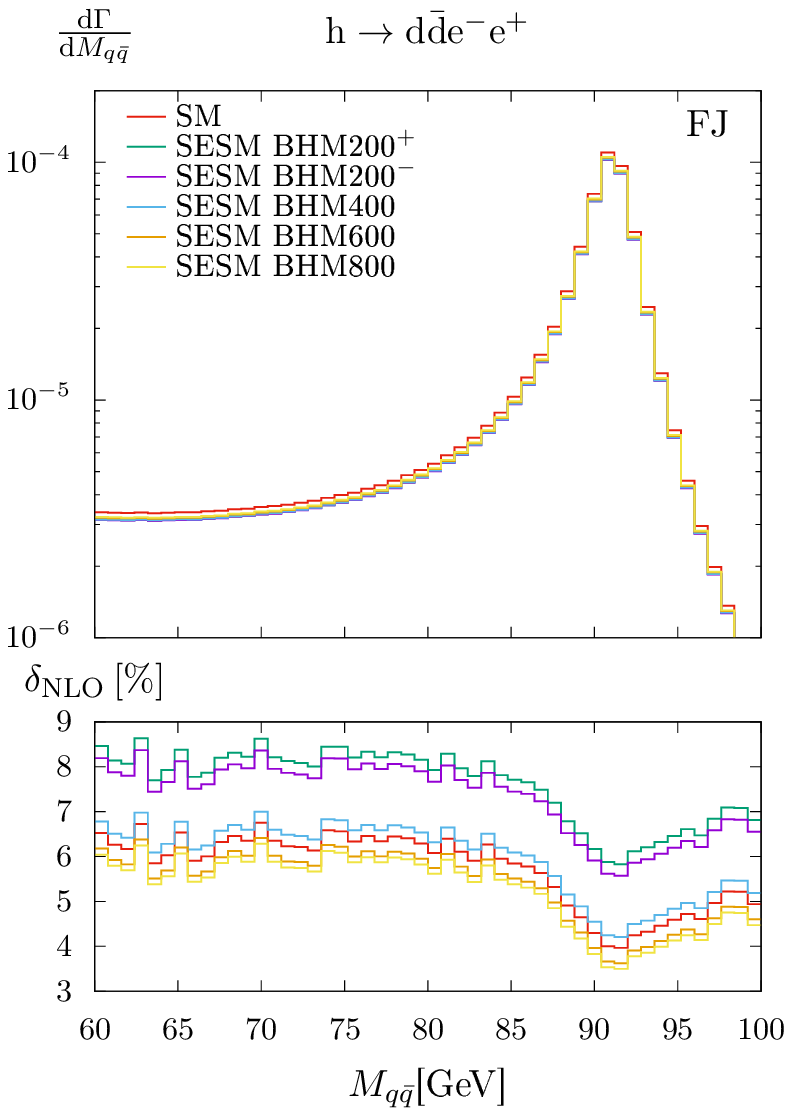}
\quad
\includegraphics[scale=0.85]{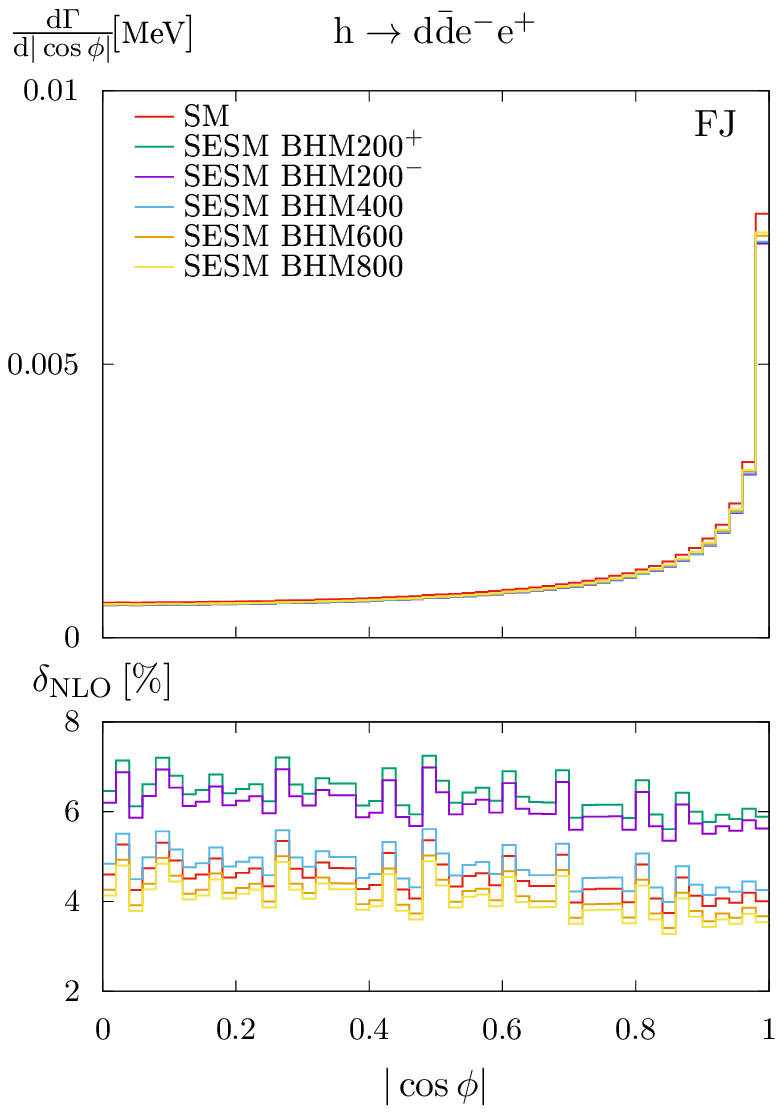}
\caption{Invariant-mass and angular distributions for the neutral-current semi-leptonic decay $\Ph \to \Pd \Pdb \Pem \Pep$, in the FJ renormalization scheme for the various SESM scenarios.
}
\label{fig:diffdistr_qqee}
\end{figure}

In \cref{fig:diffdistr_qqee} we depict differential distributions for the neutral-current final state $\Pd \Pdb \Pem \Pep$, including both EW and QCD corrections.
The left panel shows the differential width with respect to the invariant mass of the quark pair around the $\PZ$-boson peak.
Below the peak, positive NLO corrections are driven by photon radiation from the final-state quarks.
Compared to the leptonic case, the radiative tail is less pronounced due to the smaller charge factor of the quarks.
Since all the gluons are recombined with the quark pair, gluon radiation does not contribute to the tail~\cite{Bredenstein:2006ha}.
The presence of the singlet does not induce any shape distortion with respect to the SM distribution, and the difference between each SESM scenario and the SM equals the difference obtained for the corresponding integrated result.

The right panel of \cref{fig:diffdistr_qqee} shows the angular distribution in the cosine of the angle $\phi$ between the two $\PZ$-boson decay planes.
For events without gluon radiation, the final-state quarks are identified with two jets.
When gluon radiation occurs, the two QCD partons with the smallest invariant mass are recombined into a single jet, so that we always obtain events with two outgoing jets.
Since the jets cannot be distinguished, any observable must be invariant under the exchange of the two jets, and the cosine of $\phi$ can be reconstructed only up to a sign.
Thus, in the figure, it is defined by~\cite{Bredenstein:2006ha}
\begin{equation}
|\cos \phi|=\left| \frac{\left(\left(\mathbf{k}_{\text{jet}_1}+\mathbf{k}_{\text{jet}_2}\right) \times \mathbf{k}_1\right)\left(\mathbf{k}_{\text{jet}_1}\times\mathbf{k}_{\text{jet}_2}\right) }{|\left(\mathbf{k}_{\text{jet}_1}+\mathbf{k}_{\text{jet}_2}\right) \times \mathbf{k}_1| \,|\mathbf{k}_{\text{jet}_1}\times\mathbf{k}_{\text{jet}_2}|} \right|.
\end{equation}
Note that, in the binning of the distribution, $\cos \phi$ is used instead of $\phi$, so that the result looks different from the leptonic case reported in \cref{fig:diffdistr_mumuee}.
The difference between the SESM and the SM is given by an offset, which depends on the considered scenario  and is equal to the difference obtained for the corresponding integrated results.

\begin{figure}
\centering
\includegraphics[scale=0.85]{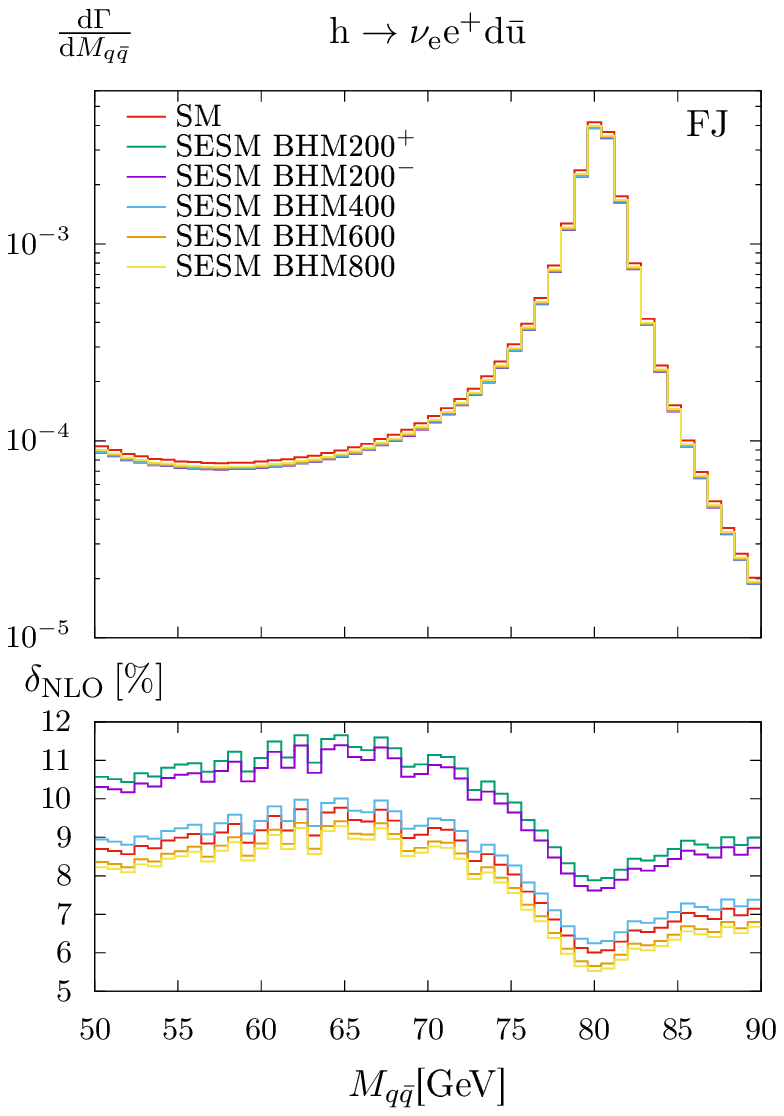}
\quad
\includegraphics[scale=0.85]{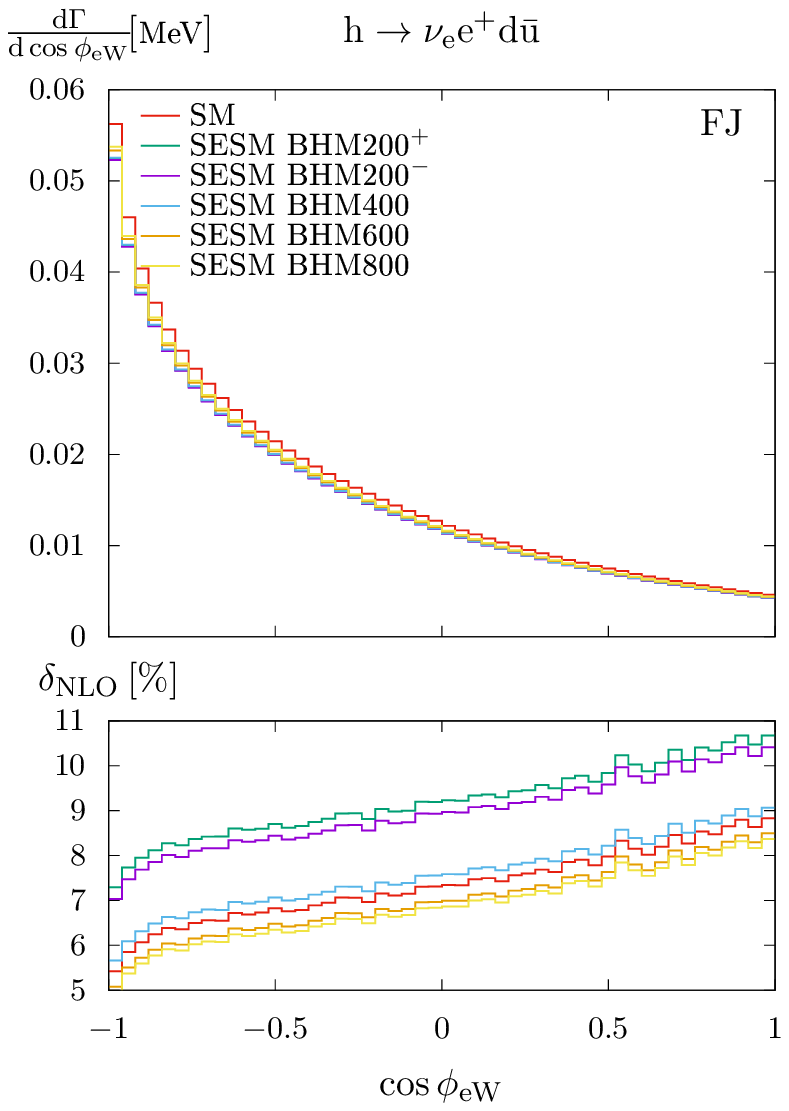}
\caption{Invariant-mass and angular distributions for the charged-current semi-leptonic decay $\Ph \to \Pnue \Pep \Pd \Pub$, in the FJ renormalization scheme for the various SESM scenarios.
}
\label{fig:diffdistr_veqq}
\end{figure}

In \cref{fig:diffdistr_veqq} invariant-mass and angular distributions for the charged-current final state $\Pnue \Pep \Pd \Pub$ are illustrated, including both EW and QCD corrections.
In the left panel, the invariant-mass distribution displays the $\PW$-boson peak, and the radiative tail induced by photon radiation can be observed.
As observed in the leptonic case, there is no shape difference between the SM and the considered SESM distributions, and the difference in the normalization equals the relative difference obtained for the integrated decay widths.
The angular distribution, reported in the right panel of \cref{fig:diffdistr_veqq}, shows the cosine of the angle between the electron and the hadronically decaying $\PW$ boson, in the Higgs rest frame. The NLO corrections slightly deform the distribution, but there is no difference in the distortion induced by the SESM and the SM. The difference between the two equals the difference encountered for the integrated results.

In general, we observe that the presence of the singlet does not change the shape of the SM distributions. Consequently, the study of differential distributions is not helpful to discriminate the SESM from the SM.  The difference between the models is given by a different normalization in the distributions and equals the relative difference observed for the integrated decay widths.
Similar results were obtained for the THDM in \citere{Altenkamp:2017kxk}.

%
\section{Conclusions}
\label{sec:conclusions}

To explore the nature of \ac{EW} symmetry breaking at the \ac{LHC}, precise theory predictions are needed. This is valid not only within the \ac{SM}, but also for its extensions, since \ac{BSM} physics might show small deviations from \ac{SM} predictions, below the~$10 \%$ level.
Precision is also required in case of new discoveries, as different \ac{BSM} theories lead to comparable effects, and a very high accuracy would be necessary to tell the right theoretical framework underlying the newly-observed states.

In this paper, we have considered a \ac{SESM} characterized by a $\mathbb{Z}_2$-invariant Lagrangian and non-vanishing \acp{vev} both for the $\SUtwo$ doublet and 
the real singlet scalar field.
The model provides two Higgs bosons, which couple to the \ac{SM} fields with the same couplings of the \ac{SM} Higgs, rescaled by the sine or cosine of a mixing angle~$\al$.
We parametrize the extended Higgs sector by the mass $\MH$ of the heavy Higgs boson, the mixing angle~$\al$, and the scalar self-coupling~$\lsd$ connecting the scalar singlet and doublet.
The renormalization of the theory has been performed adopting two schemes, in which the renormalized parameters~$\al$ and~$\lsd$ are defined by \MSb{} conditions,
since these parameters are not directly experimentally accessible.
In both schemes, all the quantities other than~$\al$ and~$\lsd$ are renormalized using \ac{OS} conditions.
In the first scheme the renormalized tadpoles are set to zero, as it is customary in \ac{OS} renormalization schemes.
This scheme introduces gauge dependences in the parametrization of observables by input parameters if \MSb{} parameters are involved, and thus should be used only within a fixed gauge.
In the second scheme, following a prescription suggested by Fleischer and Jegerlehner, gauge dependences are avoided by setting unrenormalized (bare) tadpoles to zero, so that all relations between bare parameters of the theory are gauge independent.

Identifying the lighter Higgs boson~$\Ph$ with the observed Higgs boson of mass~$125 \, \GeV$, we have computed \ac{NLO} \ac{EW} and \ac{QCD} corrections to the decays $\Ph \to \PW\PW/\PZ\PZ \to 4\,$fermions.
Using the \mathematica{} package \feynrules{}, we have implemented the \ac{SESM} into a \feynarts{} model file including the expressions of the renormalization constants, so that the model file can be used to perform \ac{NLO} computations within the two considered renormalization schemes.
Employing \feynarts{} and \formcalc{}, the model file has been used to produce \fortran{} code for the numerical computation of the matrix elements for the decays $\Ph \to \PW\PW/\PZ\PZ \to 4\,$fermions. The \fortran{} routines have been embedded in \prophecy{} to extend the capabilities of the Monte Carlo generator, which allows now for the computation of observables relevant for the Higgs decays to four fermions in the \ac{SESM} at \ac{NLO}.

The class of decay processes $\Ph \to \PW\PW/\PZ\PZ \to 4\,$fermions played a central role in the discovery of the Higgs boson and is important for the accurate characterization of the Higgs particle.
We have analyzed the decays 
for some 
\ac{SESM} benchmark scenarios proposed in the literature.
For each scenario, we have computed the total decay width $\Gamma_{\Ph \to 4 \Pf}$ for the decay of the light Higgs boson into four fermions and studied the dependence of the results on the renormalization scale~$\mur$, solving the \acp{RGE} for the \MSb{} parameters~$\al$ and~$\lsd$.
We observe that the inclusion of \ac{NLO} corrections drastically reduces the scale dependence and, 
consequently, the related theoretical uncertainty.
Changing the scale up and down by factors of two, reduces the scale uncertainty of 
$\Gamma_{\Ph \to 4 \Pf}$ typically from
$\lsim3{-}4\%$ at LO to only $\lsim0.5\%$ at NLO.

All these analyses have been performed using both renormalization schemes, 
properly converting the numerical input values between the two schemes in order 
to ensure a consistent comparison between the predictions obtained for specific scenarios.
To this end, we have investigated the conversion of the mixing angle~$\al$ between renormalization 
schemes and found 
sizeable effects which become large when approaching non-perturbative regimes.
The inclusion of \ac{NLO} corrections improves the agreement between the results computed 
in the two schemes, i.e.\ the renormalization scheme dependence is reduced at \ac{NLO}.
In the considered scenarios, the scheme dependence at the central scale typically reduces from
$\sim1\%$ at LO to $\lsim0.1\%$ at NLO.
Note that the inclusion of conversion effects is essential in a consistent 
comparison of \ac{NLO} predictions obtained in the two schemes.

Comparing the \ac{NLO} decay widths $\Gamma_{\Ph \to 4 \Pf}$ in the \ac{SESM} with the corresponding quantities in the \ac{SM}, we find deviations from the \ac{SM} that reach 
about~$-7 \%$ in the scenarios \BHMapm{} with a heavy Higgs boson of mass $200 \, \GeV$. 
The NLO corrections are typically about $5{-}10\%$, but only $1{-}2\%$
of those are due to effects beyond the SM.
For higher values of the heavy mass $\MH$, the (absolute values of the) deviations from the SM 
become smaller.
Differential distributions have been produced for the \ac{SESM} scenarios and compared to the \ac{SM} case, and no distortions
are observed on top of the \ac{SM} shapes. The only observed difference is given by a constant offset, implying that differential distributions are not helpful to observe traces of the \ac{SESM}.

Both the \feynarts{} model file and the new version of \prophecy{} are ready for further applications and can be obtained from the authors upon request.

\subsection*{Acknowledgements}
%
We thank the Research Training Group GRK~2044 of the German Research Foundation~(DFG) for funding and support and acknowledge support by the state of Baden-Württemberg through bwHPC and the DFG through grant no INST 39/963-1 FUGG.
Moreover, M.B.\ and S.D.\ acknowledge the Research Executive Agency (REA) of the European Union for funding this work through the Grant Agreement PITN-GA-2012-316704 (``HiggsTools'').

\section*{Appendix}
\appendix

\section{Results for the scenario \BHMam{}}
\label{app:resBHMam}
The results for 
benchmark scenario \BHMam{}, defined by the input values given in \cref{tab:benchmarkPoints}, are very similar to the results obtained for the scenarios \BHMap{} and \BHMc{} (see \cref{ssec:resultsBHM200,ssec:resultsBHM600});
the most important of them are collected in this appendix.

\begin{figure}
\centering
\includegraphics[scale=1.]{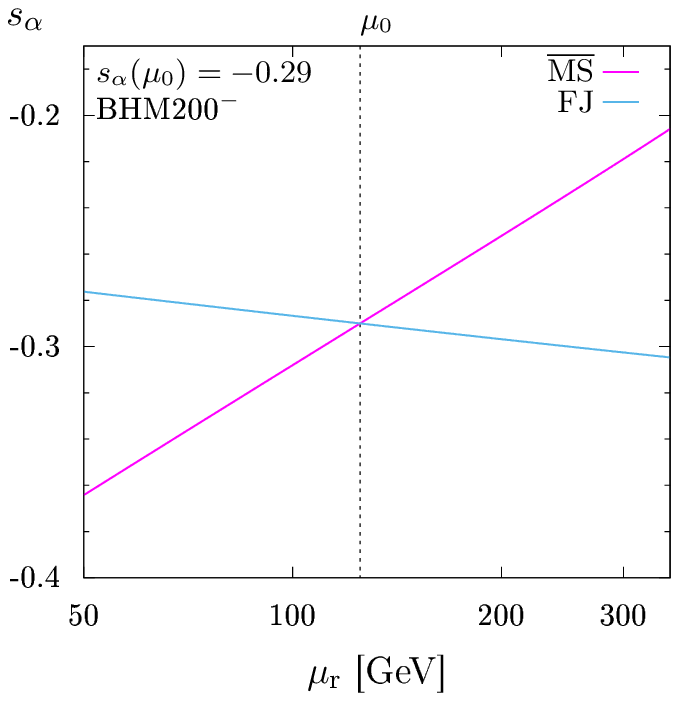}
\qquad
\includegraphics[scale=1.]{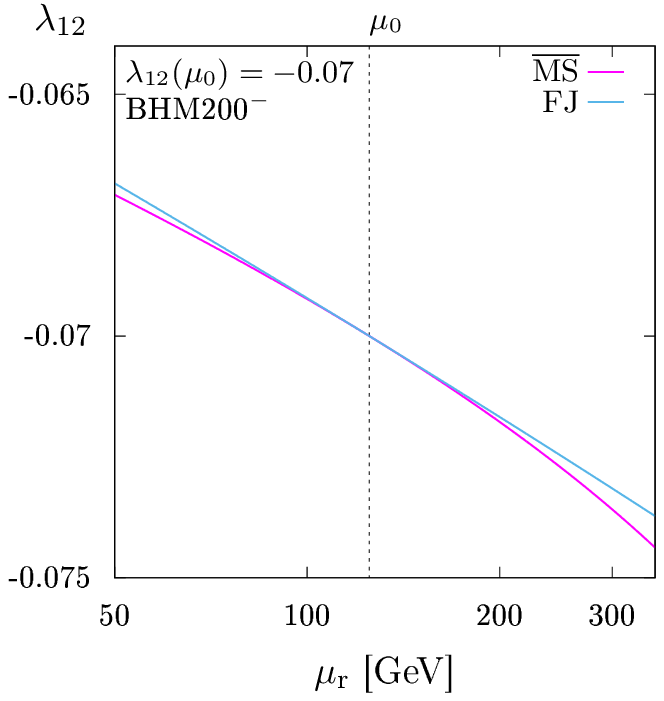}
\caption{As in \cref{fig:murscanparam2BHM200}, but for benchmark scenario \BHMam.}
\label{fig:murscanparam2BHM200n}
\vspace{2em}
\centering
\includegraphics[scale=1.]{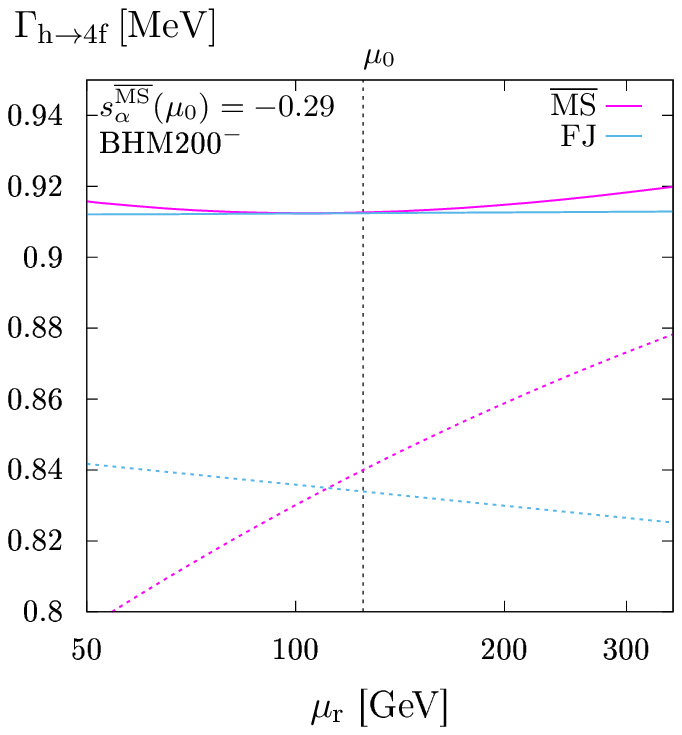}
\qquad
\includegraphics[scale=1.]{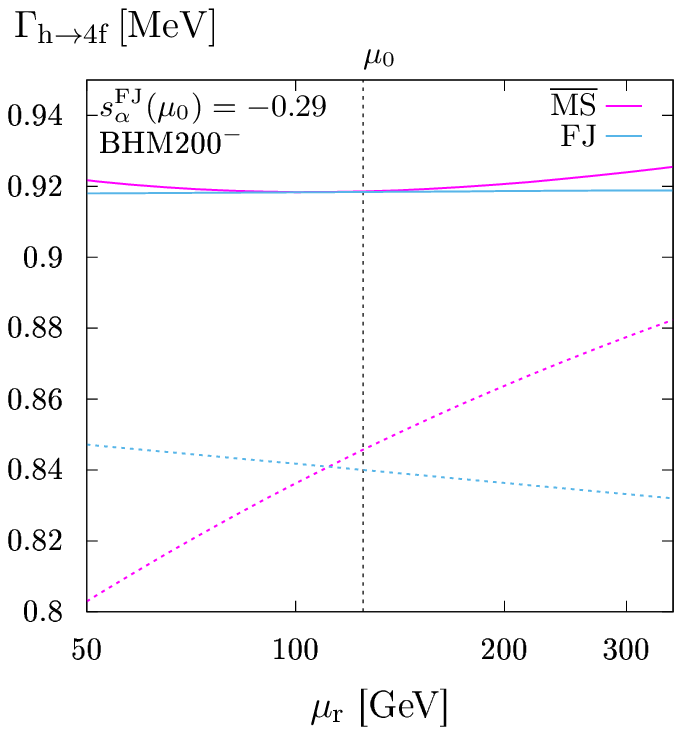}
\caption{As in \cref{fig:murscan2BHM200}, but for benchmark scenario \BHMam.}
\label{fig:murscan2BHM200n}
\end{figure}

In \cref{fig:murscanparam2BHM200n} we show the running of the parameters defined by \MSb{} conditions. Compared to the scenario \BHMap{}, the scale dependence of~$\sa$ in the FJ scheme is reduced.
The scale dependence of the inclusive decay width $\Gamma_{\Ph \to 4 \Pf}$ is shown in \cref{fig:murscan2BHM200n}.
The plot on the left is obtained using \MSb{} (FJ) input parameters converted to the FJ (\MSb{}) scheme at the scale $\mu_0 = \Mh$.
Dashed and solid lines correspond to \ac{LO} and \ac{NLO} (\ac{EW} + \ac{QCD}) results, respectively.
Including \ac{NLO} corrections, scale and scheme dependence are strongly reduced.

\begin{figure}
\centering
\includegraphics[scale=1.]{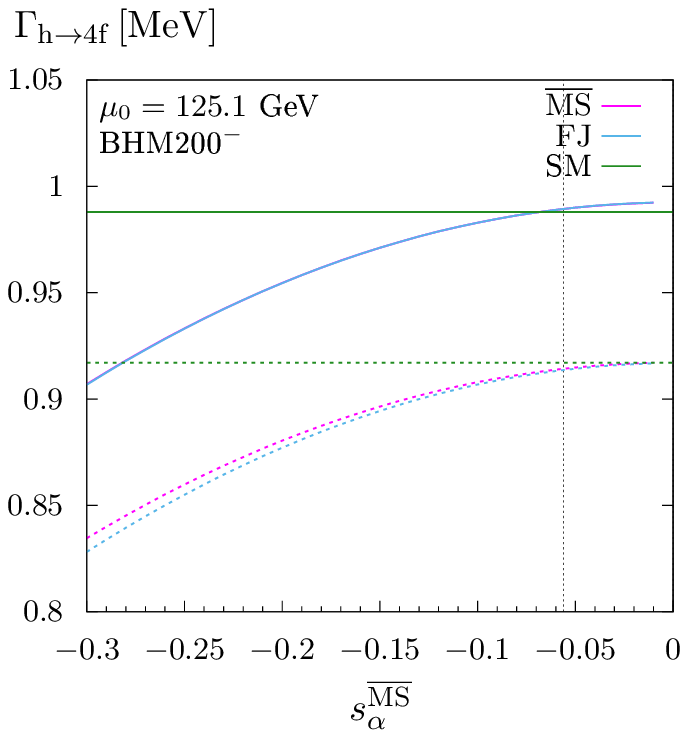}
\quad
\includegraphics[scale=1.]{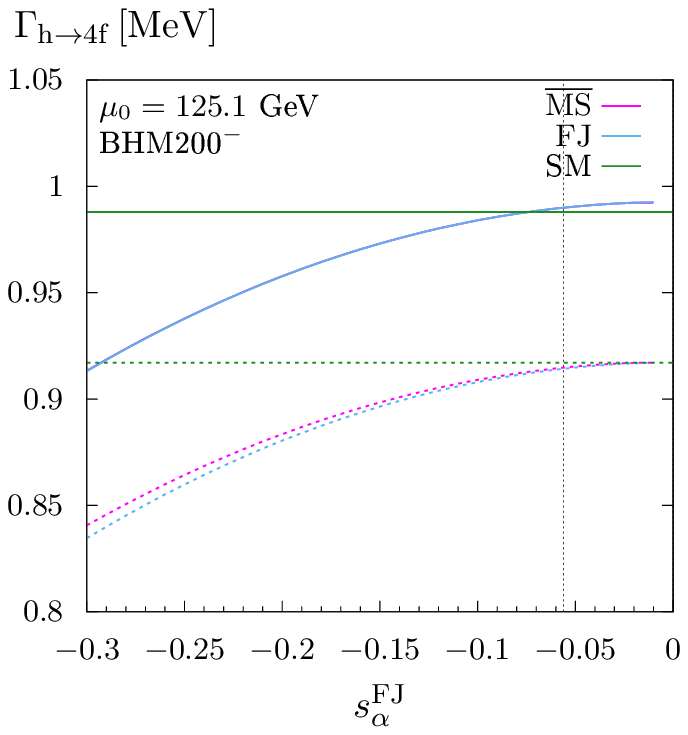}
\caption{As in \cref{fig:sascanBHM200}, but for benchmark scenario \BHMam.}
\label{fig:sascanBHM200n}
\vspace{2em}
\centering
\includegraphics[scale=0.9]{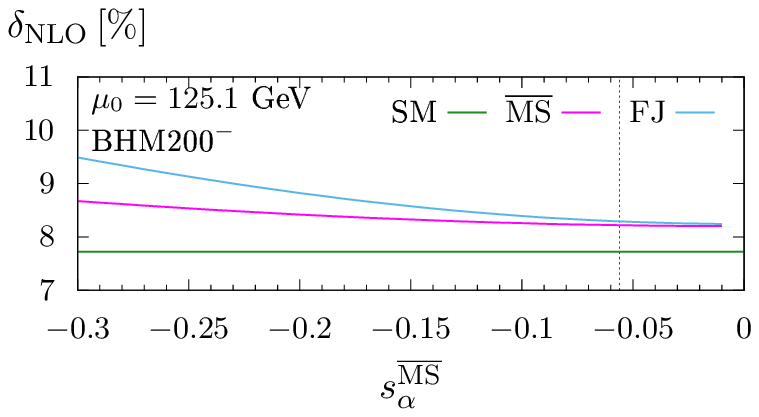}
\quad
\includegraphics[scale=0.9]{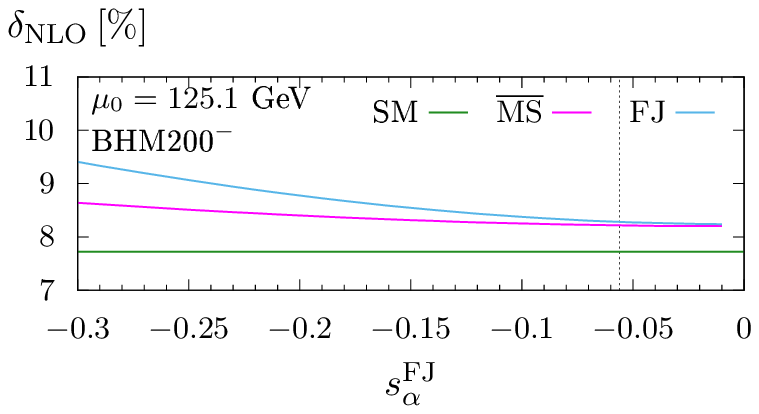}
\caption{As in \cref{fig:sascandeltaNLOBHM200}, but for benchmark scenario \BHMam.}
\label{fig:sascandeltaNLOBHM200n}
\vspace{2em}
\centering
\includegraphics[scale=1.]{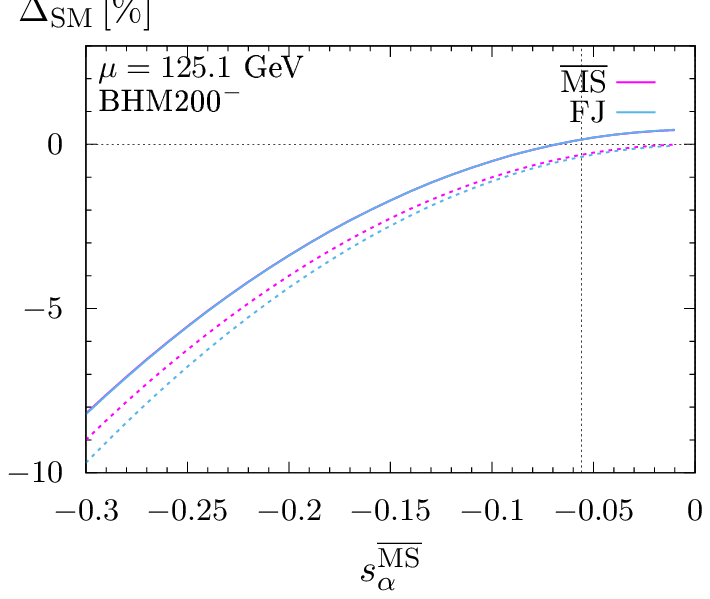}
\quad
\includegraphics[scale=1.]{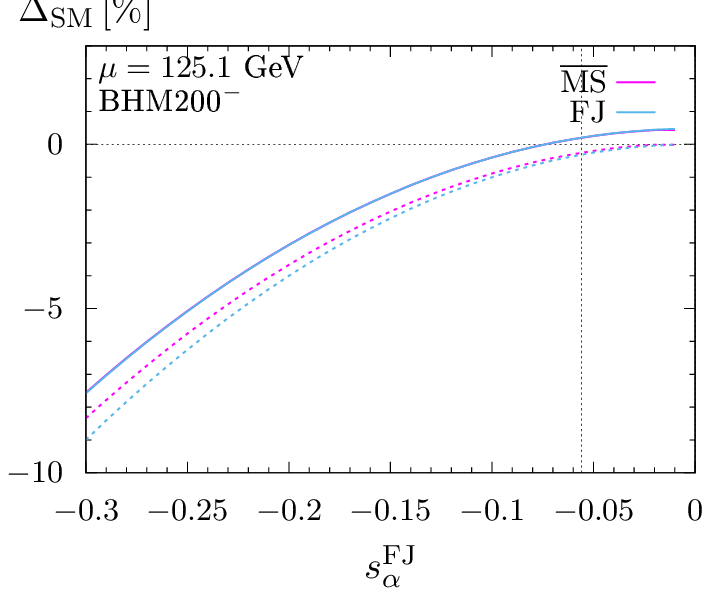}
\caption{As in \cref{fig:sascanDeltaSMBHM200}, but for benchmark scenario \BHMam.}
\label{fig:sascanDeltaSMBHM200n}
\end{figure}

The total decay width~$\Gamma_{\Ph \to 4 \Pf}$, the relative \ac{NLO} corrections, and the deviations from the \ac{SM} are shown respectively in Figs.\ \ref{fig:sascanBHM200n}, \ref{fig:sascandeltaNLOBHM200n}, and \ref{fig:sascanDeltaSMBHM200n} as functions of~$\sa$.
The values of~$\MH$ and~$\lsd$ are fixed according to scenario \BHMam.
The plots on the left (right) are obtained using the \MSb{} (FJ) input scheme and converting~$\sa$ to the FJ (\MSb{}) scheme. \ac{LO} and \ac{NLO} (\ac{EW} + \ac{QCD}) results are represented, respectively, by dashed and solid lines. 
Where relevant, green lines represent the \ac{SM} value.
The vertical dashed lines at~$\sa \sim -0.05$ mark the maximal $\sa$ value for which the perturbativity constraints~\eqref{eq:BSMconstrPert} are fulfilled.
The agreement between the two schemes improves after including \ac{NLO} corrections.
Quantitatively, the scale dependence of $\Gamma_{\Ph \to 4 \Pf}$ (by scale variations of factors $2$ and $1/2$)
reduces from $\sim3{-}4\%$ at LO to $\sim0.4\%$ at NLO in the $\MSbar$ scheme, 
and from $\sim0.7\%$ at LO to $\lsim0.1\%$ at NLO in the FJ scheme, reflecting the difference
in the running of $\alpha$ in this scenario.
The scheme dependence at the central scale
reduces from $\sim0.7\%$ at LO to $\lsim0.1\%$ at NLO.
%
%

\clearpage
\section{Results for the scenario \BHMb{}}
\label{app:resBHMb}
In this appendix we show results for the scenario \BHMb{}, defined by the input values given in \cref{tab:benchmarkPoints}. The calculations and the results are very similar to the ones described in \cref{ssec:resultsBHM200,ssec:resultsBHM600} for the scenarios \BHMap{} and \BHMc{}, respectively, thus we do not discuss the details again.

\begin{figure}
\centering
\includegraphics[scale=1.]{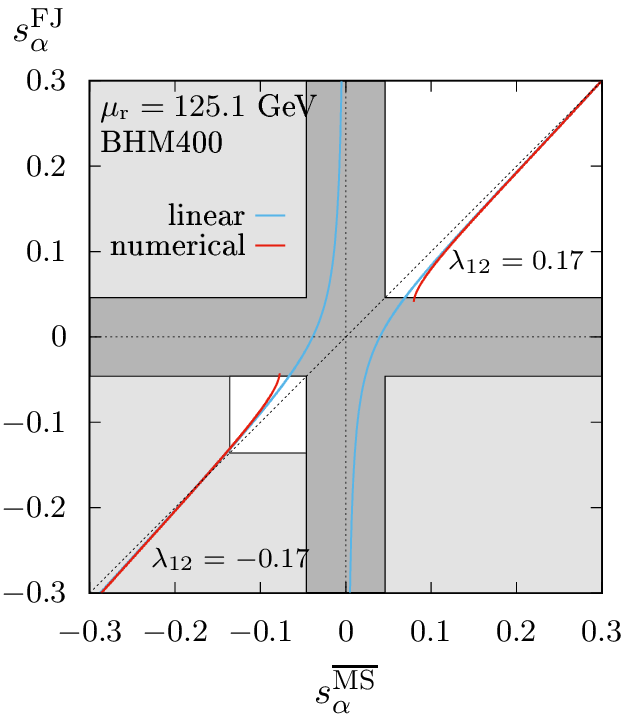}
\qquad
\includegraphics[scale=1.]{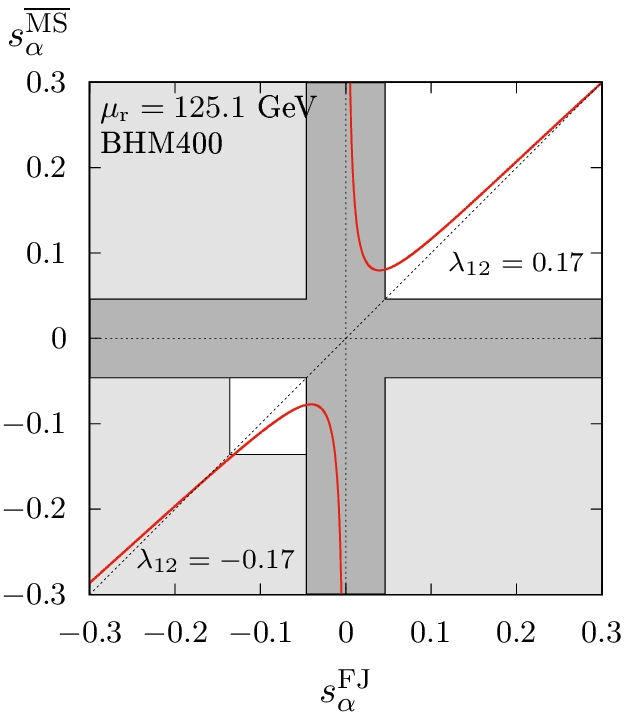}
\caption{As in \cref{fig:schemeconversion2BHM200}, but for benchmark scenario \BHMb.}
\label{fig:schemeconversion2BHM400}
\vspace*{1em}
\centering
\includegraphics[scale=1.]{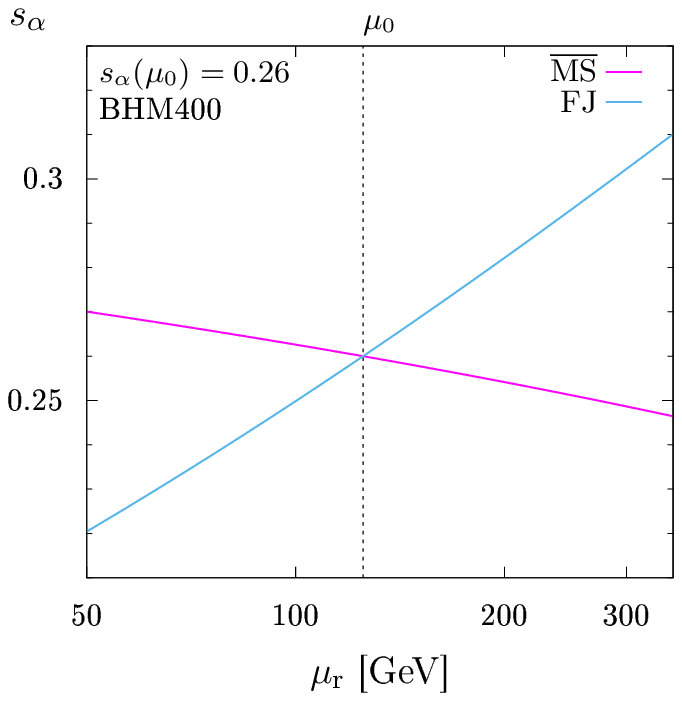}
\qquad
\includegraphics[scale=1.]{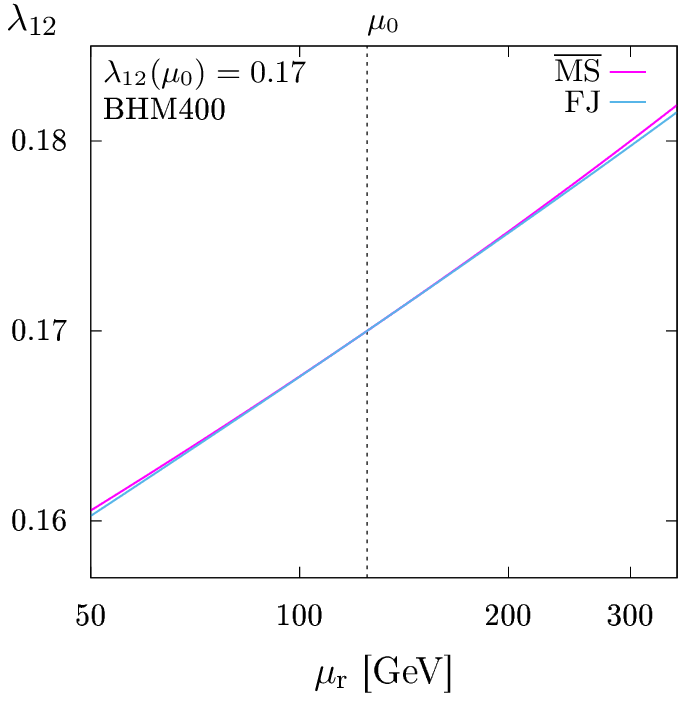}
\caption{As in \cref{fig:murscanparam2BHM200}, but for benchmark scenario \BHMb.}
\label{fig:murscanparam2BHM400}
\vspace*{1em}
\centering
\includegraphics[scale=1.]{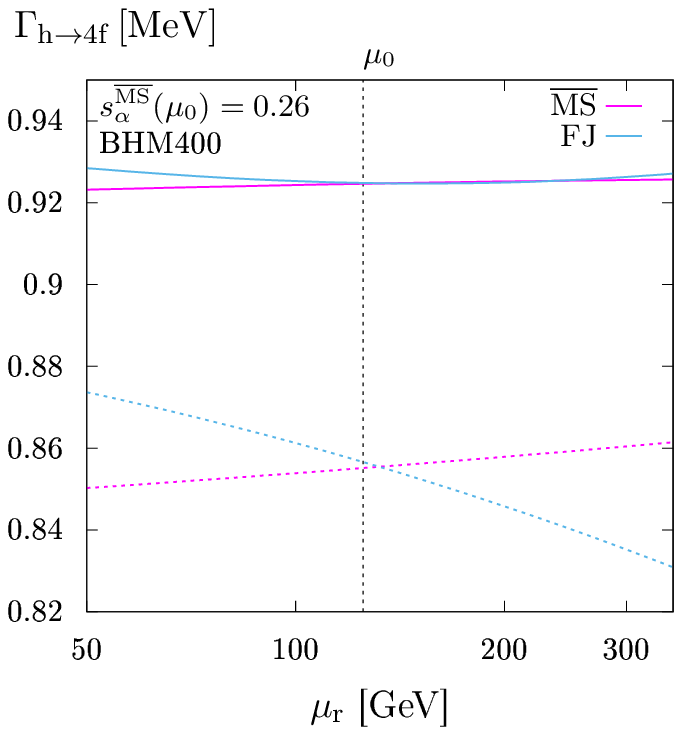}
\qquad
\includegraphics[scale=1.]{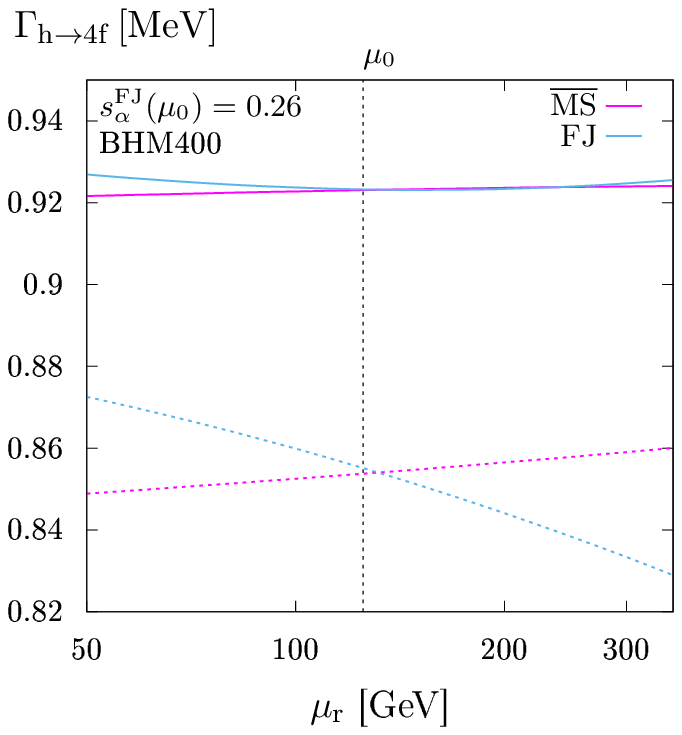}
\caption{As in \cref{fig:murscan2BHM200}, but for benchmark scenario \BHMb.}
\label{fig:murscan2BHM400}
\end{figure}

The conversions of the mixing angle from the \MSb{} to the FJ scheme, and vice versa, are computed for $\MH$ and $|\lsd|$ values corresponding to the scenario \BHMb{} and shown in \cref{fig:schemeconversion2BHM400}.
The sign of~$\lsd$ is fixed according to the sign of~$\sa$ on the $x$-axis,
as indicated in the figure.
Red and blue lines correspond, respectively, to the complete and linearized solutions of \cref{eq:FJtoMRconversion}.
For values of $\sa$ in  the dark-gray area, the perturbativity constraints \eqref{eq:BSMconstrPert} are violated.
The light-gray areas denote where the vacuum stability condition \eqref{eq:vacuumStabCond2} is violated, 
or where the sign of~$\sa$ is flipped by the conversion (and becomes inconsistent with the sign of~$\lsd$).
The conversion effects are in general small and become larger when approaching the non-perturbative region.

The running of the parameters defined by $\MSbar$ renormalization conditions is shown, for the scenario \BHMb{}, in \cref{fig:murscanparam2BHM400}.
The scale dependence of the mixing angle is more accentuate in the FJ scheme, while the running of $\lsd$ is very similar in the two renormalization schemes.
%

In \cref{fig:murscan2BHM400}, we show the renormalization scale dependence of the $\Ph \to 4 \Pf$ decay width.
The left (right) panel is obtained using \MSb{} (FJ) input parameters and converting to the FJ (\MSb{}) scheme at the scale $\mu_0 = \Mh$.
Dashed lines correspond to \ac{LO} results, solid lines include \ac{NLO} \ac{EW}+\ac{QCD} contributions.
Both scale and scheme dependence are reduced by the inclusion of NLO corrections.
The scale dependence of the width
reduces from $\sim0.5\%$ at LO to $\sim0.1\%$ at NLO in the $\MSbar$ scheme, 
and from $\sim2\%$ at LO to $\sim0.2\%$ at NLO in the FJ scheme, while
the scheme dependence at the central scale
is $\sim0.2\%$ at LO and $\lsim0.1\%$ at NLO.

\begin{figure}
\centering
\includegraphics[scale=1.]{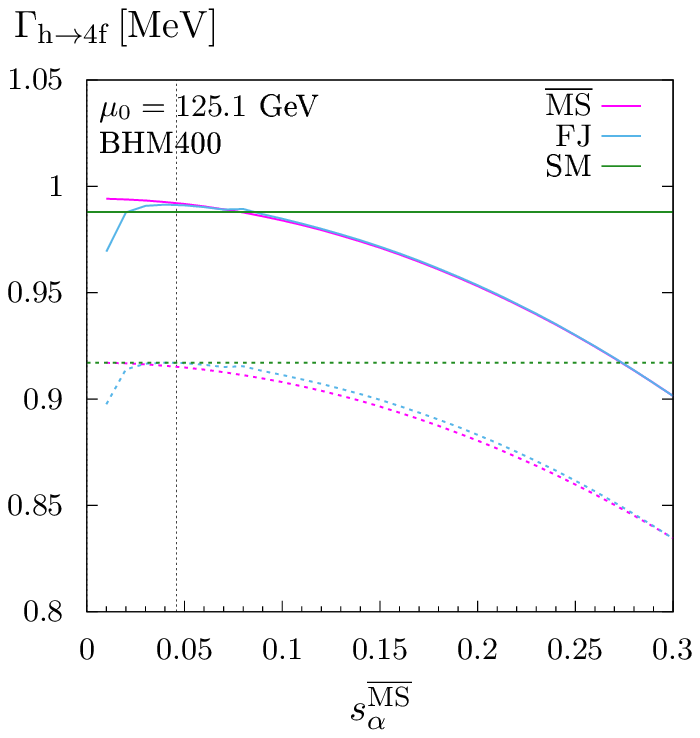}
\quad
\includegraphics[scale=1.]{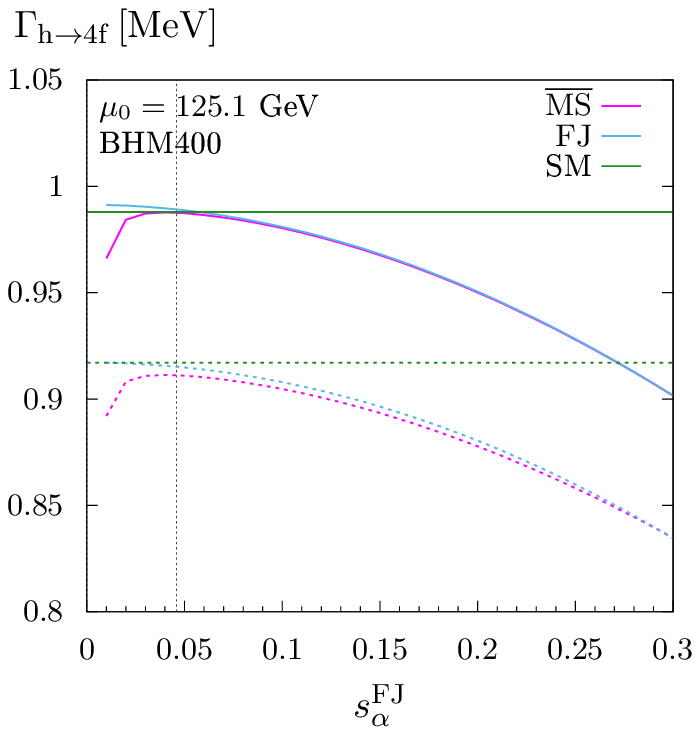}
\caption{As in \cref{fig:sascanBHM200}, but for benchmark scenario \BHMb.}
\label{fig:sascanBHM400}
\vspace*{2em}
\centering
\includegraphics[scale=0.9]{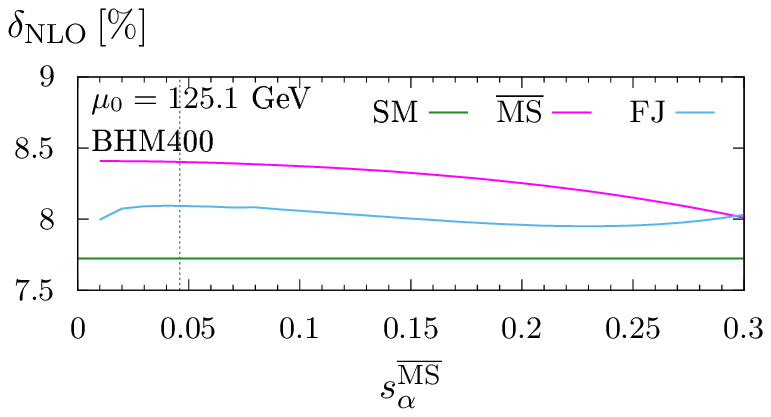}
\quad
\includegraphics[scale=0.9]{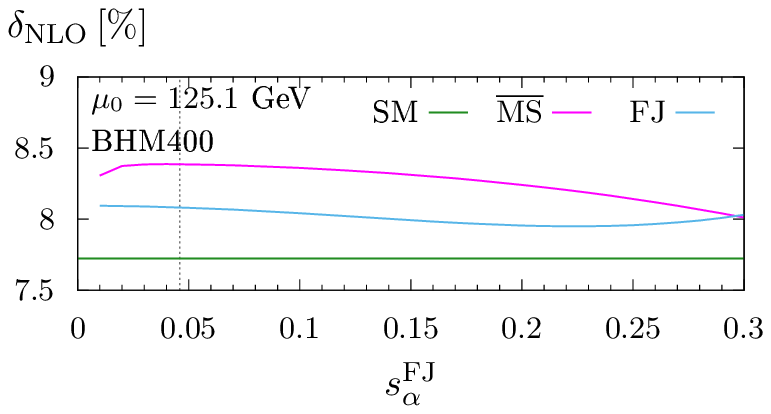}
\caption{As in \cref{fig:sascandeltaNLOBHM200}, but for benchmark scenario \BHMb.}
\label{fig:sascandeltaNLOBHM400}
\vspace*{2em}
\centering
\includegraphics[scale=1.]{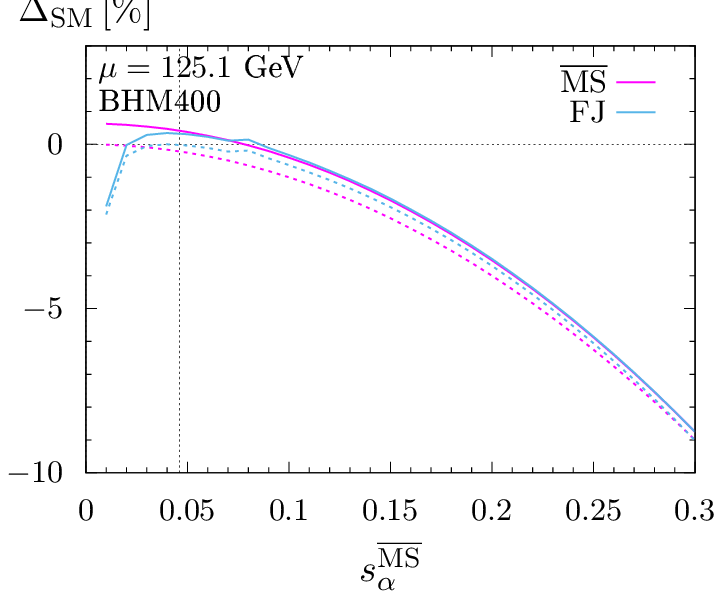}
\quad
\includegraphics[scale=1.]{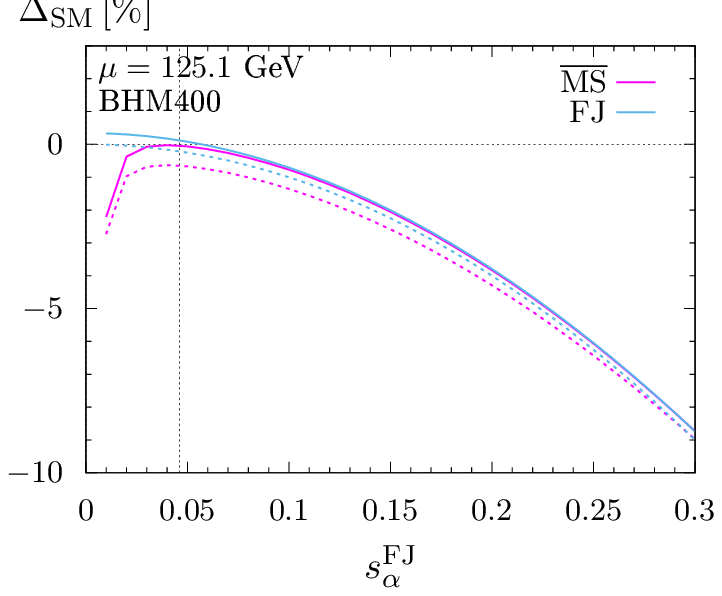}
\caption{As in \cref{fig:sascanDeltaSMBHM200}, but for benchmark scenario \BHMb.}
\label{fig:sascanDeltaSMBHM400}
\end{figure}

Figures \ref{fig:sascanBHM400}, \ref{fig:sascandeltaNLOBHM400}, and \ref{fig:sascanDeltaSMBHM400} show, respectively, the absolute values, the relative NLO corrections, and the deviations from the SM for the decay width $\Gamma_{\Ph \to 4 \Pf}$ in the SESM as functions of $\sa$. The parameters $\MH$ and $\lsd$ are fixed according to scenario \BHMb{}.
The plots on the left (right) are obtained using the \MSb{} (FJ) input scheme and converting $\sa$ to the FJ (\MSb{}) scheme.
As usual, dashed and solid lines represent, respectively, \ac{LO} and \ac{NLO} \ac{EW}+\ac{QCD} results.
Where relevant, the \ac{SM} result is reported in green.
The dashed vertical line marks the minimal $\sa$ value for which the perturbativity constraints \eqref{eq:BSMconstrPert} are fulfilled.
The inclusion of NLO corrections improves the agreement between the results computed in the two schemes.
For $\sa = 0.26$, corresponding to the scenario \BHMb{}, the decay width deviates $6{-}7\%$ from the \ac{SM} value, slightly depending on the input scheme used.
%

\clearpage
\bibliographystyle{JHEPmod}            
\bibliography{bibliography}

\end{document}

%% file: diagrams/tadpoles.tex
\begin{picture}(80,30)(0,0)
\DashLine(0,15)(40,15){4}
\Text(0,16)[lb]{$\small{\Fh, \FH}$}
\Vertex(40,15){2}
\ArrowArc(55,15)(15,-180,180)
\Text(73,15)[l]{$\small{\Ff}$}
\end{picture}
\quad
\begin{picture}(80,30)(0,0)
\DashLine(0,15)(40,15){4}
\Text(0,16)[lb]{$\small{\Fh, \FH}$}
\Vertex(40,16){2}
\DashCArc(55,15)(15,-180,180){4}
\Text(73,15)[l]{$\small{\FS}$}
\end{picture}
\quad
\begin{picture}(80,30)(0,0)
\DashLine(0,15)(40,15){4}
\Text(0,16)[lb]{$\small{\Fh, \FH}$}
\Vertex(40,15){2}
\DashArrowArc(55,15)(15,-180,180){1.2}
\Text(73,15)[l]{$\small{u}$}
\end{picture}
\quad
\begin{picture}(80,30)(0,0)
\DashLine(0,15)(40,15){4}
\Text(0,16)[lb]{$\small{\Fh, \FH}$}
\Vertex(40,15){2}
\PhotonArc(55,15)(15,-180,180){1.6}{13}
\Text(73,15)[l]{$\small{\FV}$}
\end{picture}

%% file: diagrams/selfenergies.tex
\begin{picture}(100,70)(0,0)
\DashLine(0,30)(100,30){4}
\Text(50,63)[b]{$\small{\FS}$}
\Vertex(50,30){2}
\Text(0,33)[lb]{$\small{\Fh}$}
\DashCArc(50,45)(15,-90,270){4}
\Text(100,33)[rb]{$\small{\Fh}$}
\end{picture}
\quad
\begin{picture}(100,70)(0,0)
\DashLine(0,30)(100,30){4}
\Text(50,63)[b]{$\small{\FV}$}
\Vertex(50,30){2}
\Text(0,33)[lb]{$\small{\Fh}$}
\PhotonArc(50,45)(15,-90,270){1.6}{13}
\Text(100,33)[rb]{$\small{\Fh}$}
\end{picture}
\quad
\begin{picture}(100,70)(0,0)
\DashLine(0,30)(30,30){4}
\Vertex(30,30){2}
\Text(0,33)[lb]{$\small{\Fh}$}
\ArrowArc(50,30)(20,0,180)
\ArrowArc(50,30)(20,180,360)
\Text(50,53)[b]{$\small{\Ff}$}
\Text(50,7)[t]{$\small{\Ff}$}
\Vertex(70,30){2}
\DashLine(70,30)(100,30){4}
\Text(100,33)[rb]{$\small{\Fh}$}
\end{picture}
\quad
\begin{picture}(100,70)(0,0)
\DashLine(0,30)(30,30){4}
\Vertex(30,30){2}
\Text(0,33)[lb]{$\small{\Fh}$}
\DashCArc(50,30)(20,0,360){4}
\Text(50,53)[b]{$\small{\FS}$}
\Text(50,7)[t]{$\small{\FS}$}
\Vertex(70,30){2}
\DashLine(70,30)(100,30){4}
\Text(100,33)[rb]{$\small{\Fh}$}
\end{picture}
\\
\begin{picture}(100,60)(0,0)
\DashLine(0,30)(30,30){4}
\Vertex(30,30){2}
\Text(0,33)[lb]{$\small{\Fh}$}
\DashArrowArc(50,30)(20,0,180){1.2}
\DashArrowArc(50,30)(20,180,360){1.2}
\Text(50,53)[b]{$\small{u}$}
\Text(50,7)[t]{$\small{u}$}
\Vertex(70,30){2}
\DashLine(70,30)(100,30){4}
\Text(100,33)[rb]{$\small{\Fh}$}
\end{picture}
\quad
\begin{picture}(100,60)(0,0)
\DashLine(0,30)(30,30){4}
\Vertex(30,30){2}
\Text(0,33)[lb]{$\small{\Fh}$}
\PhotonArc(50,30)(20,0,180){1.6}{7}
\PhotonArc(50,30)(20,180,360){1.6}{7}
\Text(50,53)[b]{$\small{\FV}$}
\Text(50,7)[t]{$\small{\FV}$}
\Vertex(70,30){2}
\DashLine(70,30)(100,30){4}
\Text(100,33)[rb]{$\small{\Fh}$}
\end{picture}
\quad
\begin{picture}(100,60)(0,0)
\DashLine(0,30)(30,30){4}
\Vertex(30,30){2}
\Text(0,33)[lb]{$\small{\Fh}$}
\PhotonArc(50,30)(20,0,180){1.6}{7}
\DashCArc(50,30)(20,180,360){4}
\Text(50,53)[b]{$\small{\FV}$}
\Text(50,7)[t]{$\small{\FS}$}
\Vertex(70,30){2}
\DashLine(70,30)(100,30){4}
\Text(100,33)[rb]{$\small{\Fh}$}
\end{picture}

%% file: diagrams/triangles.tex
\scalebox{.8}{
\begin{picture}(90,90)(0,0)
\DashLine(0,12)(30,25){4}
\DashLine(0,78)(30,65){4}
\Text(10,2)[b]{$\small{\Fh}$}
\Text(10,88)[t]{$\small{\Fh}$}
\ArrowLine(30,25)(60,45)
\ArrowLine(60,45)(30,65)
\ArrowLine(30,65)(30,25)
\Text(46,34)[lt]{$\small{\Ff}$}
\Text(46,57)[lb]{$\small{\Ff}$}
\Text(27,45)[r]{$\small{\Ff}$}
\Vertex(30,25){2}
\Vertex(60,45){2}
\Vertex(30,65){2}
\DashLine(60,45)(90,45){4}
\Text(90,48)[rb]{$\small{\Fh}$}
\end{picture}
\quad
\begin{picture}(90,90)(0,0)
\DashLine(0,12)(30,25){4}
\DashLine(0,78)(30,65){4}
\Text(10,2)[b]{$\small{\Fh}$}
\Text(10,88)[t]{$\small{\Fh}$}
\DashLine(30,25)(60,45){4}
\DashLine(60,45)(30,65){4}
\DashLine(30,65)(30,25){4}
\Text(46,34)[lt]{$\small{\FS}$}
\Text(46,57)[lb]{$\small{\FS}$}
\Text(27,45)[r]{$\small{\FS}$}
\Vertex(30,25){2}
\Vertex(60,45){2}
\Vertex(30,65){2}
\DashLine(60,45)(90,45){4}
\Text(90,48)[rb]{$\small{\Fh}$}
\end{picture}
\quad
\begin{picture}(90,90)(0,0)
\DashLine(0,12)(30,25){4}
\DashLine(0,78)(30,65){4}
\Text(10,2)[b]{$\small{\Fh}$}
\Text(10,88)[t]{$\small{\Fh}$}
\Photon(30,25)(60,45){1.6}{4.5}
\DashLine(60,45)(30,65){4}
\DashLine(30,65)(30,25){4}
\Text(46,34)[lt]{$\small{\FV}$}
\Text(46,57)[lb]{$\small{\FS}$}
\Text(27,45)[r]{$\small{\FS}$}
\Vertex(30,25){2}
\Vertex(60,45){2}
\Vertex(30,65){2}
\DashLine(60,45)(90,45){4}
\Text(90,48)[rb]{$\small{\Fh}$}
\end{picture}
\quad
\begin{picture}(90,90)(0,0)
\DashLine(0,12)(30,25){4}
\DashLine(0,78)(30,65){4}
\Text(10,2)[b]{$\small{\Fh}$}
\Text(10,88)[t]{$\small{\Fh}$}
\DashLine(30,25)(60,45){4}
\Photon(60,45)(30,65){1.6}{4.5}
\DashLine(30,65)(30,25){4}
\Text(46,34)[lt]{$\small{\FS}$}
\Text(46,57)[lb]{$\small{\FV}$}
\Text(27,45)[r]{$\small{\FS}$}
\Vertex(30,25){2}
\Vertex(60,45){2}
\Vertex(30,65){2}
\DashLine(60,45)(90,45){4}
\Text(90,48)[rb]{$\small{\Fh}$}
\end{picture}
\quad
\begin{picture}(90,90)(0,0)
\DashLine(0,12)(30,25){4}
\DashLine(0,78)(30,65){4}
\Text(10,2)[b]{$\small{\Fh}$}
\Text(10,88)[t]{$\small{\Fh}$}
\DashLine(30,25)(60,45){4}
\DashLine(60,45)(30,65){4}
\Photon(30,65)(30,25){1.6}{4.5}
\Text(46,34)[lt]{$\small{\FS}$}
\Text(46,57)[lb]{$\small{\FS}$}
\Text(27,45)[r]{$\small{\FV}$}
\Vertex(30,25){2}
\Vertex(60,45){2}
\Vertex(30,65){2}
\DashLine(60,45)(90,45){4}
\Text(90,48)[rb]{$\small{\Fh}$}
\end{picture}
}
\\
\vspace{15pt}
\scalebox{.8}{
\begin{picture}(90,90)(0,0)
\DashLine(0,12)(30,25){4}
\DashLine(0,78)(30,65){4}
\Text(10,2)[b]{$\small{\Fh}$}
\Text(10,88)[t]{$\small{\Fh}$}
\Photon(30,25)(60,45){1.6}{4.5}
\Photon(60,45)(30,65){1.6}{4.5}
\DashLine(30,65)(30,25){4}
\Text(46,34)[lt]{$\small{\FV}$}
\Text(46,57)[lb]{$\small{\FV}$}
\Text(27,45)[r]{$\small{\FS}$}
\Vertex(30,25){2}
\Vertex(60,45){2}
\Vertex(30,65){2}
\DashLine(60,45)(90,45){4}
\Text(90,48)[rb]{$\small{\Fh}$}
\end{picture}
\quad
\begin{picture}(90,90)(0,0)
\DashLine(0,12)(30,25){4}
\DashLine(0,78)(30,65){4}
\Text(10,2)[b]{$\small{\Fh}$}
\Text(10,88)[t]{$\small{\Fh}$}
\Photon(30,25)(60,45){1.6}{4.5}
\Photon(60,45)(30,65){1.6}{4.5}
\Photon(30,65)(30,25){1.6}{4.5}
\Text(46,34)[lt]{$\small{\FV}$}
\Text(46,57)[lb]{$\small{\FV}$}
\Text(27,45)[r]{$\small{\FV}$}
\Vertex(30,25){2}
\Vertex(60,45){2}
\Vertex(30,65){2}
\DashLine(60,45)(90,45){4}
\Text(90,48)[rb]{$\small{\Fh}$}
\end{picture}
\quad
\begin{picture}(90,90)(0,0)
\DashLine(0,12)(30,25){4}
\DashLine(0,78)(30,65){4}
\Text(10,2)[b]{$\small{\Fh}$}
\Text(10,88)[t]{$\small{\Fh}$}
\DashArrowLine(30,25)(60,45){1.2}
\DashArrowLine(60,45)(30,65){1.2}
\DashArrowLine(30,65)(30,25){1.2}
\Text(46,34)[lt]{$\small{u}$}
\Text(46,57)[lb]{$\small{u}$}
\Text(27,45)[r]{$\small{u}$}
\Vertex(30,25){2}
\Vertex(60,45){2}
\Vertex(30,65){2}
\DashLine(60,45)(90,45){4}
\Text(90,48)[rb]{$\small{\Fh}$}
\end{picture}
\quad
\begin{picture}(90,90)(0,0)
\DashLine(0,12)(45,35){4}
\DashLine(0,78)(45,65){4}
\Text(10,2)[b]{$\small{\Fh}$}
\Text(10,88)[t]{$\small{\Fh}$}
\DashCArc(45,50)(15,-90,90){4}
\DashCArc(45,50)(15,90,-90){4}
\Text(27,50)[rt]{$\small{\FS}$}
\Text(63,50)[lb]{$\small{\FS}$}
\Vertex(45,35){2}
\Vertex(45,65){2}
\DashLine(45,35)(90,45){4}
\Text(90,48)[rb]{$\small{\Fh}$}
\end{picture}
\quad
\begin{picture}(90,90)(0,0)
\DashLine(0,12)(45,25){4}
\DashLine(0,78)(45,55){4}
\Text(10,2)[b]{$\small{\Fh}$}
\Text(10,88)[t]{$\small{\Fh}$}
\DashCArc(45,40)(15,-90,90){4}
\DashCArc(45,40)(15,90,-90){4}
\Text(27,40)[rb]{$\small{\FS}$}
\Text(63,40)[tl]{$\small{\FS}$}
\Vertex(45,25){2}
\Vertex(45,55){2}
\DashLine(45,55)(90,45){4}
\Text(90,50)[rb]{$\small{\Fh}$}
\end{picture}
}
\\
\vspace{10pt}
\scalebox{.8}{
\begin{picture}(90,80)(0,0)
\DashLine(0,12)(30,40){4}
\DashLine(0,68)(30,40){4}
\Text(10,2)[b]{$\small{\Fh}$}
\Text(10,78)[t]{$\small{\Fh}$}
\DashCArc(45,40)(15,0,180){4}
\DashCArc(45,40)(15,180,360){4}
\Text(45,58)[b]{$\small{\FS}$}
\Text(45,22)[t]{$\small{\FS}$}
\Vertex(30,40){2}
\Vertex(60,40){2}
\DashLine(60,40)(90,40){4}
\Text(90,43)[rb]{$\small{\Fh}$}
\end{picture}
\quad
\begin{picture}(90,90)(0,0)
\DashLine(0,12)(45,35){4}
\DashLine(0,78)(45,65){4}
\Text(10,2)[b]{$\small{\Fh}$}
\Text(10,88)[t]{$\small{\Fh}$}
\PhotonArc(45,50)(15,-90,90){1.6}{6}
\PhotonArc(45,50)(15,90,-90){1.6}{6}
\Text(27,50)[rt]{$\small{\FV}$}
\Text(63,50)[lb]{$\small{\FV}$}
\Vertex(45,35){2}
\Vertex(45,65){2}
\DashLine(45,35)(90,45){4}
\Text(90,48)[rb]{$\small{\Fh}$}
\end{picture}
\quad
\begin{picture}(90,90)(0,0)
\DashLine(0,12)(45,25){4}
\DashLine(0,78)(45,55){4}
\Text(10,2)[b]{$\small{\Fh}$}
\Text(10,88)[t]{$\small{\Fh}$}
\PhotonArc(45,40)(15,-90,90){1.6}{6}
\PhotonArc(45,40)(15,90,-90){1.6}{6}
\Text(27,40)[rb]{$\small{\FV}$}
\Text(63,40)[tl]{$\small{\FV}$}
\Vertex(45,25){2}
\Vertex(45,55){2}
\DashLine(45,55)(90,45){4}
\Text(90,50)[rb]{$\small{\Fh}$}
\end{picture}
\quad
\begin{picture}(90,80)(0,0)
\DashLine(0,12)(30,40){4}
\DashLine(0,68)(30,40){4}
\Text(10,2)[b]{$\small{\Fh}$}
\Text(10,78)[t]{$\small{\Fh}$}
\PhotonArc(45,40)(15,0,180){1.6}{6}
\PhotonArc(45,40)(15,180,360){1.6}{6}
\Text(45,58)[b]{$\small{\FV}$}
\Text(45,22)[t]{$\small{\FV}$}
\Vertex(30,40){2}
\Vertex(60,40){2}
\DashLine(60,40)(90,40){4}
\Text(90,43)[rb]{$\small{\Fh}$}
\end{picture}
}

%% file: diagrams/ssppdiagrams.tex
\scalebox{.8}{
\begin{picture}(115,80)(-10,0)
\DashLine(0,12)(45,25){4}
\DashLine(0,68)(45,55){4}
\Text(-5,8)[r]{$\small{\sigma}$}
\Text(-5,72)[r]{$\small{\sigma}$}
\DashCArc(45,40)(15,0,360){4}
\Text(25,40)[r]{$\small{\sigma}$}
\Text(65,40)[l]{$\small{\phi}$}
\Vertex(45,55){2}
\Vertex(45,25){2}
\DashLine(45,25)(90,12){4}
\DashLine(45,55)(90,68){4}
\Text(97,8)[l]{$\small{\phi^-}$}
\Text(97,72)[l]{$\small{\phi^+}$}
\end{picture}
\qquad
\begin{picture}(115,80)(-10,0)
\DashLine(0,12)(30,40){4}
\DashLine(0,68)(30,40){4}
\Text(-5,8)[r]{$\small{\sigma}$}
\Text(-5,72)[r]{$\small{\sigma}$}
\DashCArc(45,40)(15,0,360){4}
\Text(45,60)[b]{$\small{\sigma}$}
\Text(45,20)[t]{$\small{\sigma}$}
\Vertex(30,40){2}
\Vertex(60,40){2}
\DashLine(60,40)(90,12){4}
\DashLine(60,40)(90,68){4}
\Text(97,8)[l]{$\small{\phi^-}$}
\Text(97,72)[l]{$\small{\phi^+}$}
\end{picture}
\qquad
\begin{picture}(115,80)(-10,0)
\DashLine(0,12)(30,40){4}
\DashLine(0,68)(30,40){4}
\Text(-5,8)[r]{$\small{\sigma}$}
\Text(-5,72)[r]{$\small{\sigma}$}
\DashCArc(45,40)(15,0,360){4}
\Text(45,60)[b]{$\small{\phi,\chi,h_2}$}
\Text(45,20)[t]{$\small{\phi,\chi,h_2}$}
\Vertex(30,40){2}
\Vertex(60,40){2}
\DashLine(60,40)(90,12){4}
\DashLine(60,40)(90,68){4}
\Text(97,8)[l]{$\small{\phi^-}$}
\Text(97,72)[l]{$\small{\phi^+}$}
\end{picture}
\qquad
\begin{picture}(115,80)(-10,0)
\DashLine(0,12)(30,40){4}
\DashLine(0,68)(30,40){4}
\Text(-5,8)[r]{$\small{\sigma}$}
\Text(-5,72)[r]{$\small{\sigma}$}
\DashLine(30,40)(60,25){4}
\DashLine(30,40)(60,55){4}
\Photon(60,25)(60,55){2}{4}
\Text(45,60)[b]{$\small{\phi,\chi,h_2}$}
\Text(45,20)[t]{$\small{\phi,\chi,h_2}$}
\Text(73,42)[t]{$\small{V}$}
\Vertex(30,40){2}
\Vertex(60,25){2}
\Vertex(60,55){2}
\DashLine(60,25)(90,12){4}
\DashLine(60,55)(90,68){4}
\Text(97,8)[l]{$\small{\phi^-}$}
\Text(97,72)[l]{$\small{\phi^+}$}
\end{picture}
}

%% file: diagrams/diagramsDecay_LO.tex
\scalebox{1.8}{
\begin{picture}(55,50)(0,0)
\DashLine(0,25)(17.5,25){2}
\Text(-3,25)[r]{\scalebox{.56}{$\small{\Fh}$}}
\Photon(17.5,25)(30,40){1}{4}
\Text(21,33)[rb]{\scalebox{.56}{$\small{\FW}$}}
\Photon(30,10)(17.5,25){1}{4}
\Text(21,17)[rt]{\scalebox{.56}{$\small{\FW}$}}
\Vertex(17.5,25){1}
\ArrowLine(50,30)(30,40)
\ArrowLine(30,40)(50,50)
\Text(52,30)[l]{\scalebox{.56}{$\small{\bar{\Ff}}$}}
\Text(52,50)[l]{\scalebox{.56}{$\small{\Ff^\prime}$}}
\Vertex(30,40){1}
\ArrowLine(50,0)(30,10)
\ArrowLine(30,10)(50,20)
\Text(52,0)[l]{\scalebox{.56}{$\small{\bar{\FF}}$}}
\Text(52,20)[l]{\scalebox{.56}{$\small{\FF^\prime}$}}
\Vertex(30,10){1}
\end{picture}
}
\qquad \,
\scalebox{1.8}{
\begin{picture}(55,50)(0,0)
\DashLine(0,25)(17.5,25){2}
\Text(-3,25)[r]{\scalebox{.56}{$\small{\Fh}$}}
\Photon(17.5,25)(30,40){1}{4}
\Text(21,33)[rb]{\scalebox{.56}{$\small{\FZ}$}}
\Photon(30,10)(17.5,25){1}{4}
\Text(21,17)[rt]{\scalebox{.56}{$\small{\FZ}$}}
\Vertex(17.5,25){1}
\ArrowLine(50,30)(30,40)
\ArrowLine(30,40)(50,50)
\Text(52,30)[l]{\scalebox{.56}{$\small{\bar{\Ff}}$}}
\Text(52,50)[l]{\scalebox{.56}{$\small{\Ff}$}}
\Vertex(30,40){1}
\ArrowLine(50,0)(30,10)
\ArrowLine(30,10)(50,20)
\Text(52,0)[l]{\scalebox{.56}{$\small{\bar{\FF}}$}}
\Text(52,20)[l]{\scalebox{.56}{$\small{\FF}$}}
\Vertex(30,10){1}
\end{picture}
}

%% file: diagrams/diagramsDecay_NLO.tex
\begin{picture}(70,50)(0,0)
\DashLine(0,25)(15,25){2}
\Vertex(15,25){1}
\Photon(15,25)(25,30){1}{2}
\Vertex(25,30){1}
\DashCArc(30,32.5)(5.59,35,215){2}
\DashCArc(30,32.5)(5.59,215,35){2}
\Vertex(35,35){1}
\Photon(35,35)(45,40){1}{2}
\Photon(45,10)(15,25){1}{6}
\ArrowLine(70,30)(45,40)
\ArrowLine(45,40)(70,50)
\Vertex(45,40){1}
\ArrowLine(70,0)(45,10)
\ArrowLine(45,10)(70,20)
\Vertex(45,10){1}
\end{picture}
\quad
\begin{picture}(70,50)(0,0)
\DashLine(0,25)(15,25){2}
\Vertex(15,25){1}
\Photon(15,25)(25,30){1}{2}
\Vertex(25,30){1}
\DashCArc(30,32.5)(5.59,35,215){2}
\PhotonArc(30,32.5)(5.59,215,35){1}{3}
\Vertex(35,35){1}
\Photon(35,35)(45,40){1}{2}
\Photon(45,10)(15,25){1}{6}
\ArrowLine(70,30)(45,40)
\ArrowLine(45,40)(70,50)
\Vertex(45,40){1}
\ArrowLine(70,0)(45,10)
\ArrowLine(45,10)(70,20)
\Vertex(45,10){1}
\end{picture}
\quad
\begin{picture}(70,50)(0,0)
\DashLine(0,25)(15,25){2}
\Vertex(15,25){1}
\Photon(15,25)(30,32.5){1}{3}
\Photon(30,32.5)(45,40){1}{3}
\Vertex(30,32.5){1}
\DashCArc(27.5,37.5)(5.59,0,360){2}
\Photon(45,10)(15,25){1}{6}
\ArrowLine(70,30)(45,40)
\ArrowLine(45,40)(70,50)
\Vertex(45,40){1}
\ArrowLine(70,0)(45,10)
\ArrowLine(45,10)(70,20)
\Vertex(45,10){1}
\end{picture}
\quad
\begin{picture}(70,50)(0,0)
\DashLine(0,25)(15,25){2}
\Vertex(15,25){1}
\DashLine(30,40)(30,10){2}
\DashLine(30,10)(15,25){2}
\DashLine(15,25)(30,40){2}
\Vertex(30,40){1}
\Vertex(30,10){1}
\Photon(30,40)(45,40){1}{3}
\ArrowLine(70,30)(45,40)
\ArrowLine(45,40)(70,50)
\Vertex(45,40){1}
\Photon(30,10)(45,10){1}{3}
\ArrowLine(70,0)(45,10)
\ArrowLine(45,10)(70,20)
\Vertex(45,10){1}
\end{picture}
\quad
\begin{picture}(70,50)(0,0)
\DashLine(0,25)(15,25){2}
\Vertex(15,25){1}
\Photon(30,40)(30,10){1}{5}
\DashLine(30,10)(15,25){2}
\DashLine(15,25)(30,40){2}
\Vertex(30,40){1}
\Vertex(30,10){1}
\Photon(30,40)(45,40){1}{3}
\ArrowLine(70,30)(45,40)
\ArrowLine(45,40)(70,50)
\Vertex(45,40){1}
\Photon(30,10)(45,10){1}{3}
\ArrowLine(70,0)(45,10)
\ArrowLine(45,10)(70,20)
\Vertex(45,10){1}
\end{picture}
\\
\vspace{15pt}
\begin{picture}(70,50)(0,0)
\DashLine(0,25)(15,25){2}
\Vertex(15,25){1}
\DashLine(30,40)(30,10){2}
\DashLine(30,10)(15,25){2}
\Photon(15,25)(30,40){1}{4}
\Vertex(30,40){1}
\Vertex(30,10){1}
\Photon(30,40)(45,40){1}{3}
\ArrowLine(70,30)(45,40)
\ArrowLine(45,40)(70,50)
\Vertex(45,40){1}
\Photon(30,10)(45,10){1}{3}
\ArrowLine(70,0)(45,10)
\ArrowLine(45,10)(70,20)
\Vertex(45,10){1}
\end{picture}
\quad
\begin{picture}(70,50)(0,0)
\DashLine(0,25)(15,25){2}
\Vertex(15,25){1}
\DashLine(30,40)(30,10){2}
\Photon(30,10)(15,25){1}{4}
\DashLine(15,25)(30,40){2}
\Vertex(30,40){1}
\Vertex(30,10){1}
\Photon(30,40)(45,40){1}{3}
\ArrowLine(70,30)(45,40)
\ArrowLine(45,40)(70,50)
\Vertex(45,40){1}
\Photon(30,10)(45,10){1}{3}
\ArrowLine(70,0)(45,10)
\ArrowLine(45,10)(70,20)
\Vertex(45,10){1}
\end{picture}
\quad
\begin{picture}(70,50)(0,0)
\DashLine(0,25)(15,25){2}
\Vertex(15,25){1}
\DashLine(30,40)(30,10){2}
\Photon(30,10)(15,25){1}{4}
\Photon(15,25)(30,40){1}{4}
\Vertex(30,40){1}
\Vertex(30,10){1}
\Photon(30,40)(45,40){1}{3}
\ArrowLine(70,30)(45,40)
\ArrowLine(45,40)(70,50)
\Vertex(45,40){1}
\Photon(30,10)(45,10){1}{3}
\ArrowLine(70,0)(45,10)
\ArrowLine(45,10)(70,20)
\Vertex(45,10){1}
\end{picture}
\quad
\begin{picture}(70,50)(0,0)
\DashLine(0,25)(15,25){2}
\Vertex(15,25){1}
\Photon(30,40)(30,10){1}{5}
\DashLine(30,10)(15,25){2}
\Photon(15,25)(30,40){1}{4}
\Vertex(30,40){1}
\Vertex(30,10){1}
\Photon(30,40)(45,40){1}{3}
\ArrowLine(70,30)(45,40)
\ArrowLine(45,40)(70,50)
\Vertex(45,40){1}
\Photon(30,10)(45,10){1}{3}
\ArrowLine(70,0)(45,10)
\ArrowLine(45,10)(70,20)
\Vertex(45,10){1}
\end{picture}
\quad
\begin{picture}(70,50)(0,0)
\DashLine(0,25)(15,25){2}
\Vertex(15,25){1}
\Photon(30,40)(30,10){1}{5}
\Photon(30,10)(15,25){1}{4}
\DashLine(15,25)(30,40){2}
\Vertex(30,40){1}
\Vertex(30,10){1}
\Photon(30,40)(45,40){1}{3}
\ArrowLine(70,30)(45,40)
\ArrowLine(45,40)(70,50)
\Vertex(45,40){1}
\Photon(30,10)(45,10){1}{3}
\ArrowLine(70,0)(45,10)
\ArrowLine(45,10)(70,20)
\Vertex(45,10){1}
\end{picture}
\\
\vspace{15pt}
\begin{picture}(70,50)(0,0)
\DashLine(0,25)(15,25){2}
\Vertex(15,25){1}
\DashCArc(25,25)(10,0,360){2}
\Photon(35,25)(50,40){-1}{4.5}
\Photon(50,10)(35,25){1}{4.5}
\Vertex(35,25){1}
\ArrowLine(50,40)(70,50)
\ArrowLine(70,30)(50,40)
\Vertex(50,40){1}
\ArrowLine(50,10)(70,20)
\ArrowLine(70,0)(50,10)
\Vertex(50,10){1}
\end{picture}
\quad
\begin{picture}(70,50)(0,0)
\DashLine(0,25)(25,25){2}
\Photon(25,25)(45,10){1}{6}
\Photon(35,35)(45,40){1}{3}
\Vertex(25,25){1}
\Vertex(35,35){1}
\DashCArc(30,30)(7.07,225,45){2}
\PhotonArc(30,30)(7.07,45,225){1}{4}
\ArrowLine(70,30)(45,40)
\ArrowLine(45,40)(70,50)
\Vertex(45,40){1}
\ArrowLine(70,0)(45,10)
\ArrowLine(45,10)(70,20)
\Vertex(45,10){1}
\end{picture}
\quad
\begin{picture}(70,50)(0,0)
\DashLine(0,25)(25,25){2}
\Photon(35,15)(45,10){1}{3}
\Photon(25,25)(45,40){1}{6}
\Vertex(25,25){1}
\Vertex(35,15){1}
\DashCArc(30,20)(7.07,315,135){2}
\PhotonArc(30,20)(7.07,135,315){1}{4}
\ArrowLine(70,30)(45,40)
\ArrowLine(45,40)(70,50)
\Vertex(45,40){1}
\ArrowLine(70,0)(45,10)
\ArrowLine(45,10)(70,20)
\Vertex(45,10){1}
\end{picture}

%% file: diagrams/diagramsDecay_NLOreal.tex
\scalebox{1.8}{
\begin{picture}(55,50)(0,0)
\DashLine(0,25)(17.5,25){2}
\Text(-3,25)[r]{\scalebox{.56}{$\small{\Fh}$}}
\Photon(17.5,25)(30,40){1}{4}
\Text(21,33)[rb]{\scalebox{.56}{$\small{\FV}$}}
\Photon(30,10)(17.5,25){1}{4}
\Text(21,17)[rt]{\scalebox{.56}{$\small{\FV}$}}
\Vertex(17.5,25){1}
\ArrowLine(50,30)(40,35)
\ArrowLine(40,35)(30,40)
\ArrowLine(30,40)(50,50)
\Photon(40,35)(50,40){1}{2.5}
\Vertex(30,40){1}
\Vertex(40,35){1}
\ArrowLine(50,0)(30,10)
\ArrowLine(30,10)(50,20)
\Vertex(30,10){1}
\end{picture}
}
\qquad \,
\scalebox{1.8}{
\begin{picture}(55,50)(0,0)
\DashLine(0,25)(17.5,25){2}
\Text(-3,25)[r]{\scalebox{.56}{$\small{\Fh}$}}
\Photon(17.5,25)(30,40){1}{4}
\Text(21,33)[rb]{\scalebox{.56}{$\small{\FV}$}}
\Photon(30,10)(17.5,25){1}{4}
\Text(21,17)[rt]{\scalebox{.56}{$\small{\FV}$}}
\Vertex(17.5,25){1}
\ArrowLine(50,30)(40,35)
\ArrowLine(40,35)(30,40)
\ArrowLine(30,40)(50,50)
\Gluon(40,35)(50,40){1.5}{2.5}
\Vertex(30,40){1}
\Vertex(40,35){1}
\ArrowLine(50,0)(30,10)
\ArrowLine(30,10)(50,20)
\Vertex(30,10){1}
\end{picture}
}